# Saturn's icy satellites and rings investigated by Cassini - VIMS. III. Radial compositional variability

G. Filacchione[a,*], F. Capaccioni[a], M. Ciarniello[a], R. N. Clark[b], J. N. Cuzzi[c], P. D. Nicholson[d], D. P. Cruikshank[c], M.M. Hedman[d], B. J. Buratti[e], J. I. Lunine[d], L. A. Soderblom[f], F. Tosi[a], P. Cerroni[a], R. H. Brown[e], T. B. McCord[h], R. Jaumann[i], K. Stephan[i], K. H. Baines[e], E. Flamini[j]

[a]*INAF-IAPS, Istituto di Astrofisica e Planetologia Spaziali, Area di Ricerca di Tor Vergata, via del Fosso del Cavaliere, 100, 00133, Rome, Italy*
[b]*US Geological Survey, Federal Center, Denver, CO, 80228, USA*
[c]*NASA Ames Research Center, Moffett Field, CA 94035-1000, USA*
[d]*Cornell University, Astronomy Department, 418 Space Sciences Building, Ithaca, NY 14853, USA*
[e]*Jet Propulsion Laboratory, California Institute of Technology, 4800 Oak Groove Drive, Pasadena, CA 91109, USA*
[f]*US Geological Survey, Flagstaff Station, 2255 N. Gemini Drive, Flagstaff, AZ 86001, USA*
[g]*Lunar and Planetary Laboratory and Steward Observatory, University of Arizona, Tucson, AZ 85721, USA*
[h]*Bear Fight Center, 22 Fiddlers Rd., Winthrop, WA 98862, USA*
[i]*Institute for Planetary Exploration, DLR, Rutherfordstaße 2,12489, Berlin, Germany*
[j]*ASI, Agenzia Spaziale Italiana, viale Liegi 26, 00198, Rome, Italy*

**Abstract**

In the last few years Cassini-VIMS, the Visible and Infared Mapping Spectrometer, returned to us a comprehensive view of the Saturn's icy satellites and rings. After having analyzed the satellites' spectral properties (Filacchione et al. (2007a), paper I) and their distribution across the satellites' hemispheres (Filacchione et al. (2010), paper II), we proceed in this paper to investigate the radial variability of icy satellites (principal and minor) and main rings average spectral properties. This analysis is done by using 2,264 disk-integrated observations of the satellites and a $12 \times 700$ pixels-wide rings radial mosaic acquired with a spatial resolution of about 125 km/pixel. The comparative analysis of these data allows us to retrieve the amount of both water ice and red contaminant materials distributed across Saturn's system and the typical surface regolith grain sizes. These measurements highlight very striking differences in the population here analyzed, which vary from the almost uncontaminated and water ice-rich surfaces of Enceladus and Calypso to the metal/organic-rich and red surfaces of Iapetus' leading hemisphere and Phoebe. Rings spectra appear more red than the icy satellites in the visible range but show more intense 1.5-2.0 $\mu m$ band depths. Although their orbits are close to the F-ring, Prometheus and Pandora are different in surface composition: Prometheus in fact appears very water ice-rich but at the same time very red at VIS wavelengths. These properties make it very similar to A-B ring particles while Pandora is bluer. Moving outwards, we see the effects of E ring particles, generated by Enceladus plumes, which contaminate satellites surfaces from Mimas out to Rhea. We found some differences between Tethys lagrangian moons, Calypso being much more water ice-rich and bluer than Telesto. Among outer satellites (Hyperion, Iapetus and Phoebe) we observe a linear trend in both water ice decrease and in reddening, Hyperion being the reddest object of the population. The correlations among spectral slopes, band depths, visual albedo and phase permit us to cluster the saturnian population in different spectral classes which are detected not only among the principal satellites and rings but among co-orbital minor moons as well. These bodies are effectively the "connection" elements, both in term of composition and evolution, between the principal satellites and main rings. Finally, we have applied Hapke's theory to retrieve the best spectral fits to Saturn's inner regular satellites (from Mimas to Dione) using the same methodology applied previously for Rhea data discussed in Ciarniello et al. (2011).

---

[*]Corresponding author
  *Email address:* `gianrico.filacchione@iaps.inaf.it` (G. Filacchione)

# 1. Introduction

Among the external planets of the Solar System, Saturn has one of the most complex system of icy satellites and rings. These objects are among the main scientific objectives of the VIMS, Visual and Infrared Mapping Spectrometer, investigation aboard the NASA-ESA-JPL-ASI Cassini-Huygens mission (Matson et al., 2002). The VIMS experiment (Brown et al., 2004), provided by US, Italian and French teams, performs imaging spectroscopy by using two separate optical channels: the Italian-made VIS channel operates in the 0.35-1.05 $\mu m$ region with a spectral sampling of $\Delta\lambda$=7.3 nm/band using 96 bands; the US-made IR channel covers the 0.8-5.1 $\mu m$ range with a spectral sampling of $\Delta\lambda$=16.6 nm/band using 256 bands. Both channels have a nominal field of view (FOV) of 1.86° (square) with an instantaneous field of view (IFOV) of 500 $\mu rad$ (square) which decreases to 166 $\mu rad$ (square, VIS channel) and to 250 × 500 $\mu rad$ (rectangular, IR channel) in high spatial resolution mode. The two channels are commanded and powered by a common US-made Data Processing and Power Distribution Unit which includes French-made data compressors.

Taking advantage of the fly-bys that occurred during the Cassini's nominal (2004-2008) and equinox (2008-2010) mission phases, VIMS has harvested a large number of surface-resolved observations of the satellites that led to the identification of crystalline water ice, in particular on Enceladus south pole "tiger stripes" features (Brown et al. (2006), Jaumann et al. (2008)); $CO_2$ and organics (CH and possibly CN) on Hyperion (Cruikshank et al. (2007), Cruikshank et al. (2010)), Iapetus (Cruikshank et al. (2008), Pinilla-Alonso et al. (2011), Clark et al. (2012)) and Phoebe (Clark et al. (2005), Buratti et al. (2008), Coradini et al. (2008)); the composition and distribution of the exogenic dark material dispersed on Dione's trailing hemisphere (Clark et al. (2008), Stephan et al. (2010)) and across Rhea (Stephan et al., 2012).

Many VIMS low spatial resolution data allowed us to retrieve disk-integrated spectra of the icy satellites that were used to 1) infer the spectral differences between the leading and trailing hemispheres of the regular satellites (Filacchione et al. (2007a), Filacchione et al. (2010)); 2) retrieve photometric parameters, including phase and light curves (Pitman et al. (2010), Ciarniello (2012)); 3) infer spectral compositions of minor satellites (Filacchione et al. (2006), Buratti et al. (2010a), Filacchione et al. (2010)); 4) determine Rhea's surface properties through Hapke (1993) modeling, including the size of water ice particles covering the surface, the amount of organic contaminants, the large scale surface roughness and the opposition effect surge (Ciarniello et al., 2011).

The composition of the main rings is inferred from specific observations taken at low solar phase angles and in regions where the contamination of Saturnshine is negligible. In this way it is possible to minimize the complications induced by multiple and forward scattering. The main rings are composed of 90-95% pure crystalline water ice particles (Cuzzi et al., 2009). At wavelengths <0.52 $\mu m$, rings show a strong spectral slope resulting in a red color pointing out to the presence of a non-icy component that must represent few percent by mass (Filacchione et al., 2007b). Any IR absorption features of $CO_2$, $CH_4$, $NH_3$ ices, silicates or C-H organics (i.e. tholins) are not detected on VIMS spectra (Cuzzi et al., 2010). Sub-micron size iron or hematite particles were suggested as possible contaminant mixed in water ice (Cuzzi et al. (2009), Clark et al. (2012)); a few percents of hematite mixed in water ice can reproduce the average rings optical properties showing a strong UV absorption without changing the 1-5 $\mu m$ spectral range. The presence of hematite is supported by the oxidation of nanophase iron particles by the oxygen atmosphere around the rings. The radial composition of the rings is correlated with local surface mass density and/or optical depth (Nicholson et al., 2008) and is retrieved from radial variations of ring color, particle albedo, and water ice band depths: in general C ring and Cassini Division appear more contaminated by non-icy material than the A and B rings (Cuzzi and Estrada , 1998). The degree of visual redness is highly correlated with the water ice band strengths, suggesting that the UV absorber is intimately mixed within the ice grains (Filacchione et al. (2007b), Cuzzi et al. (2009), Hedman et al. (submitted)). Finally, D'Aversa et al. (2010) have applied a radiative transfer code to model VIMS observations of the spokes that occurred in 2008 on the B ring.

In general, the icy surfaces are susceptible to alterations induced by exogenic processes (Schenk et al., 2011); among these, important roles are played by: 1) the accretion of ring particles on inner minor satellites (Charnoz et al., 2007); 2) the irradiation by charged particles trapped in radiation belts (Krupp et al., 2009); 3) the contamination of E ring particles, continuously fed by Enceladus' plumes, affects a region spanning from the orbit of Mimas out to that of Rhea (Showalter et al., 1991); 4) the flux of dark material (Clark et al., 2008), falling mostly on the outermost icy satellites, that moves inward from Phoebe and/or from other retrograde irregular satellites (Tosi et al., 2010) and is supported by the existence of the dusty Phoebe's ring (Verbiscer et al., 2009). All these processes contribute to the evolution of the chemical composition and physical properties of the Saturnian satellites and rings.

The main aim of this work is to trace the spectral variability of the very different objects populating the Saturn's system using specific spectrophotometric quantities (visible spectral slopes and water ice infrared band depths) to infer the amount of surface water ice and contaminants for each object orbiting in the Saturnian system, from the inner C ring at 1.13 $R_S$ (Saturn's radii) up to Phoebe at $\approx 215\ R_S$.

In section 2 we describe the method used to select the different VIMS datasets for satellites and rings, the data processing and calibration techniques, and the definition of the indicators applied to the VIS-IR spectra. Two subsections, one for the VIS range and one for the IR range, are devoted to explain in detail how to correlate the variation of the spectral indicators with the different water ice grain sizes. The analysis of the variations of spectral indicators across the Saturnian system is the subject of section 3, where, for each satellite and ring region, we report a description of the VIMS results. Although we are considering only disk-integrated observations, the large statistical dataset processed allows us to monitor very different aspects: hemispherical differences, illumination geometry variations (like the opposition surge) and exogenic processes, e.g., interactions with E ring and magnetospheric particles, are described for each object. Sections 4-5 contain, respectively, an update of the classification of VIS spectral slopes and IR water ice band depths for the entire population of objects here considered. In section 6 we cross-compare VIS and IR indicators in order to define the different spectral classes. Section 7 is completely devoted to the discussion of spectral modeling of Saturn's inner satellites (Mimas to Dione) adopting Hapke's modeling. For this analysis we have used the same method recently applied by Ciarniello et al. (2011) to infer the spectrophotometric parameters of Rhea. In particular we present here the best fit solutions to VIMS disk-integrated spectra in terms of water ice and contaminants fraction, grain size and their radial trends.

## 2. VIMS spectra, data processing and spectral indicators

### 2.1. Saturn's rings and satellites disk-integrated VIS-IR spectra

In this work we have analyzed a statistically significant dataset consisting of 2,264 disk-integrated observations of the regular (Mimas, Enceladus, Tethys, Dione, Rhea, Hyperion and Iapetus) and minor (Atlas, Prometheus, Pandora, Janus, Epimetheus, Telesto, Calypso, Helene, Phoebe) satellites corresponding to 147,300 VIS and 184,459 IR single-point spectra. Table 1 contains a summary of the processed observations for each satellite, including the number of resulting VIS and IR spectra.

With respect to the previous analysis (Filacchione et al., 2010), we have enlarged the dataset to observations taken during the Cassini Equinox mission until June 2010 adding: 1) more than 800 observations of the icy satellites, including 4 for Helene which was not considered in our previous works; 2) a radial mosaic of the main rings spanning across C, B, Cassini division (CD) and A rings imaged on both east and west ansae; this mosaic, built from 34 VIMS cubes and composed of 8,400 spectra, has an average spatial resolution of about 125 km/pixel.

The selection of the icy satellites observations is done with the same method previously explained, e.g., by considering only data-cubes showing complete coverage of a satellite's disk in both VIS-IR channels and with an unsaturated signal across the entire spectral range. Because VIMS observations are carried out from very different distances and solar phase angles, the resulting satellites' images span from one to about 32 pixels in radius. At the lowest spatial scale the images are influenced by the instrumental point spread function; at the largest scale the images are limited by the instrumental FOV (64 pixels or 32 mrad in nominal mode). Moreover, since VIMS-V and IR channels can operate with different spatial resolutions, in many cases the same image is acquired in high spatial resolution mode with one channel and in nominal mode with the other. For this reason, the number of spectra processed for each satellite, as reported in Table 1, can be different between the two channels.

### 2.2. Data calibration and processing

VIMS reflectance data are stored in 3D hyperspectral cubes coded in BIL format (Band Interleaved by Lines) with size (sample, bands, line)=(s, b, l). Samples and lines are variable in the 1-64 range while bands identify the 352 spectral channels (1-96 for the VIS range, 97-352 for the IR) (Brown et al., 2004).

The final spectrum is built from VIS wavelengths $\lambda(b_{VIS} = 1) = 0.350\ \mu m$ to $\lambda(b_{VIS} = 87) = 0.981\ \mu m$ and IR wavelengths $\lambda(b_{IR} = 7) = 0.983\ \mu m$ to $\lambda(b_{IR} = 256) = 5.125\ \mu m$. The average spectra coming from the two VIMS channels overlap around 0.981- 0.983 $\mu m$ with the IR channel spectra scaled to the same reflectance level as the VIS.

| Target | $R_{Saturn}$ | # observations | # VIS spectra | # IR spectra |
|---|---|---|---|---|
| Rings | 1.24-2.27 | 34 | 8,400 | 8,400 |
| (S15) Atlas | 2.28 | 1 | 4 | 4 |
| (S16) Prometheus | 2.31 | 9 | 94 | 388 |
| (S17) Pandora | 2.35 | 3 | 29 | 26 |
| (S10) Janus | 2.51 | 98 | 1,080 | 1,236 |
| (S11) Epimetheus | 2.51 | 1 | 6 | 4 |
| (S1) Mimas | 3.08 | 265 | 8,302 | 10,190 |
| (S2) Enceladus | 3.95 | 298 | 13,862 | 17,599 |
| (S3) Tethys | 4.88 | 278 | 17,456 | 26,109 |
| (S13) Telesto | 4.88 | 2 | 48 | 28 |
| (S14) Calypso | 4.88 | 11 | 87 | 124 |
| (S4) Dione | 6.26 | 268 | 17,081 | 21,185 |
| (S12) Helene | 6.26 | 4 | 17 | 50 |
| (S5) Rhea | 8.74 | 511 | 53,140 | 77,962 |
| (S7) Hyperion | 24.55 | 207 | 11,592 | 8,389 |
| (S8) Iapetus | 59.03 | 298 | 22,292 | 17,478 |
| (S9) Phoebe | 214.69 | 10 | 2,280 | 3,687 |
| TOTAL | | 2,298 | 155,700 | 192,859 |

Table 1: Summary of Saturn's rings and satellites disk-integrated observations processed in this work.

The calibration of VIMS data is described in several papers (Brown et al. (2003), Brown et al. (2004), McCord et al. (2004)); VIMS-VIS channel raw data are calibrated in reflectance by using the method discussed in Filacchione et al. (2006). The IR channel data are calibrated by using the most recent RC17 transfer function (Clark et al., 2012) applying an improved spectral calibration (Cruikshank et al., 2010).

The reflectance of the icy satellites and rings is obtained starting from instrumental raw data number (*DN*) and integration time ($t_{exp}$, in seconds) through the following equation:

$$\frac{I}{F}(s, \lambda, l) = \frac{RC(s, b, l) \cdot \pi \cdot DN(s, b, l) \cdot D_{\odot}^2}{t_{exp} \cdot SI(\lambda)} \quad (1)$$

where RC(s,b,l) is the instrumental transfer function for sample=s, band=b and line=l; $D_{\odot}$ is the distance of the target form the Sun measured in AU and $SI(\lambda)$ is the solar irradiance measured at 1 AU (Thekekara, 1973). During the Cassini mission $D_{\odot}$ varied from 9.04 AU in 2004 to 9.52 AU in 2010. The conversion from bands to wavelengths, $\lambda(b)$, is given by the spectral calibration table. For VIMS, the flat field corresponds to a 2D-array for the VIS channel (along sample and bands directions, constant for each line) while it is a 3D-array for the IR (along samples, bands and lines directions): both are included in the responsivity RC in the previous eq. 1.

As explained in Filacchione et al. (2007a), VIS channel data are further corrected for readout noise and dark current by subtracting the sky level signal measured at a sufficient distance from the satellite's disk. On the IR channel spectra we have applied a linear interpolation on the three intervals corresponding to the order sorting filters gaps (between bands $44 < b < 49$, $125 < b < 129$ and $178 < b < 182$). In some cases it becomes necessary to correct some hot IR pixels, in particular at wavelengths 1.22 and 4.74 μm. The resulting VIS-IR spectra are processed with a standard despiking filter to remove high frequency noise and negative counts. After having selected all illuminated pixels on the satellite's disk, the average reflectance is computed for each observation. In such a way it is possible to boost the signal-to-noise ratio (SNR) on the resulting spectra.

As far as the geometrical reconstruction of the observations is concerned, we have retrieved the latitude and longitude for both sub-spacecraft and sub-solar points, the solar phase angle and the Cassini-satellite distance by using the on-line ephemeris calculator (Cassini/Saturn Viewer 1.6[1]). Due to its length, the list of the 2,298 disk-integrated observations processed with the relative geometric parameters is included as online supplemental material.

---
[1]Tool by M. Showalter, available on-line at *http : //pds − rings.seti.org/tools/viewer2_satc.html*.

The spectral variability of the saturnian satellites is evident when the average disk-integrated spectra are compared among them. Figures 1-2 contain stack plots of the average spectra for the principal and minor satellites. These averages are computed after having normalized at 0.55 $\mu m$ each single disk-integrated observation. Rings average VIS-IR spectra, taken across uniform radial regions of A, B, C rings and Cassini division, are shown in Fig. 3. We discuss the rings observations and data processing method in the next section 3.9.

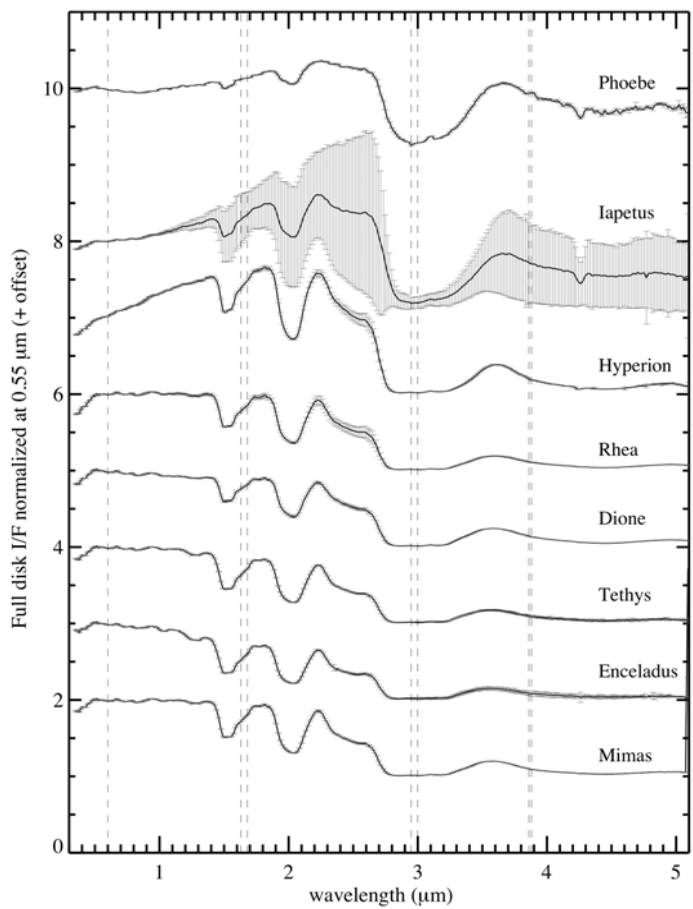

Figure 1: Average disk-integrated reflectance spectra ($\pm 1\sigma$) of the principal satellites and Phoebe. Spectra are normalized at 0.55 $\mu m$ and stacked with an offset. Vertical lines indicate order sorting filters gaps (one in the VIS and three in the IR range).

*2.3. A summary of icy satellites and rings spectral characteristics as observed by VIMS*

Icy satellite spectra (see Fig.1-2) show the following spectral characteristics:

- a general positive (or red) slope in the 0.35-0.55 $\mu m$ range caused by the presence of contaminants and chromophores dispersed on the surfaces;

- a more variable slope between 0.55-0.95 $\mu m$ ($< 0$ for Enceladus, $\approx 0$ for Mimas, Tethys and Rhea, $> 0$ for Iapetus and Hyperion) is indicative of both water ice grain size and amount of contaminants;

- in some observations a 0.55 $\mu m$ peak is evident; this could be an indication of Rayleigh scattering caused by iron or carbon nanophase particles embedded in water ice (Clark et al. (2008), Clark et al. (2012));

- a very steep, red slope in the VIS up to the 1.5 $\mu m$ water ice band on Hyperion and Iapetus;

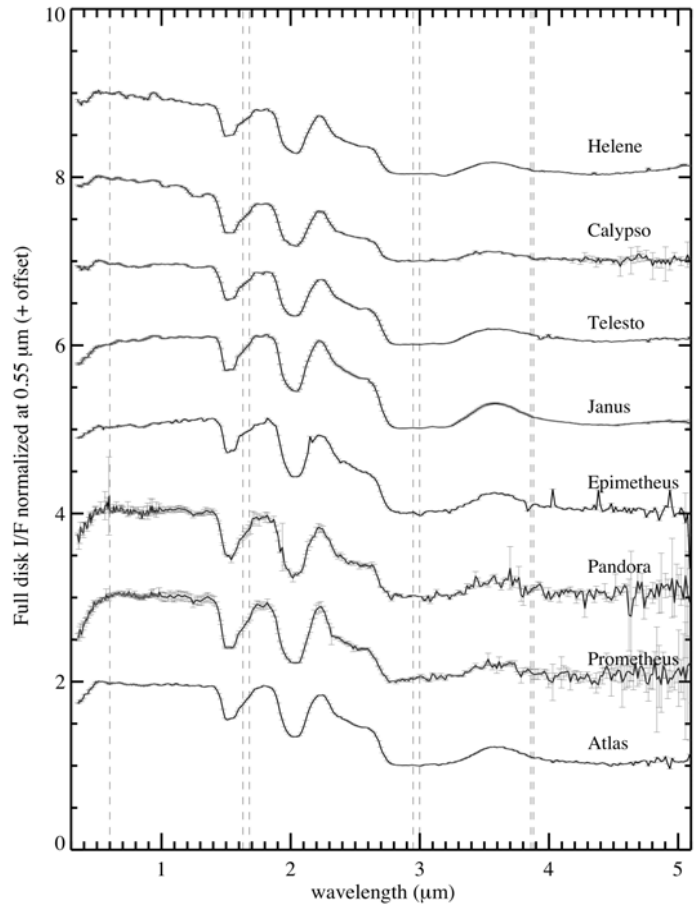

Figure 2: Average disk-integrated reflectance spectra (±1$\sigma$) of the minor satellites. Spectra are normalized at 0.55 $\mu m$ and stacked with an offset. Vertical lines indicate order sorting filters gaps (one in the VIS and three in the IR range).

- well defined, despite faint, water ice absorption bands at 1.05 - 1.25 $\mu m$ are detected on Mimas, Enceladus, Tethys, Calypso and Rhea; sometimes calibration artifacts persist in this spectral region;

- sharp water ice band at 1.5, 2.05, 3.0 $\mu m$ are seen on every satellite. The presence of the first IR order-sorting filter around 1.65 $\mu m$ partially precludes the detection of the secondary crystalline water ice band. A discussion about the discrimination between amorphous vs. crystalline phase is included in Filacchione et al. (2010).

- a variable slope in the 1.11-2.25 $\mu m$ range is an indicator of the water ice grain size and amount of contaminants;

- in absence of contaminants, the 3.6 $\mu m$ peak is an indicator of the water ice grain size (higher peak corresponds to smaller grains);

- the 4.26 $\mu m$ $CO_2$ ice and 3.1 $\mu m$ water ice Fresnel peak are evident mainly on the three outermost satellites Hyperion, Iapetus and Phoebe.

Saturn's rings (Fig. 3) are characterized by the following spectral properties:

- A strong reddening, caused by a decrease of the I/F in the 0.35-0.52 $\mu m$ region, is an indication of the presence of contaminant/darkening material mixed in the water ice particles. This effect is stronger across the regions having higher optical depth, e.g., A and B rings, while is less strong across C ring and Cassini division.

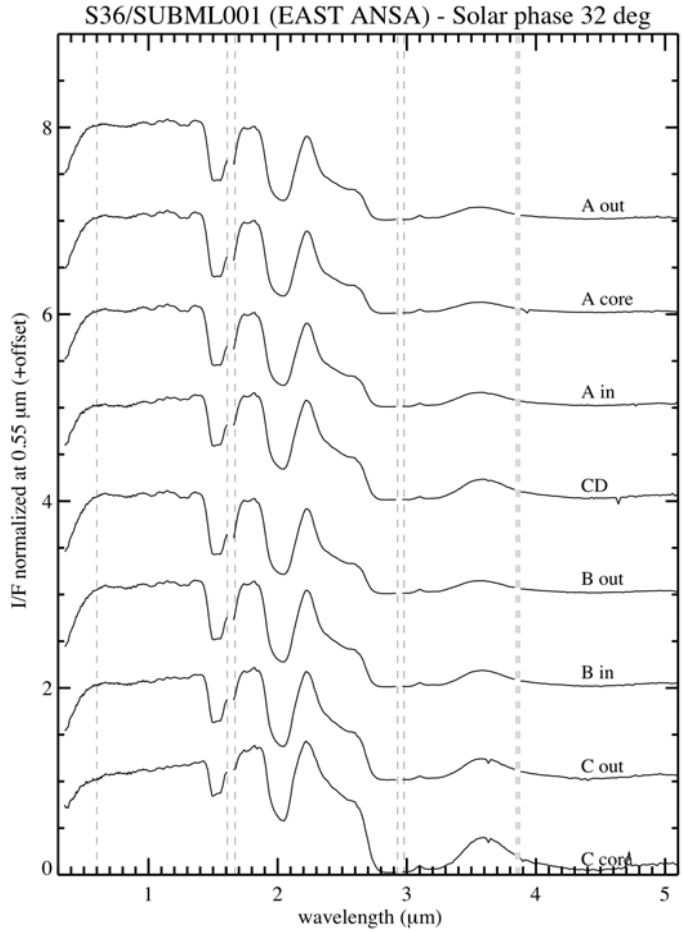

Figure 3: Rings average VIS-IR spectra taken across uniform radial regions of the A, B, C rings and Cassini division. Vertical lines indicate the instrumental order sorting filters.

- The water ice absorption bands at 1.05, 1.25, 1.5, 2.05, 3.0 $\mu m$ are the most intense measured in the saturnian system;

- the 3.1 $\mu m$ Fresnel peak is resolved across the entire rings system.

- On this VIMS dataset (and on many other taken at similar spatial resolution) we cannot discriminate any other absorption features which might help to decipher the nature of contaminant/darkening agent.

Since we are considering a wide population of objects having very different properties, it becomes necessary to use a standard processing of the dataset to retrieve specific spectral indicators, or markers, for each spectrum of the satellites and rings. In this way, rather than analyzing the 352 spectral variables (e.g. the VIMS channels), we reduce the number of parameters through an accurate selection of physical parameters, like spectral slopes and water ice band depths.

2.4. Water ice: spectral properties in the visible range

Pure water ice frost seen at visible wavelengths has high reflectance (about 1 at 0.5 $\mu m$) with a slight positive (or red) $S_{0.35-0.55}$ slope and negative (or blue) $S_{0.55-0.95}$ slope (see Fig. 4; measurements by Jack Salisbury at JHU, with data available on-line via anonymous ftp at rocky.eps.jhu.edu).

*Visible spectral slopes* $S_{ij}$ $[\mu m^{-1}]$ are defined as the best fit linear trend to the spectra computed according to the following equations (Cuzzi et al. (2009), Filacchione et al. (2010)):

$$S_{0.35-0.55} = \frac{I/F_{0.55} - I/F_{0.35}}{0.2\,\mu m \cdot I/F_{0.55}} \quad (2)$$

$$S_{0.55-0.95} = \frac{I/F_{0.95} - I/F_{0.55}}{0.4\,\mu m \cdot I/F_{0.55}} \quad (3)$$

In order to remove possible dependence of the color on albedo and solar phase, the slopes are measured after the spectra are normalized at 0.55 $\mu m$.

Increasing the regolith grain size from fine snow to coarse, three different effects are seen: 1) the max reflectance decreases by about 3 % at 0.5 $\mu m$; 2) the reddening of the $S_{0.35-0.55}$ slope increases; 3) the blueing of the $S_{0.55-0.95}$ increases. The wavelength at which the two sloped lines intersect, or $\lambda_{crossing}$, remains almost identical for the different grains, e.g. between 0.51-0.52 $\mu m$.

Therefore, if the composition were pure water ice, the two visible spectral slopes could be used as markers to retrieve the regolith grain size (Table 2). However at these wavelengths water ice is extremely sensitive to any amount of contaminants and chromophores distributed or mixed within ice: very small amounts (< 1% in intraparticle and intramolecular mixing) of these substances can introduce a significant reddening especially in the 0.35-0.55 $\mu m$ range. With different strengths, as discussed later, this effect is seen across the entire saturnian system.

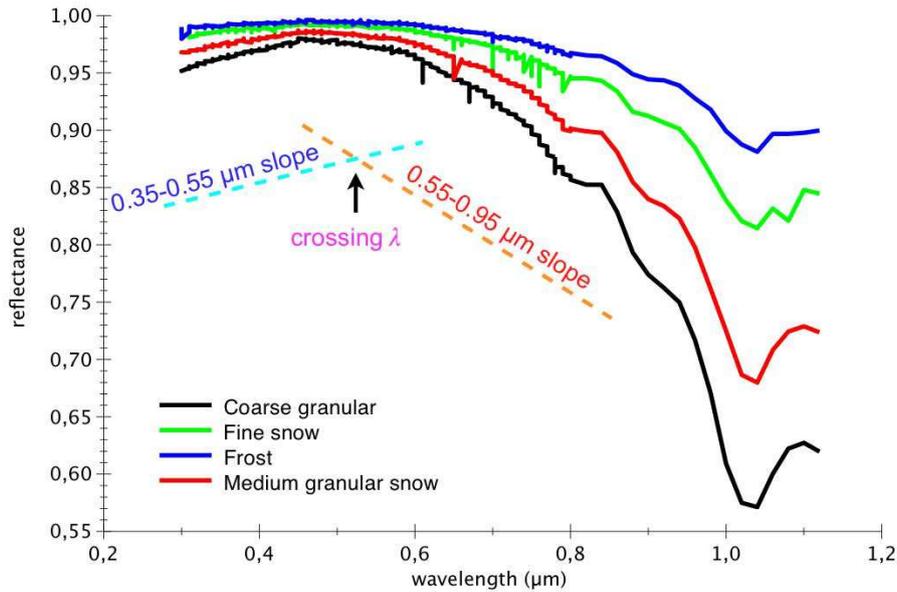

Figure 4: Water ice reflectance spectra in the visible range for different grain sizes. A schematics of two slopes ($S_{0.35-0.55}$, $S_{0.55-0.95}$) and the resulting $\lambda_{crossing}$ are indicated.

| Grain size | $S_{0.35-0.55}$ ($\mu m^{-1}$) | $S_{0.55-0.95}$ ($\mu m^{-1}$) | $\lambda_{crossing}$ ($\mu m$) |
|---|---|---|---|
| Coarse granular | 0.1428 | -0.3562 | 0.5157 |
| Medium granular | 0.0992 | -0.2581 | 0.5235 |
| Fine snow | 0.0550 | -0.1447 | 0.5193 |
| Frost | 0.0330 | -0.0874 | 0.5223 |

Table 2: Water ice spectral indicators in the visible range.

## 2.5. Water ice: spectral properties in the infrared range

Several vibrational combinations and overtones characterize pure water ice spectra in the infrared range. As a result, the most diagnostic absorption bands are those located at 1.05, 1.25, 1.5, 2.0 and 3.0 $\mu m$. By using lab-measured optical constants of crystalline water ice (Mastrapa et al. (2008), Mastrapa et al. (2009)) in a radiative transfer code, it is possible to retrieve synthetic spectra at different grain sizes. As an example, in Fig. 5 are shown synthetic spectra for 10 to 100 $\mu m$ grain diameters simulated at 40° phase angle by using the Hapke method discussed in Ciarniello et al. (2011).

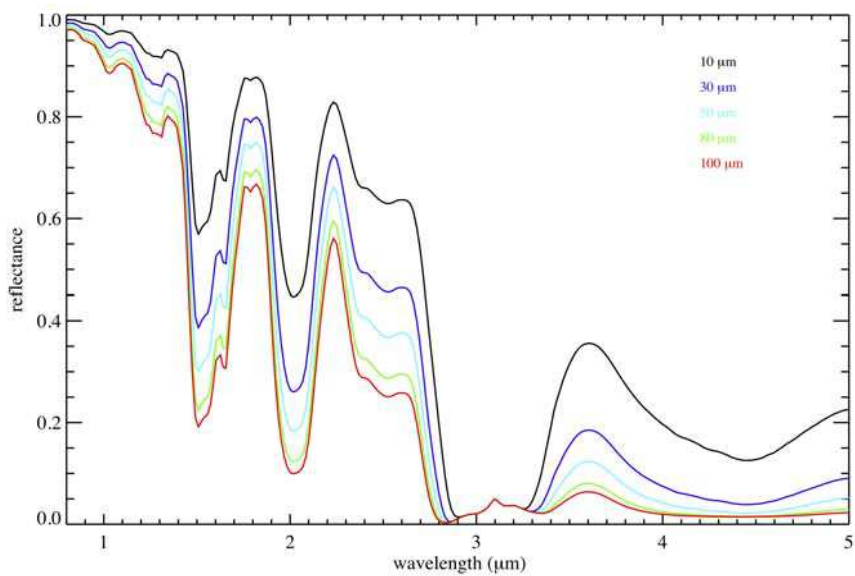

Figure 5: Synthetic spectra of water ice reflectance in the infrared range for different grain diameters (10 to 100 $\mu m$) computed for 40° phase angle. Spectra are sampled at VIMS spectral resolution.

In general, band depth (BD) is function of material composition, surface grain size and illumination phase (Hapke, 1993). For a fixed phase angle, small grains (10 $\mu m$) have high reflectance at continuum level but small band depth. On contrary, large grains (100 $\mu m$) are characterized by a lower reflectance and larger band depth.

*Water ice band depths* are defined in Clark et al. (1999) as:

$$BD_i = 1 - \frac{I/F_i}{I/F_{continuum}} \qquad (4)$$

The BD are calculated for the i=1.05, 1.25, 1.5 and 2.0 $\mu m$ bands; for each one, the local level of the continuum, $I/F_{continuum}$, is evaluated as the best fit line passing through the band's wings taken at the following wavelengths:

- b=85-86 (0.973-0.981 $\mu m$) and b=94-95 (1.098-1.114 $\mu m$) for the band at $\lambda$ = 1.05 $\mu m$;

- b=97-98 (1.147-1.164 $\mu m$) and b=110-111 (1.360-1.377 $\mu m$) for the band at $\lambda$ = 1.25 $\mu m$;

- b=111-112 (1.377-1.393 $\mu m$) and b=136-137 (1.787-1.804 $\mu m$) for the band at $\lambda$ = 1.5 $\mu m$;

- b=140-141 (1.853-1.869 $\mu m$) and b=163-164 (2.233-2.250 $\mu m$) for the band at $\lambda$ = 2.05 $\mu m$.

Increasing the grain size, or scattering length, we observe a gradual decrease of the continuum level joined to a corresponding increase of the band depths. This effect is caused by the longer optical path of the incident ray in the particle.

As Clark & Lucey (1984) have demonstrated, the 1.25, 1.5 and 2.0 $\mu m$ band depths can be synergistically used to retrieve the regolith grain diameter. In fact, as shown in Figure 5, these bands have a characteristic behavior when measured for different grain sizes. In the 1.5-2.0 μm band depth space we observe a linear increase of the band depths as the grain size increases from 1 μm to about 0.1 cm. For larger grains the band depths start to decrease moving towards minima for centimeter-size grains. The two branches overlap for BD(1.5 μm)≈0.6, BD(2.0 μm)≈0.67 for grain diameters of ≈50 μm and 0.5 cm. This "degeneracy" is removed if we consider the 1.5-1.25 μm band depth space where the trend becomes almost circular: while the BD(1.5 μm) increases from 1 μm to 0.1 cm and then decreases from 0.1 cm to centimeter-size particles, the BD(1.25 μm) increases linearly from 1 μm to about 5 cm size particles. The combination of these band depths is used later to retrieve the regolith grain sizes on the icy satellites and ring particles.

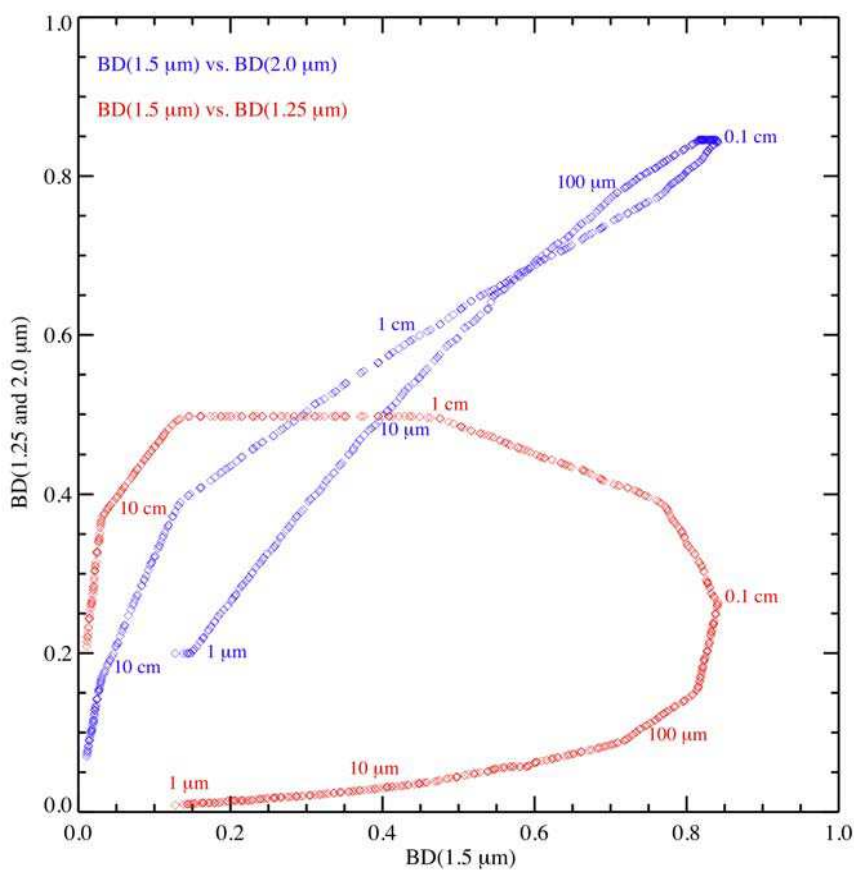

Figure 6: Variations of the water ice band depths (1.5 vs 1.25 and 1.5 vs. 2.0 μm) as function of the grain diameter. Reference grain diameters (from 1 μm to 1 cm) are indicated. Data inferred from Clark & Lucey (1984).

Two additional spectral indicators, both correlated with the grain size, can be considered for pure water ice. The first is the $1.11 - 2.25$ μm infrared slope, retrieved through a linear fit passing through the continuum points at 1.113, 1.359, 1.820, 2.249 μm on spectra normalized at 0.55 μm. This slope can be used as an indicator for small ($< 100$ μm) grains. As reported in Table 3, the slope decreases monotonically from 10 up to 80 μm, increasing again starting from 100 μm grains.

The second indicator is the ratio of the reflectances at 3.6/1.82 μm after having normalized the spectra at 0.55 μm. As shown in Fig. 5, the continuum level at 3.6 μm is strongly influenced by the grain size, being higher for small grains. The ratio, tabulated in Table 3, decreases from small to larger grains.

| Grain diameter ($\mu m$) | $S_{1.11-2.25}$ ($\mu m^{-1}$) | $\frac{I/F(3.6)}{I/F(1.82)}$ |
|---|---|---|
| 10 | -0.125 | 0.406 |
| 30 | -0.191 | 0.232 |
| 50 | -0.225 | 0.164 |
| 80 | -0.257 | 0.116 |
| 100 | -0.191 | 0.096 |

Table 3: Water ice spectral indicators in the infrared range for different grain diameters: spectral slope $S_{1.11-2.25}$ and continuum ratio $I/F(3.6)/I/F(1.82)$. Values retrieved from spectra shown in Fig. 5 normalized at 0.55 $\mu m$.

## 3. The variability of the spectral indicators in the Saturnian system

In this section we describe the variability, across the entire Saturnian system, of the previously discussed spectral indicators. For each regular and minor satellite, VIMS observations and resulting spectral indicators are included. A final paragraph describes the dataset processed to build a radial mosaic of the principal rings (C, B, CD, A) used to measure the radial profiles of the VIS-IR spectral indicators and to compare them with similar quantities retrieved from satellites.

*3.1. Mimas observations*

Mimas is Saturn's innermost regular satellite, orbiting at an average distance of 3.08 $R_S$ from Saturn. The satellite has a mean radius of 198.2 km and a density of 1.15 $g\ cm^{-3}$ (satellites physical properties are discussed in Jaumann et al. (2009)). Mimas' surface appears dominated by impact craters and by few tectonic features. Among the craters is the remarkable Herschel, about 110 km in diameter, located at the equator at longitude 105°.

VIMS has returned 265 disk-integrated observations for Mimas from distances ranging between $3.0537 \cdot 10^4$ to $1.19796 \cdot 10^6$ km and with solar phase angles between 1.36° and 140°. The top panel of Fig. 7 shows a summary of observational geometries, e.g. the sub-spacecraft point longitude and latitude, the solar phase and the Cassini-surface distance are reported for each observation. Along the x axis the observations are sorted according to the temporal sequence (observations in 2004 are the first, and those in 2010 the last). The dataset shows a great variability in longitude, resulting in observations aimed on both Leading/Trailing and Saturnian/Antisaturnian hemispheres [2]. As far as the coverage in latitude is concerned, VIMS observations are mainly centered along the equator and on the meridional hemisphere down to -30°.

The bottom panel of Fig. 7 contains the plots of the three spectral slopes and the four water ice band depths measured for each observation. These plots allow us to correlate spectral indicators with the corresponding viewing geometry, surface coverage and regolith properties.

Observations $\#225 - 260$, taken on the trailing hemisphere at about SSC lon=250°, show a significant water ice band depth downturn: here, in fact, BD(1.25 $\mu m$), BD(1.5 $\mu m$) and BD(2.0 $\mu m$) decrease below 0.027, 0.4 and 0.6, respectively. Similar band depth values, as inferred from Fig. 6, are compatible with grain sizes of about 20 $\mu m$. Moving on to points taken on the leading hemisphere, i.e. observations $\#205 - 225$, where (SSC lon$\approx$ 90°), we measure BD(1.25 $\mu m$)$\approx$0.035, BD(1.5 $\mu m$)$\approx$0.48 and BD(2 $\mu m$)$\approx$0.65. Such as values are compatible with 50 $\mu m$ grains. Therefore the distribution of the water ice band depth suggests that between Mimas' leading and trailing hemispheres there exists a different average distribution of regolith grain sizes, with the larger grains distinguishing the leading and the smaller the trailing. These results are in agreement with high spatial resolution data collected by VIMS during close-encounters flybys (Buratti et al., 2011). However, is important to remember here a general trend which affects the band depth across a surface observed at different solar phases (a detailed discussion is given in section 5): band depth, in fact, is not constant but decreases with phase in particular below about 20°. VIMS data point out a $\approx$ 20% reduction of the band depth when phase falls from 20° to 0°. Since Mimas leading hemisphere observations (#205 − 225) were obtained at 10 − 15° phases while trailing hemisphere observations (#225 − 260) were obtained at lower phases, 1 − 5°, we should expect that this difference in illumination condition causes a difference of about 5% in band depth between these two groups of observations. Increasing the band depth of the trailing hemisphere observations by such a percentage ought to compensate for the difference in phase from $1-5°$ to $10-15°$. For the BD(1.5 $\mu m$) this means a 5% increase of the measured value, from 0.4 to 0.42, which is however lower than 0.48 measured on the leading hemisphere at similar phase.

---

[2]For convention, leading hemisphere interval: 0° ⩽ *lon.* ⩽ 180°, center at lon.=90°. Trailing hemisphere interval: 180° ⩽ *lon.* ⩽ 360°, center at lon.=270°. Saturnian hemisphere interval 270° ⩽ *lon.* ⩽ 90°, center at lon.=0°. Anti-Saturnian interval 90° ⩽ *lon.* ⩽ 270°, center at lon.=180°.

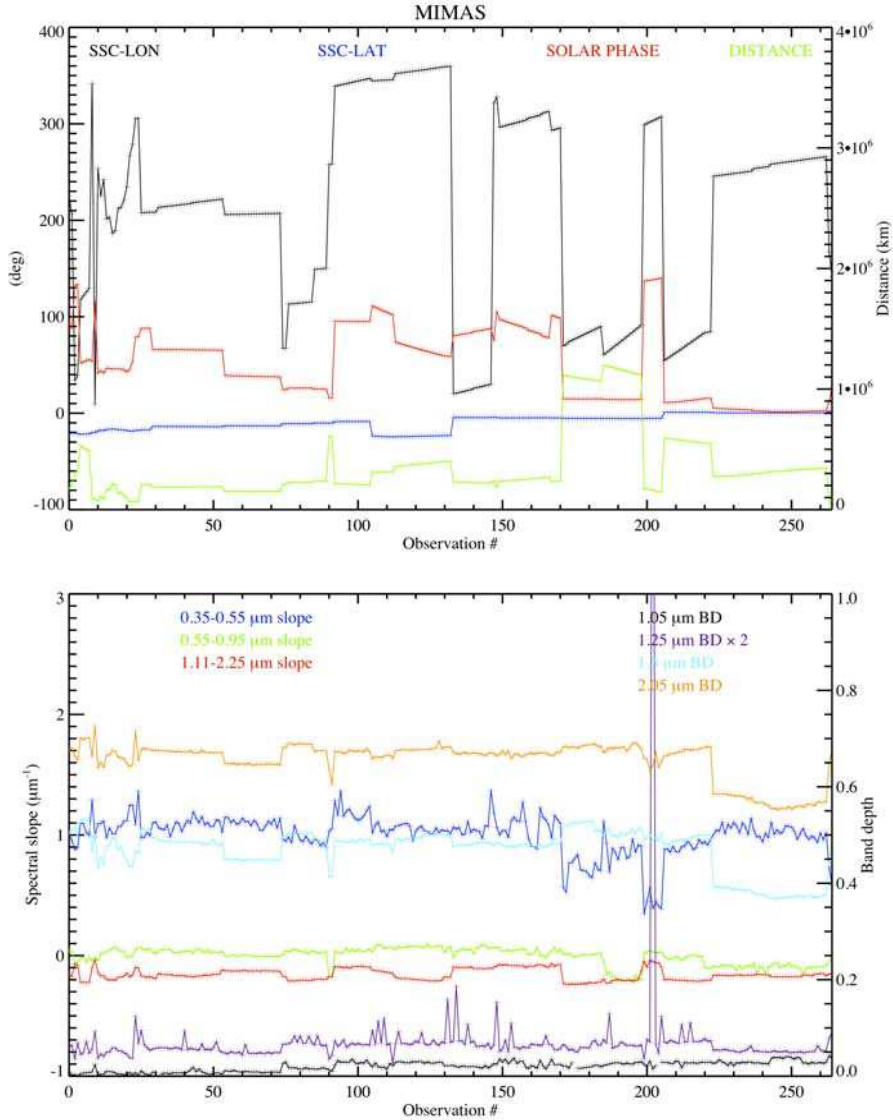

Figure 7: (S1) Mimas observations and spectral parameters. Top panel: geometry parameters: sub-spacecraft point latitude, longitude, solar phase angle and Cassini-surface distance. Bottom panel: 0.35-0.55, 0.55-0.95, 1.1-2.25 $\mu m$ spectral slopes (left vertical axis) and 1.05, 1.25, 1.5, 2.05 $\mu m$ water ice band depth (right vertical axis).

The different distribution of the grain sizes is confirmed by the spectral slope $S_{1.11-2.25}$ values too; in fact, we measure values dispersed around -0.16 $\mu m^{-1}$ on the trailing hemisphere (observations #225 − 260) and around -0.20 $\mu m^{-1}$ on the leading hemisphere (observations #205 − 225); the former value, according to Table 3, is compatible with grain sizes of 20 $\mu m$, while the latter corresponds to grains of about 50 $\mu m$. While in the IR range we are able to discriminate between the two hemispheres, the slopes measured in the VIS range don't have any significant differences thus indicating a substantial color homogeneity. This is a further confirmation of the small albedo asymmetry between the two hemispheres of the satellite (Buratti et al. (1990), Verbiscer & Veverka (1992)).

Similar differences in grain size between the two hemispheres are seen in high spatial resolution data taken by VIMS (Buratti et al., 2010b). Despite that disk-integrated observations considered in this work do not allow us to discriminate between equatorial and polar zones, nor inside the large Herschel crater, such results could be correlated with one of, or both, of the following phenomena: 1) the regional thermal anomaly observed by Cassini-CIRS on the leading side around the Herschel crater (Spencer et al., 2010). During the day, equatorial to midlatitude regions on the leading hemisphere appears to be more cold (T=77 K) than regions on the trailing (T=92 K). This anomalous thermal behavior could be a consequence of the alteration of surface particles (in grain size, surface roughness, compaction and consequently thermal inertia) that occurred during the giant impact that generated the Herschel crater; 2) the presence of a bluish equatorial region on the leading side caused by the bombardment of high-energy, magnetospheric electrons, combined with the contemporary contamination of E ring particles falling on the trailing side (Schenk et al., 2011).

*3.2. Enceladus observations*

With a density of 1.608 $g\ cm^{-3}$, Enceladus is the densest object among the regular icy satellites (and leaving aside Titan), a strong indication of an interior rocky core (Matson et al., 2009). The satellite, orbiting at 3.95 $R_S$ from Saturn, exhibits one of the largest surface water ice abundances among the Saturnian moons (Spencer et al., 2009) together with prominent tectonic and cryovolcanic activities and different surface units (ancient cratered, recent cratered and uncratered terrains). The "tiger stripes", hundred-km long depressions of tectonic origin from which are released the icy particles feeding the E-ring, are located in the south polar region (Porco et al., 2006). Enceladus's plumes are mainly composed of micron-size water ice particles with minor amounts of ammonia, carbon dioxide and silicates (Spahn et al. (2006), Hillier et al. (2007)).

A total of 298 data cubes, acquired from distances spanning between 3.5699 ·$10^4$ and 1.21234 ·$10^6$ km, are available. The coverage in latitude is wide, ranging from -21° to 53°; solar phase runs between 12° and 150° with a large number of observations taken at high phases ($\geq 90°$). As discussed for Mimas, we report in Fig. 8 geometrical and spectral parameters.

The water ice band depths measured for Enceladus are the strongest among the regular satellites reaching maximum values of about 0.08, 0.15, 0.6, 0.7 for the 1.05, 1.25, 1.5 and 2.0 $\mu m$ bands, respectively. These values, according to Fig. 6, correspond to water ice grains diameters of about 70 $\mu m$. A similar indication is given by the 1.1-2.25 $\mu m$ slope which oscillates around -0.24 $\mu m^{-1}$ (see Table 3). In the visible range the 0.35-0.55 $\mu m$ slope is variable in the range 0.5-1.0 $\mu m^{-1}$, values which are too high to be compatible with pure ice spectra (see Table 2). In this range, in fact, reflectance spectra are extremely sensitive to small amount of contaminants mixed with ice and whose presence causes reddening. Among the icy satellites this reddening is minimum on Enceladus but becomes progressively stronger on the other icy satellites, reaching the maximum value on Rhea, and on the A and B rings.

The 0.55-0.95 $\mu m$ slope, more influenced by the presence of contaminants in intimate and areal mixing, runs in the -0.12 to -0.3 $\mu m^{-1}$ range, compatible with fine to coarser water ice grains (Table 2). The effects of the different mixing modes of the contaminants in water ice as seen through albedo and $S_{0.35-0.55} - S_{0.55-0.95}$ slopes variations is discussed in section 4.

*3.3. Tethys observations*

One of the most prominent surface features of Tethys is Ithaca Chasma, a canyon-like structure of a few km depth, running from north to south crossing the equator at longitude = 0°. This formation is probably caused by endogenic tectonic processes (Giese et al., 2007). The 400 km-wide Odysseus crater, centered at lon=130°, lat=30° is the most prominent impact feature on the leading hemisphere. With a density of 0.97 $g\ cm^{-3}$, Tethys can be considered as an undifferentiated object.

The VIMS dataset consists of 278 observations taken from 6.7465 ·$10^4$ to 3.94831 ·$10^6$ km distances; the coverage is more complete on the southern hemisphere (between -22° to 2° in latitude) while the solar phase changes between 0.5° to 147° with two different groups of observations taken close to the opposition effect: observations #130 − 133 at solar phase 0.9 − 1.0° about on the leading hemisphere and #243 − 278 at 0.5 − 1.6° centered on the trailing (Fig. 9).

Observations #10 − 70, acquired at intermediate phase (> 100°) on the saturnian hemisphere, have the highest water ice band depth values, up to 0.05, 0.60 and 0.68 for the 1.25, 1.5 and 2.0 $\mu m$ bands, respectively. This means that on average the largest particles across the satellite's surface, 50 to 60 $\mu m$ in size, are found in the Ithaca Chasma region. The leading regions (observations #155 − 238) have large band depths compatible with grains of 50 $\mu m$, while the trailing hemisphere (observations #244 − 278) shows the lowest band depths, indicative of small grains (10 $\mu m$ size).

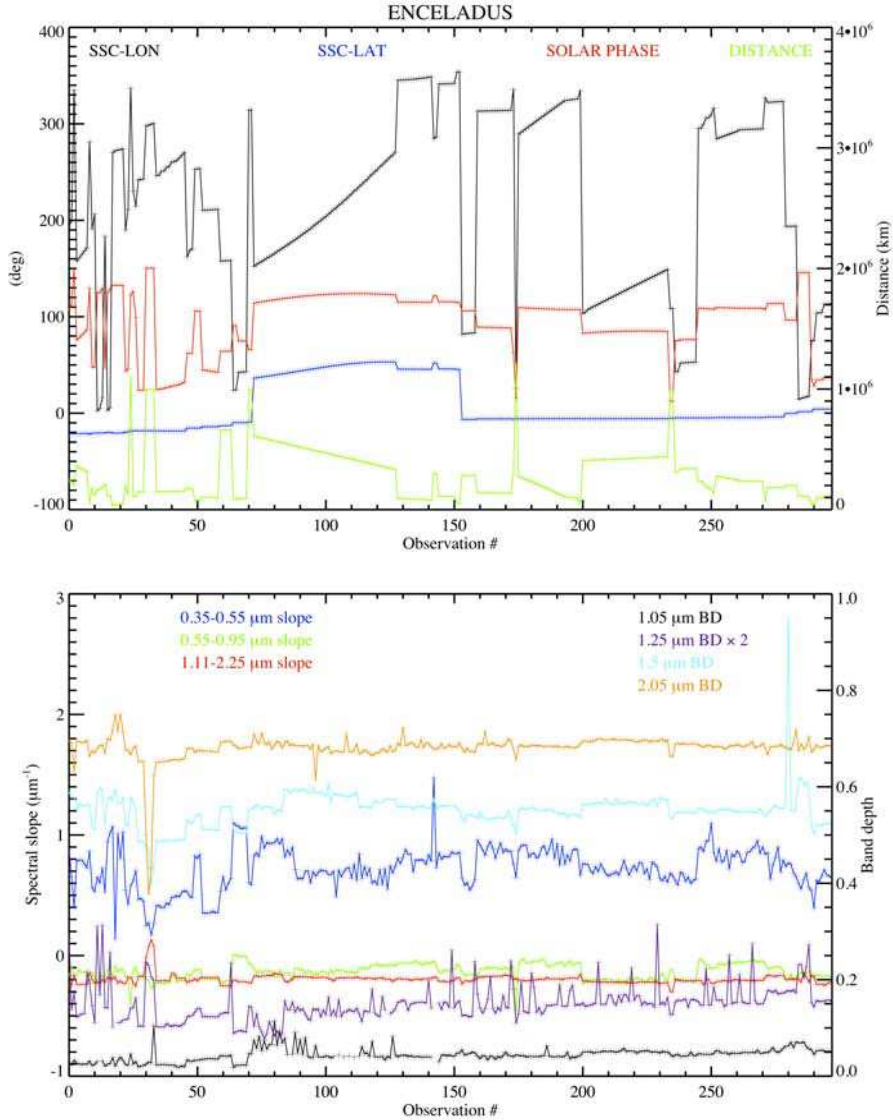

Figure 8: (S2) Enceladus observations and spectral parameters. Same caption as Fig. 7.

Concerning the 0.35-0.55 $\mu m$ spectral slope, which is almost constant across the entire dataset and which oscillates around 1.0 $\mu m^{-1}$, the same considerations previously described for Enceladus are valid.

## 3.4. Dione observations

At an average distance of 6.3 $R_S$ from Saturn, Dione orbits inside the E ring and Saturn's magnetosphere environment; as a result the surface is altered by both E ring and magnetospheric particle bombardment. The moon has a radius of about 562 km and a density of 1.48 $g\ cm^{-3}$, making it the densest regular satellite after Enceladus (and Titan). Dione surface's morphology is characterized by the presence of wispy terrains, which are the result of tectonic processes active across the trailing hemisphere; on the same hemisphere the presence of non-icy dark material is evident (Clark et al. (2008), Stephan et al. (2010)), probably caused by the interaction with hundreds-KeV magnetospheric electrons (Carbary et al., 2009), energetic enough to allow the formation of tholins and to alter the surface composition. The composition of the leading hemisphere is much more icy thanks to the interaction and deposition of E ring particles.

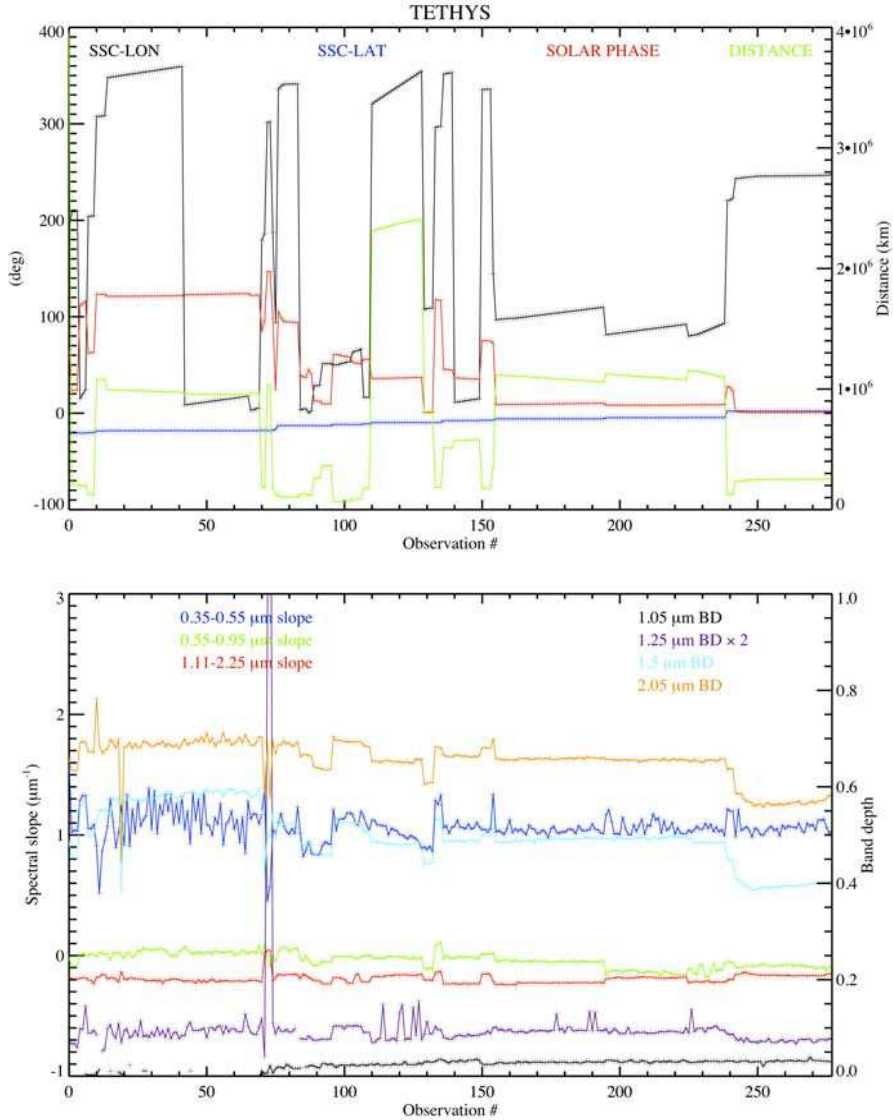

Figure 9: (S3) Tethys observations and spectral parameters. Same caption as Fig. 7.

We have selected and analyzed 268 disk-integrated observations acquired from distances between $6.3758 \cdot 10^4$ and $1.35747 \cdot 10^6$ km and with solar phase between $0.1°$ and $125°$. Apart from observations #101 − 108, which were taken when the sub-spacecraft point was at $40°$ latitude, the majority of the data were collected along the equator down to about $-20°$ latitudes. As far as the illumination conditions are concerned, Dione was imaged on average at very low phase angles.

Three different sequences, corresponding to observations #37 − 67, #113 − 118, #235 − 239, have less than $1°$ in solar phase thus allowing us to monitor the opposition surge effect (see top panel in Fig. 10). In general, under equal or comparable illumination conditions we measure the strongest water ice band depth on the leading hemisphere. In fact, as shown in the bottom panel in Fig. 10, points #113 − 118 taken on the leading hemisphere (sub-spacecraft point at 100° longitude) have band depths higher than points #37 − 67 (270° longitude) and #235 − 259 (340° longitude).

In order to infer the average grain size dimensions we use the first 25 observations of the series that were taken at moderate solar phase angle (between 20° and 40°) and allow us to make a comparison with lab data (Fig. 6). We infer grain diameters larger than 50 $\mu m$ across the leading face and smaller than 30 $\mu m$ across the trailing face.

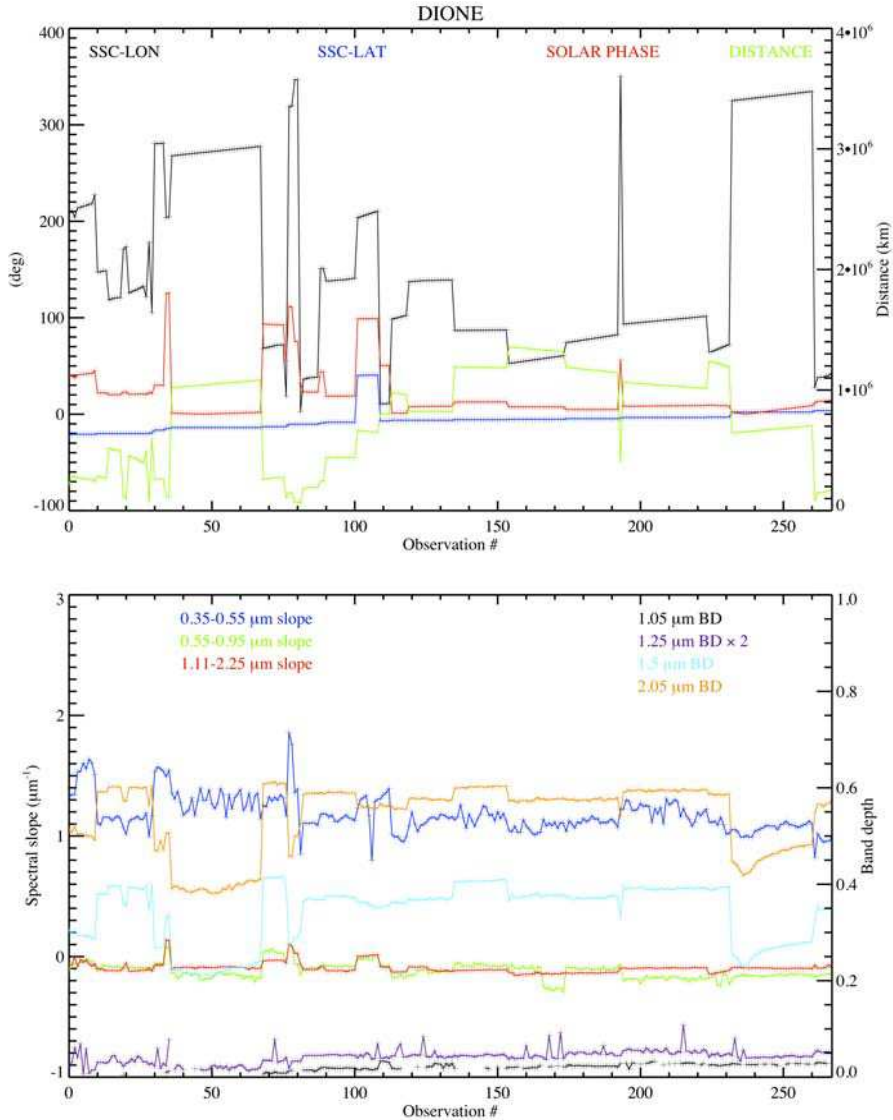

Figure 10: (S4) Dione observations and spectral parameters. Same caption as Fig. 7.

*3.5. Rhea observations*

With a mean radius of 764 km and a mean density of 1.23 $g\ cm^{-3}$, Rhea is the largest Saturnian icy satellite (after Titan). The moon orbits at 8.74 $R_S$ from Saturn, in a region located at the outer edge of the E ring and magnetosphere border. The presence of "megascarps", large ridges extending along the north-south direction, is considered as an indication of endogenic geological processes (Moore et al., 1985).

With 511 data cubes available, Rhea is the icy satellite observed the most times by VIMS, from distances ranging between $7.1926 \cdot 10^4$ and $1.65550 \cdot 10^6$ km and solar phase angles between 0.3° and 161° (Fig. 11). The coverage in latitude runs between mid-southern zones (−22.5°) to the equator (4.5°).

Three different sequences taken at low phase angles allow us to monitor the variation of the spectral indicators induced by the opposition surge. The first starts at point #189 where the solar phase is at 2.82°, reaches the minimum angle (0.67°) at #219 and then increases up to 1.63° at #230; the second sequence corresponds to points #290 − 348 where solar phase decreases from 4° to 0.33°; the third is between #428 − 460 for which solar phase moves from 2° to 0.59°. Observations #189 − 230 are located on the antisaturnian hemisphere (sub-spacecraft longitude ≈ 150°) while #290 − 348 and #428 − 460 are aimed to the leading hemisphere. As shown in Fig. 11, bottom panel, the 1.5-2.0 $\mu m$ water ice band depths as well as the visible 0.35-0.55 $\mu m$ spectral slope have an evident decrease in correspondence of the minimum solar phase angle.

Despite the large number of available observations, the Rhea dataset does not provide us an optimal coverage of the trailing hemisphere at low phases. The only observations taken close to 270° in longitude (i.e. points around #60 and #170), have in fact phases larger than 100°. On the other hand, it is possible to look at the differences between the saturnian (observations #267 − 289 and #349 − 358) and antisaturnian (#96 − 148) hemispheres. These points, taken at similar illumination conditions, i.e. at solar phase angles between 30° and 50°, show a slight decrease of the water ice bands depths moving from the antisaturnian to the saturnian hemisphere. The 1.25, 1.5 and 2.0 $\mu m$ band depth are, in fact equal to 0.03, 0.46, 0.66 on the antisaturnian and to 0.02, 0.39, 0.60 on the saturnian points, respectively: these values are compatible with water ice grains diameters larger than 30 $\mu m$ on the antisaturnian hemisphere and smaller than 10 $\mu m$ on the saturnian. For the 1.1-2.25 $\mu m$ slope we measure values of about -0.13 which, according to Table 3 are compatible with very fine grains.

### 3.6. Hyperion observations

A midsize radius (133 km) joined to an irregular shape and a low (0.569 $g\ cm^{-3}$) density (Jaumann et al., 2009), a state of chaotic rotation (Wisdom et al., 1984) and the presence of $CO_2$ ice distributed on the bottom of cup-like craters (Cruikshank et al., 2007) are the main physical characteristics of Hyperion.

A total of 207 VIMS observations, taken from distances $2.2471 \cdot 10^4$ to $7.79571 \cdot 10^5$ km and with solar phases 12.4° to 127.7°, are processed in our analysis (Fig. 12, top panel). These observations cover the entire range of longitudes but are limited mainly to the south-equatorial hemisphere (subSC point ranging between -9.8° to 0° latitudes).

As shown in the bottom panel of Fig. 12, Hyperion's spectral indicators are almost constant, not showing significant differences induced by the regions of the surface observed. With the exclusion of points #77 − 85, taken at high phase angle (about 127°), the water ice band depths are constant and equal to 0.32, 0.57 for the 1.5 and 2.0 $\mu m$ bands, respectively. Moreover we note the lack of the 1.05 and 1.25 $\mu m$ water ice bands in Hyperion's average spectrum (Fig. 1): these bands are obliterated by the strong absorption that characterizes the VIS and near IR spectral range. This is an evidence for the presence of chromophores intimately mixed within water ice. The 1.5 vs. 2.0 $\mu m$ band depths values we have retrieved correspond to grain sizes larger than 1 cm (Fig. 6) in the case of pure water ice composition. This result however is not reliable because band depths are strongly influenced by the presence of other materials (at least $CO_2$ and possibly CN). The same problem will affect the retrieval of grain sizes in the cases of Iapetus' leading hemisphere and Phoebe. In these cases it becomes essential to use spectral mixing codes (Clark et al., 2012).

### 3.7. Iapetus observations

Orbiting at an average distance of 25.55 $R_S$ from Saturn, Iapetus is the regular satellite located at the greatest distance from the planet in the entire solar system. Moreover, while the remaining other regular satellites orbit in the equatorial plane, Iapetus' orbit is inclined by about 15°. These orbital characteristics, coupled with a radius of 735.6 km, a density of 1.08 $g\ cm^{-3}$, a very peculiar "nut-shell" shape caused by a 10 km (on average) high equatorial ridge (Porco et al., 2005), and finally the hemispheric dichotomy between leading and trailing sides, indicate that this satellite has very distinctive characteristics and a unique evolutionary history (Jaumann et al., 2009).

We have selected and processed 298 VIMS observations of Iapetus, taken from distances ranging between $6.3655 \cdot 10^4$ and $3.42731 \cdot 10^6$ km and with solar phase angle between 0.04° and 145.5°. The dataset covers the entire range in longitudes allowing us to observe both hemispheres, but is limited to ±7.5° in latitude (Fig. 13).

At least three groups of observations allow us to evaluate Iapetus' spectral variability: 1) points #1−137 correspond to the leading dark hemisphere imaged at solar phase angles from 38° to 144°. Points #10 − 40, taken at phase 38°, show the minimum water ice 1.5-2.0 $\mu m$ band depths, equal to 0.1 and 0.2, respectively. This is the condition where the band depths reach the minimum value for the entire population of the regular satellites. Moving towards the last point (#137), at increasing solar phase we observe an increase of the two band depths. A similar behavior is seen for the $S_{1.11-2.25}$ slope, while the two visible slopes have an opposite trend, becoming smaller at higher phases; this means that the surface appears less red at high solar phase angles; 2) points #170 − 255 are aimed to the center of the Saturn-facing hemisphere (longitude ≈ 0°) and therefore are a 50-50% average of both leading and trailing hemispheres (dark material on leading and almost pure water ice on the trailing). Moreover, these observations are taken at small solar phases (min 0.4°) and are influenced by the opposition effect. On average we measure 1.5-2.0 $\mu m$ band depths equal to 0.25, 0.45, respectively. At these points the visible reddening is maximum, with the $S_{0.35-0.55}$ reaching about 1 $\mu m^{-1}$; 3) points #256 − 298 cover the trailing hemisphere observed at phases between 33° and 144°. Seen at these longitudes, Iapetus has one of the deepest water ice band depths, up to 0.55, 0.7 for the 1.5 and 2.0 $\mu m$ bands, respectively. The visible slope $S_{0.35-0.55}$ is on average about 0.5 $\mu m^{-1}$: similar values are observed on Enceladus and demonstrate the extreme spectral variability of the two hemispheres of Iapetus in the infrared range. The typical regolith grain size on the trailing hemisphere is in the range 80 $\mu m$-0.5 cm.

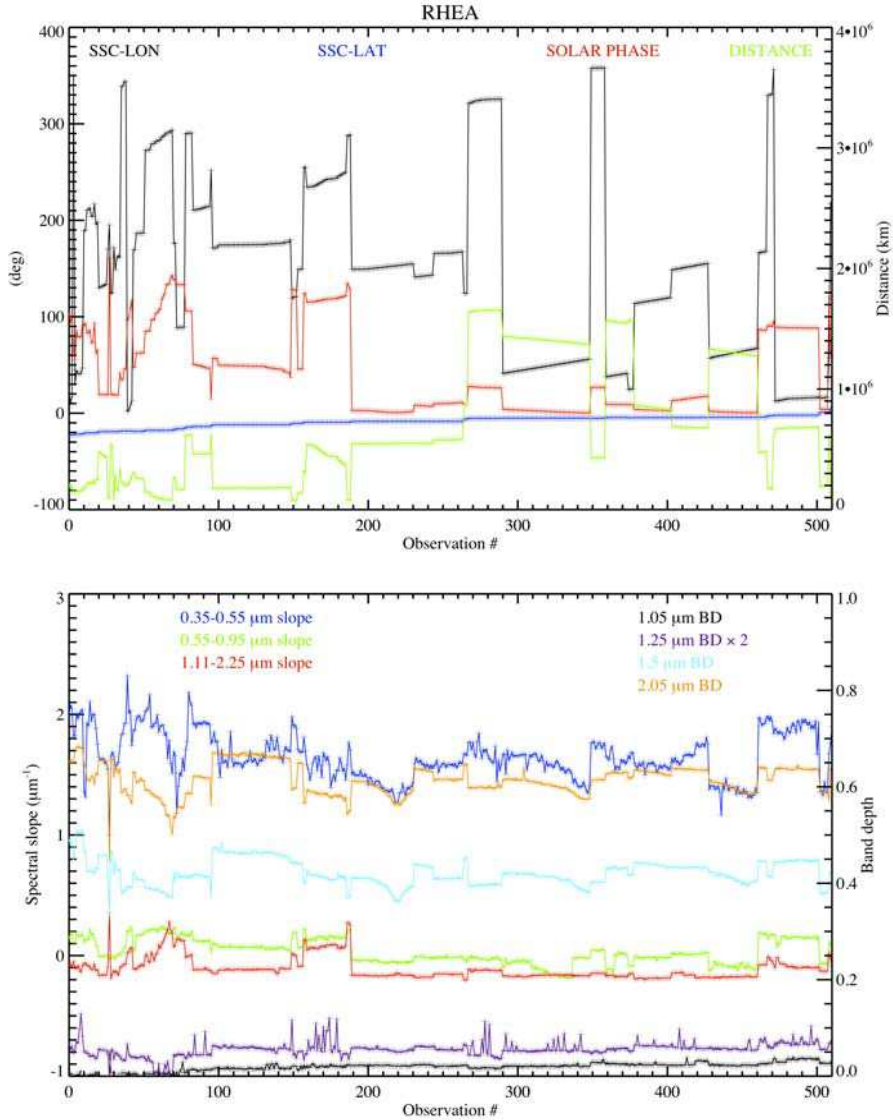

Figure 11: (S5) Rhea observations and spectral parameters. Same caption as Fig. 7.

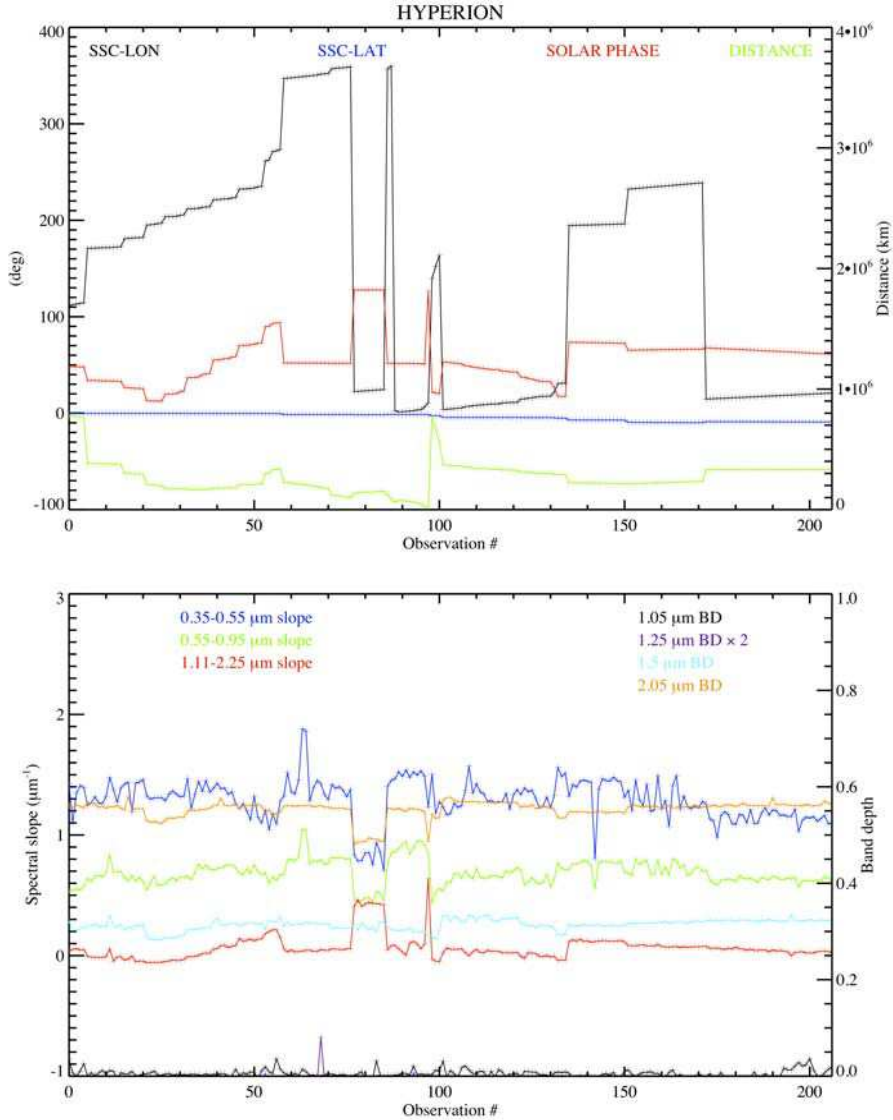

Figure 12: (S7) Hyperion observations and spectral parameters. Same caption as Fig. 7.

*3.8. Observations of minor moons*

Apart from serendipitous observations, e.g., observations planned for other targets where the passage of a minor moon is also recorded, we have included in this work all available VIMS pointed observations of the minor moons (Atlas, Pandora, Prometheus, Janus, Epimetheus, Calypso, Telesto and Helene). A summary of the observations available and the corresponding spectral parameters are included in Table 4 (and Fig. 14 for Janus).

Minor moons can be classified in two groups: inner moons orbiting close to Saturn and Lagrange-point moons of the regular satellites. The first group includes Atlas, Pandora, Prometheus, Janus, Epimetheus; the second is formed by Calypso, Telesto (Tethys' lagrangian moons) and Helene (Dione's lagrangian).

Since these moons have small radii, ranging from 10.6 km for Calypso to 89.6 km for Janus, VIMS images are only few pixels wide and do not allow the resolution of surface features. On the contrary, Cassini-ISS has returned high resolution images (up to 1 km/pixel) from which it is possible to recognize surface structures as well as to determine the bodies' densities, which are in the range 0.46-0.69 *g cm$^{-3}$* (Porco et al., 2007). Both ISS images and dynamical models (Charnoz et al., 2007) indicate that minor inner satellites are porous bodies, formed through accretion of particles scattered along the equatorial plane; in some cases, as for Atlas and Pan, equatorial bulges where the accretion occurs in a preferential mode are evident (Porco et al., 2007). VIMS can test this idea, searching for compositional correlations between minor moons and ring particles.

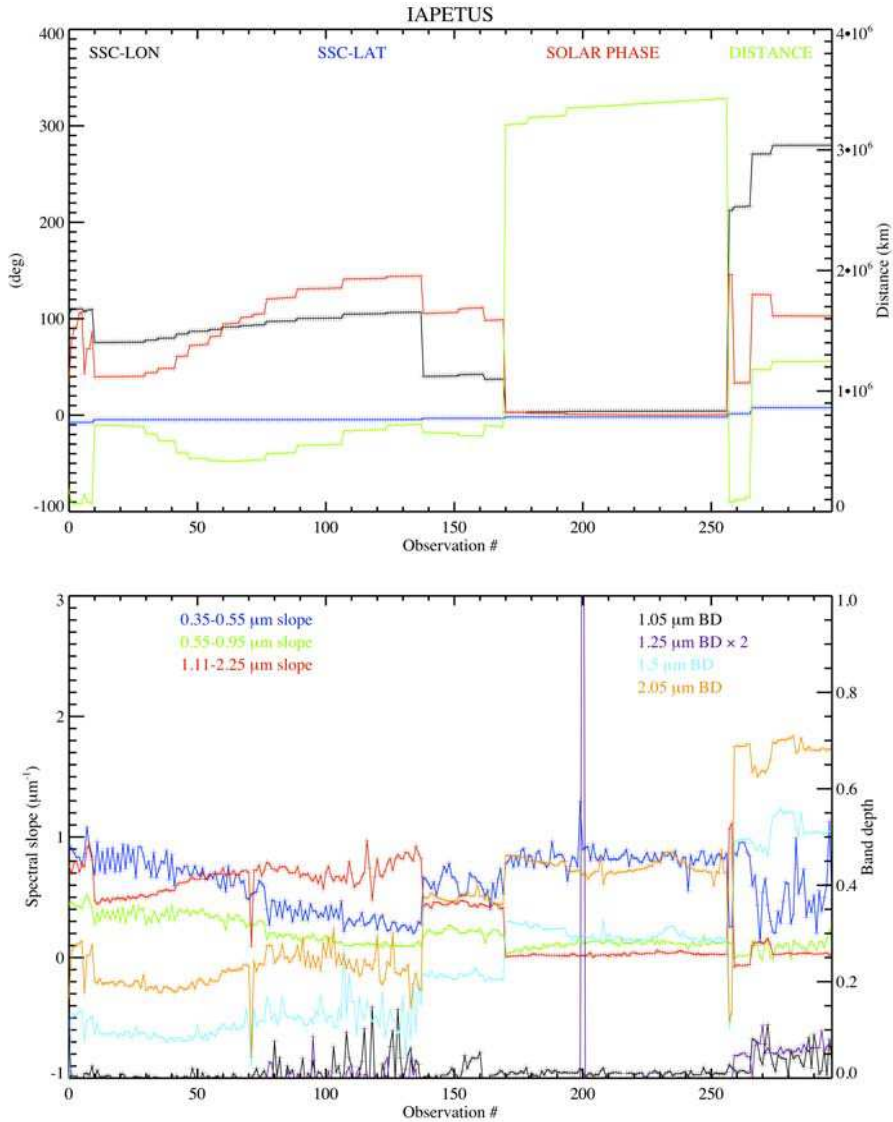

Figure 13: (S8) Iapetus observations and spectral parameters. Same caption as Fig. 7.

Analyzing the average spectra of the minor satellites (shown in Fig. 2) some striking differences are evident in the visible spectral range: the inner moons in general appear "red" while the lagrangians are in general "blue". As we'll discuss in the next sections, the reddening is a spectral characteristic of particles in the main rings. These results may provide evidence of the rings' material transport and a mechanism of accretion on the minor moons. In the infrared range we do not observe any spectral features apart from the usual 1.5-2.0 $\mu m$ water ice bands. The 1.05-1.25 $\mu m$ bands are evident only on Calypso and Pandora, while completely absent on the remaining satellites. The same is observed for the 3.1 $\mu m$ Fresnel peak.

Some results are biased by the limited number of available observations (Table 1). With 98 observations, Janus has the largest available dataset among the small satellites. Janus' water ice band depths are on average equal to 0.35-0.6 for 1.5-2.0 $\mu m$ bands, respectively (Fig. 14). These values are almost constant for observations taken at SSC longitudes between 300°-350°. A small increase in the band depths is measured in spectra corresponding to observations #47 − 55, at SSC longitude 130°. Since all observations are taken across the equatorial region this result indicates that the distribution of ring material is not uniform across the surface. Beyond this general conclusion, the lack of more spatially resolved VIMS data prevents us from investigating further.

Visible slope $S_{0.35-0.55}$ is almost constant for observations taken at phases < 100°. Points #22−37 have a reddening reduction by a factor 2 that is probably caused by the concurrence of the high solar phase (> 100°) and large Cassini-Janus distance; as a consequence of these conditions, the satellite's image is only few pixels wide, giving less accurate results.

| Satellite | Observation | SSC lon, lat (deg) | Phase (deg) | Distance (km) | $S_{0.35-0.55}$ ($\mu m^{-1}$) | $S_{0.55-0.95}$ ($\mu m^{-1}$) | $S_{1.11-1.25}$ ($\mu m^{-1}$) | 1.05 $\mu m$ BD | 1.25 $\mu m$ BD | 1.5 $\mu m$ BD | 2.0 $\mu m$ BD |
|---|---|---|---|---|---|---|---|---|---|---|---|
| ATLAS | V1501558502 | 7.5, -11.5 | 9.1 | 868100. | 1.56607 | -0.0912785 | -0.120375 | NaN | NaN | 0.427 | 0.614 |
| PANDORA | V1501560730 | 216.3, -8.5 | 26.2 | 802520. | 1.15173 | -0.0854211 | -0.262787 | 0.003 | NaN | 0.503 | 0.706 |
| PANDORA | V1654245701 | 68.5, 4.5 | 29.0 | 100260. | 1.42286 | -0.135966 | -0.223678 | 0.019 | 0.088 | 0.646 | 0.874 |
| PANDORA | V1654246008 | 70.5, 4.5 | 27.7 | 100926. | 2.23156 | -0.0470573 | -0.228867 | 0.082 | 0.107 | 0.519 | 0.668 |
| PROMETHEUS | V1501562854 | 47.8, -7.5 | 10.9 | 944031. | 1.25982 | -0.0855611 | -0.280512 | NaN | 0.046 | 0.604 | 0.755 |
| PROMETHEUS | V1640497354 | 99.1, 2.1 | 37.0 | 56512.0 | 2.56254 | -0.110908 | -0.275166 | 0.025 | 0.083 | 0.596 | 0.836 |
| PROMETHEUS | V1640497406 | 99.5, 2.1 | 36.4 | 56502.0 | 2.62483 | 0.0657929 | -0.132438 | 0.010 | 0.107 | 0.601 | 0.770 |
| PROMETHEUS | V1640497458 | 99.9, 2.1 | 35.8 | 56498.0 | 2.79768 | 0.0748260 | -0.106781 | NaN | 0.077 | 0.660 | 0.791 |
| PROMETHEUS | V1640497510 | 100.4, 2.1 | 35.1 | 56500.0 | 2.52831 | -0.0866784 | -0.141057 | 0.181 | 0.059 | 0.652 | 0.776 |
| PROMETHEUS | V1640499316 | 112.6, 2.1 | 17.9 | 58749.0 | 2.33285 | -0.0144084 | -0.152807 | NaN | 0.107 | 0.584 | 0.760 |
| PROMETHEUS | V1640499368 | 113.0, 2.1 | 17.4 | 58884.0 | 2.45791 | 0.0312756 | -0.174791 | 0.070 | 0.106 | 0.552 | 0.819 |
| PROMETHEUS | V1640499420 | 113.4, 2.1 | 16.9 | 59022.0 | 2.85318 | -0.0369617 | -0.0751793 | 0.01 | 0.129 | 0.601 | 0.803 |
| PROMETHEUS | V1640499472 | 113.8, 2.1 | 16.4 | 59163.0 | 2.25813 | -0.100951 | -0.0643012 | 0.092 | 0.117 | 0.587 | 0.741 |
| EPIMETHEUS | V1501559616 | 178.2, -13.9 | 27.7 | 835887. | 0.907143 | 0.0794194 | -0.235520 | 0.005 | 0.009 | 0.345 | 0.576 |
| CALYPSO | V1506183157 | 3.9, -6.0 | 51.2 | 111207. | 0.951450 | -0.123490 | -0.299257 | NaN | 0.039 | 0.610 | 0.857 |
| CALYPSO | V1506184697 | 5.0, -6.1 | 54.3 | 107487. | 0.727210 | -0.107242 | -0.288513 | 0.0219 | 0.073 | 0.554 | 0.783 |
| CALYPSO | V1506186237 | 6.0, -6.2 | 57.9 | 103734. | 0.778726 | -0.0803493 | -0.200342 | 0.035 | 0.064 | 0.607 | 0.718 |
| CALYPSO | V1506187583 | 7.0, -6.3 | 61.0 | 100816. | 0.961003 | -0.269171 | -0.275596 | NaN | 0.061 | 0.601 | 0.763 |
| CALYPSO | V1644755696 | 130.0, 1.9 | 62.0 | 21560.0 | 0.747976 | -0.0752070 | -0.234935 | 0.061 | 0.095 | 0.571 | 0.714 |
| CALYPSO | V1644756003 | 130.7, 1.9 | 54.4 | 22269.0 | 0.770187 | -0.0996565 | -0.228334 | 0.095 | 0.091 | 0.558 | 0.702 |
| CALYPSO | V1644756276 | 131.3, 1.9 | 47.3 | 23348.0 | 0.813202 | -0.119237 | -0.229692 | 0.055 | 0.090 | 0.553 | 0.701 |
| CALYPSO | V1644756373 | 131.5, 1.9 | 46.0 | 23604.0 | 0.719078 | -0.117360 | -0.228506 | 0.061 | 0.135 | 0.545 | 0.704 |
| CALYPSO | V1644756439 | 131.8, 1.9 | 43.5 | 24153.0 | 0.732447 | -0.0929641 | -0.244566 | 0.082 | 0.082 | 0.537 | 0.697 |
| CALYPSO | V1644756746 | 132.4, 1.9 | 37.6 | 25723.0 | 0.721648 | -0.137931 | -0.244445 | 0.084 | 0.120 | 0.531 | 0.688 |
| CALYPSO | V1644757054 | 132.9, 1.9 | 33.5 | 27156.0 | 0.663591 | -0.0975938 | -0.241652 | 0.098 | 0.113 | 0.522 | 0.684 |
| TELESTO | V1514154679 | 298.3, -7.1 | 16.1 | 59072.8 | 0.512164 | -0.172904 | -0.166454 | NaN | 0.0247 | 0.400 | 0.583 |
| TELESTO | V1514165575 | 324.7, -6.1 | 89.4 | 20555.5 | 0.767842 | -0.0686878 | -0.191472 | 0.003 | 0.027 | 0.414 | 0.570 |
| HELENE | V1646319441 | 333.9, 3.2 | 25.2 | 17776.0 | 0.818959 | -0.145286 | -0.180331 | 0.029 | 0.029 | 0.426 | 0.632 |
| HELENE | V1646320344 | 335.3, 3.2 | 23.6 | 25977.0 | 0.980363 | -0.183070 | -0.177711 | 0.033 | 0.035 | 0.435 | 0.633 |
| HELENE | V1646320043 | 334.8, 3.2 | 24.0 | 23238.0 | 0.958341 | -0.144945 | -0.183903 | 0.045 | 0.031 | 0.423 | 0.622 |
| HELENE | V1646319742 | 334.5, 3.2 | 24.4 | 21051.0 | 0.961145 | -0.177375 | -0.183224 | 0.052 | 0.026 | 0.414 | 0.623 |

Table 4: Minor satellites observations and spectral parameters. Janus data are shown in Fig. 14.

Finally, in this paper we do not repeat the discussion for Phoebe because there have been no new observations available since those previously analyzed in Filacchione et al. (2007a) and Filacchione et al. (2010).

*3.9. Rings mosaic observation*

Among the planets of the solar system, Saturn has the most prominent and complex ring system; extending along a radial axis from 74,658 km (inner C ring) to 136,780 km (outer A) the main rings have a total area of about $41.2 \cdot 10^9$ $km^2$. Thanks to VIMS imaging capabilities it is possible to collect VIS-IR reflectance spectra of the entire ring structure at moderate spatial resolution. For the scope of this paper, *e.g.* the study of the radial composition across the rings and satellite system, we have selected a ring radial mosaic that allows us to retrieve the spectral information for both ansae. The observations we have processed were acquired by VIMS during S36-SUBML001 sequence (19-20 December 2007) contemporary to a CIRS-prime radial scan. On average during the observations the solar phase angle was about 29° and the solar elevation angle 12°; the average Cassini-Saturn distance was about 675,000 km, resulting in spatial resolution of about 112 km for the VIS channel and 170 by 340 km for the IR channel. The resulting mosaic image, shown in Fig. 15, is 12 (along the azimuthal direction) by 700 (along the radial direction) pixels wide, corresponding to 8400 spectra; Table 5 reports the list of the observations processed to obtain this mosaic.

As we are interested in retrieving spectral variations in the rings along the radial direction, a method to build spectrograms, e.g., 2D-arrays containing the full spectral (0.35 − 5.0 $\mu m$) and spatial (from 73,500 to 141,375 km) information was developed. A similar approach was used by Nicholson et al. (2008) and Hedman et al. (submitted) to correlate spectral properties with radial structures. The method is based on the following steps:

- selection of 50 tie points, corresponding to very sharp, well-defined spatial structures (edges, gaps, ringlets) placed at known radial distances as measured in Voyager's reflectance profile (Collins et al., 1984). The a priori knowledge of these points, evenly distributed across the A, B, C rings and Cassini division, allow us to define a reference radial grid of distances for the VIMS mosaic;

- for each of the 12 horizontal lines belonging to the mosaic image shown in Fig. 15, VIMS pixels are sorted along the radial axis by linear interpolation between adjacent tie points;

- as the illumination variations across the azimuthal direction (corresponding to the vertical direction in the image shown in Fig. 15) are only few percent, it is possible to average the 12 radial profiles on the same distance grid to boost signal-to-noise and improve the radial resolution; a grid with 125 km/sample resolution allows a uniform radial reconstruction of the VIS channel data without any interruption, while leaving few missing points in the IR channel (seen as vertical gaps in Fig. 16).

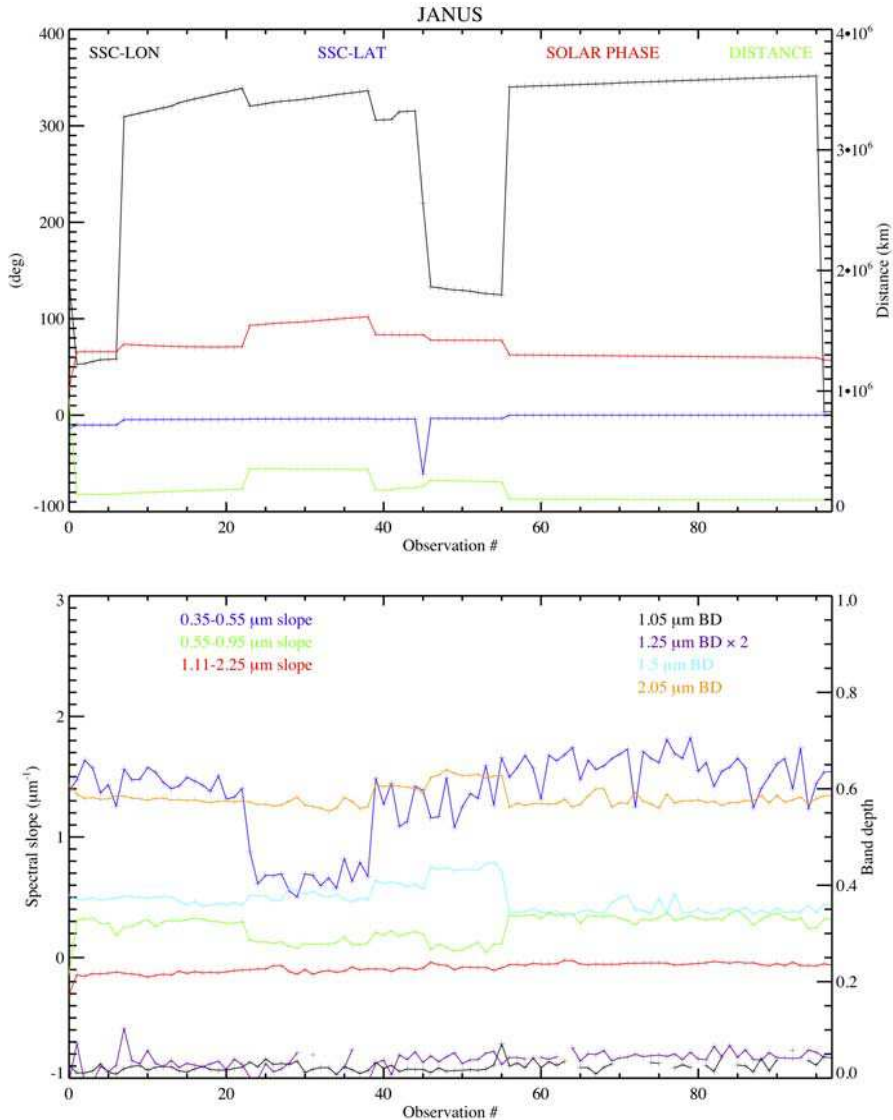

Figure 14: Janus observations and spectral parameters. Same caption as Fig. 7.

| Observation | Start time | End time | S/C distance (km) | Phase (deg) | sub S/C lat (deg) | sub S/C lon (deg) | sub sol lat (deg) | sub sol lon (deg) |
|---|---|---|---|---|---|---|---|---|
| V1576789867 | 2007-353T20:35:07.360Z | 2007-353T20:37:17.013Z | 611983. | 32.182 | -11.425 | 36.398 | -9.226 | 3.750 |
| V1576790388 | 2007-353T20:43:48.003Z | 2007-353T20:45:57.656Z | 615656. | 31.942 | -11.320 | 40.659 | -9.226 | 8.254 |
| V1576790909 | 2007-353T20:52:28.657Z | 2007-353T20:54:38.310Z | 619775. | 31.675 | -11.202 | 45.455 | -9.225 | 13.321 |
| V1576791429 | 2007-353T21:01:09.286Z | 2007-353T21:03:18.939Z | 623882. | 31.413 | -11.086 | 50.256 | -9.225 | 18.388 |
| V1576791950 | 2007-353T21:09:49.936Z | 2007-353T21:11:59.589Z | 627522. | 31.182 | -10.983 | 54.526 | -9.225 | 22.892 |
| V1576792471 | 2007-353T21:18:30.592Z | 2007-353T21:20:40.245Z | 631604. | 30.927 | -10.869 | 59.334 | -9.225 | 27.959 |
| V1576792991 | 2007-353T21:27:11.219Z | 2007-353T21:29:20.872Z | 635673. | 30.675 | -10.757 | 64.145 | -9.225 | 33.026 |
| V1576793512 | 2007-353T21:35:51.872Z | 2007-353T21:38:01.525Z | 639729. | 30.427 | -10.645 | 68.959 | -9.225 | 38.094 |
| V1576794033 | 2007-353T21:44:32.525Z | 2007-353T21:46:42.178Z | 643324. | 30.209 | -10.547 | 73.241 | -9.225 | 42.598 |
| V1576794553 | 2007-353T21:53:13.156Z | 2007-353T21:55:22.809Z | 647357. | 29.968 | -10.438 | 78.062 | -9.225 | 47.665 |
| V1576795074 | 2007-353T22:01:53.806Z | 2007-353T22:04:03.459Z | 651376. | 29.729 | -10.330 | 82.886 | -9.225 | 52.732 |
| V1576795594 | 2007-353T22:10:34.439Z | 2007-353T22:12:44.092Z | 654939. | 29.520 | -10.235 | 87.177 | -9.225 | 57.236 |
| V1576796115 | 2007-353T22:19:15.089Z | 2007-353T22:21:24.742Z | 658935. | 29.288 | -10.129 | 92.006 | -9.225 | 62.303 |
| V1576796636 | 2007-353T22:27:55.742Z | 2007-353T22:30:05.395Z | 662477. | 29.084 | -10.036 | 96.302 | -9.224 | 66.807 |
| V1576797156 | 2007-353T22:36:36.372Z | 2007-353T22:38:46.025Z | 666449. | 28.858 | -9.932 | 101.138 | -9.224 | 71.874 |
| V1576797677 | 2007-353T22:45:17.025Z | 2007-353T22:47:26.678Z | 670410. | 28.635 | -9.830 | 105.976 | -9.224 | 76.942 |
| V1576798198 | 2007-353T22:53:57.678Z | 2007-353T22:56:07.331Z | 674358. | 28.415 | -9.728 | 110.817 | -9.224 | 82.009 |
| V1576798761 | 2007-353T23:03:21.297Z | 2007-353T23:05:30.949Z | 678294. | 28.197 | -9.627 | 115.661 | -9.224 | 87.076 |
| V1576799282 | 2007-353T23:12:01.941Z | 2007-353T23:14:11.594Z | 682218. | 27.983 | -9.528 | 120.508 | -9.224 | 92.143 |
| V1576799803 | 2007-353T23:20:42.594Z | 2007-353T23:22:52.247Z | 685696. | 27.795 | -9.440 | 124.818 | -9.224 | 96.647 |
| V1576800323 | 2007-353T23:29:23.223Z | 2007-353T23:31:32.876Z | 689597. | 27.586 | -9.342 | 129.670 | -9.224 | 101.714 |
| V1576800844 | 2007-353T23:38:03.877Z | 2007-353T23:40:13.530Z | 693487. | 27.379 | -9.246 | 134.524 | -9.224 | 106.781 |
| V1576801365 | 2007-353T23:46:44.528Z | 2007-353T23:48:54.181Z | 696934. | 27.198 | -9.160 | 138.841 | -9.224 | 111.285 |
| V1576801885 | 2007-353T23:55:25.157Z | 2007-353T23:57:34.809Z | 700801. | 26.997 | -9.065 | 143.700 | -9.224 | 116.353 |
| V1576802406 | 2007-354T00:04:05.811Z | 2007-354T00:06:15.464Z | 704657. | 26.798 | -8.971 | 148.561 | -9.223 | 121.420 |
| V1576802927 | 2007-354T00:12:46.464Z | 2007-354T00:14:56.117Z | 708075. | 26.623 | -8.889 | 152.884 | -9.223 | 125.924 |
| V1576803447 | 2007-354T00:21:27.093Z | 2007-354T00:23:36.746Z | 711908. | 26.429 | -8.796 | 157.749 | -9.223 | 130.991 |
| V1576803968 | 2007-354T00:30:07.746Z | 2007-354T00:32:17.399Z | 715731. | 26.237 | -8.705 | 162.617 | -9.223 | 136.058 |
| V1576804488 | 2007-354T00:38:48.398Z | 2007-354T00:40:58.051Z | 719119. | 26.069 | -8.624 | 166.946 | -9.223 | 140.562 |
| V1576805009 | 2007-354T00:47:29.027Z | 2007-354T00:49:38.680Z | 722920. | 25.882 | -8.534 | 171.818 | -9.223 | 145.629 |
| V1576805530 | 2007-354T00:56:09.680Z | 2007-354T00:58:19.333Z | 726709. | 25.697 | -8.445 | 176.692 | -9.223 | 150.697 |
| V1576806050 | 2007-354T01:04:50.310Z | 2007-354T01:06:59.963Z | 730068. | 25.534 | -8.367 | 181.026 | -9.223 | 155.201 |
| V1576806571 | 2007-354T01:13:30.960Z | 2007-354T01:15:40.613Z | 733837. | 25.354 | -8.280 | 185.904 | -9.223 | 160.268 |
| V1576807092 | 2007-354T01:22:11.616Z | 2007-354T01:24:21.269Z | 737594. | 25.175 | -8.193 | 190.785 | -9.223 | 165.335 |

Table 5: VIMS observations used to built the S36 SUBML001 mosaic images across the two Saturn's ring ansae. Each data cube has dimensions (sample, bands, line)=(64, 12, 352). Observations were acquired in high resolution mode for both VIS and IR channels. The integration time is fixed to 10 seconds for the VIS and 160 msec for the IR. The first 17 cubes correspond to the west ansa while the last 17 cubes to the east ansa mosaics. The resulting VIS mosaic images are shown in Fig. 15.

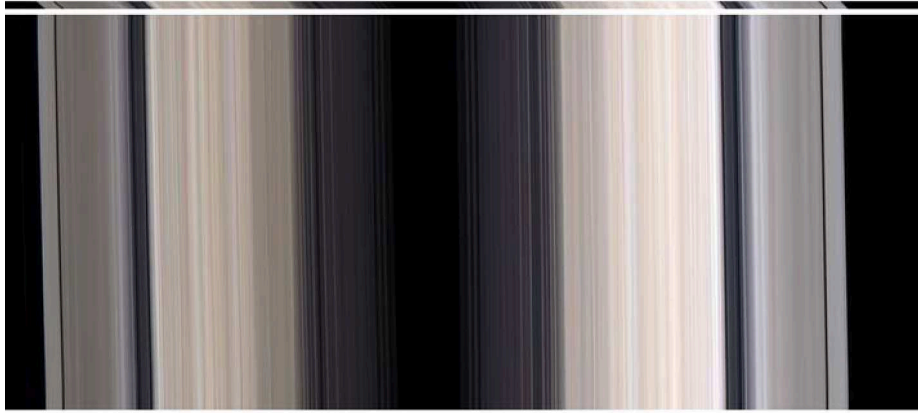

Figure 15: Rings radial mosaics obtained by VIMS-V (color images, RGB at 0.7, 0.55, 0.44 $\mu m$). Top panel: original mosaic, west ansa on the left, east ansa on the right. Bottom panel: same image resized by a factor 60 along y axis to improve visualization. The original image's height along y is 12 pixels wide while the resized image is 700 pixels. From the center to the two sides are visible the C, B, Cassini division and A rings.

- the VIS-IR channels are bridged at 980 nm by using the same method explained before for the icy satellites

- spectra;
- spectrograms (radial distance vs. wavelength) are retrieved, one for each ring ansa mosaic.

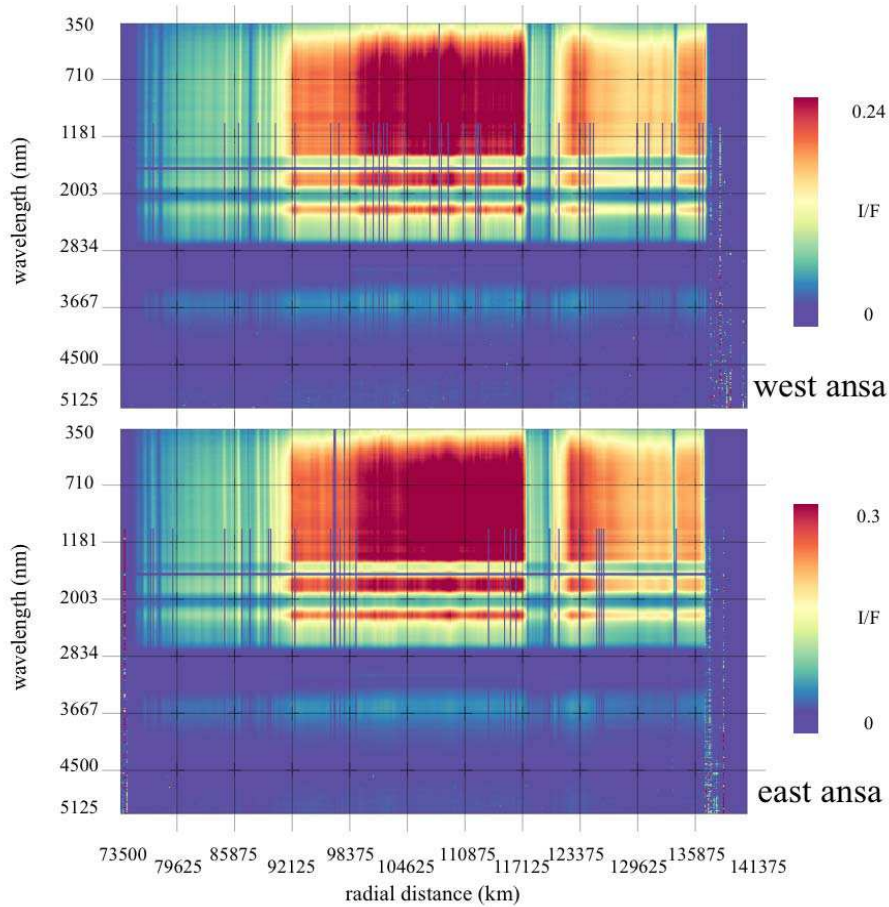

Figure 16: Rings spectrograms retrieved from the west (top) and east (bottom) mosaics. Spatial and spectral scales shown along the x and y axes.

The spectrograms shown in Fig. 16 contain the entire principal rings spectral and spatial information displaying along the x axis the radial distance (spanning from 73,500 to 141,375 km at 125 km/sample resolution) and the reflectance along the y axis; the vertical line in each spectrogram contains the average reflectance coming from one or more spectra taken in a 125 km wide radial interval while each horizontal line represent the "monochromatic" (one VIMS band) average radial profile. These spectrograms allow us to highlight the following radial features for the principal rings:

- The faint C ring, extending between 74,658 and 91,975 km, has a reflectance of few percent in the inner regions and about I/F(1 $\mu m$)=0.10-0.15 on the outer part. The Titan ringlet ($\approx$77,883 km) and Maxwell gap ($\approx$87,523 km) are resolved on both east and west spectrograms.

- With an optical depth $\tau$ running between 0.4 to greater than 2.5, the B ring is the most dense and reflective region; it extends from 91,975 to 117,507 km with a wide inner plateau ($92,000 < r < 99,000$ km) characterized by lower optical depth and smaller reflectance (I/F(1 $\mu m$)=0.2 on the west ansa, 0.25 on the east ansa). The B2 region ($100,000 < r < 105,000$ km), where significant variations in color and in water ice band depth are seen, is placed almost in the middle of the B ring. The outer part of the B ring is the brightest region of the system having I/F(1 $\mu m$)=0.24 on the west ansa and 0.3 on the east)

- The 5,000 km-wide Cassini division extends between 117,507 to 122,340 km; the Huygens gap is localized close to the inner edge at 118,005 km and is particularly evident on the west ansa spectrogram shown in Fig. 16.

- The outermost A ring spans from 122,340 to 136,780 km. It has an optical depth in the range $0.4 < \tau < 1.0$, a bowl-shaped reflectance profile and several density and vertical waves near its outer edge; the maximum reflectance is measured on the inner region where I/F(1 $\mu m$)=0.22 on the west ansa and 0.27 on the east ansa. The 300 km wide Encke gap (133,410-133,740 km) is clearly evident on both spectrograms while the narrower Keeler gap (136,510-136,550) is barely detectable on the east ansa spectrogram.

The spectral information derived from the VIS portion of the spectrograms is shown in Fig. 17 where we plot the $S_{0.35-0.55}$ and $S_{0.55-0.95}$ slope profiles retrieved across the west (top panel) and east (middle panel) ansae. As discussed in (Nicholson et al., 2008), the $S_{0.35-0.55}$ profile is closely correlated with the water ice band depth profiles (discussed later and shown in Fig. 18) indicating the presence of contaminants/darkening agents bound at intramolecular level with the water ice regolith covering the ring particles. VIMS measures maximum values of the $S_{0.35-0.55}$ across the A (outer region) and B rings; it decreases towards the inner part of the A ring, corresponding to an increase of contaminants with respect to the outer regions. Across the B ring the $S_{0.35-0.55}$ slope has a local maximum around 100,000 km and a minimum around 102,000 km, more evident on the east ansa profile; across the B2 region we observe also corresponding variations in the water ice band depths (see next Fig. 18). The "blue" peak placed at 100,000 km may identify a region with purer ices because a maximum both in $S_{0.35-0.55}$ slope and in water ice band depth is measured here. The minimum blue slope is reached in the region corresponding to the CD and C ring where more contaminated and small particles are distributed and where the optical depth and surface mass density are smaller.

$S_{0.55-0.95}$ slope is in general almost flat and fluctuating around zero across A and B ring, in accordance with more pure water ice spectral behavior, while it is anticorrelated to the $S_{0.35-0.55}$ slope in the CD and C ring, where it increases. This long-wavelength reddening effect is more evident across low optical depth regions where it is compatible with the presence of a significant fraction of a darkening agent distributed among ice grains. Therefore VIMS profiles shown in Fig. 17 are in agreement with the model proposed by Cuzzi and Estrada (1998) in which the residual dark material of cometary bombardment accumulates in the CD and C ring.

The high-frequency fluctuations observed in the $S_{0.55-0.95}$ profiles across C ring and CD are caused by an instrumental residual effect (spectral tilt), which is particularly evident across small and high contrast features like ringlets, edges and gap. This artifact is caused by a small misalignment occurring among the spectrometer's entrance slit axis, grating's grooves and focal plane's side, resulting in a spatial shift of about half pixel (along the sample direction) across the full spectral range when the instrument is operating in nominal mode. As a consequence, the last monochromatic image, at 1.05 $\mu m$, is shifted by about half pixel with respect to the first one at 0.35 $\mu m$. Since the effect is linear with wavelength, the resulting shift corresponds to about 14% of pixel in the $S_{0.35-0.55}$ range and 29% of pixel for the $S_{0.55-0.95}$ slope. For this reason in general the $S_{0.55-0.95}$ slope is noisier across regions of higher contrast. A detailed discussion of this effect is given in (Filacchione et al., 2006).

Finally, in the bottom panel of Fig. 17 are shown the two radial profiles relative to the visible slopes crossing wavelength ($\lambda_{crossing}$) which, as reported in Table 2, is a spectral indicator for regolith grain size. The wavelengths for different pure ice grains (coarse, fine, frost, medium size) as derived from lab data are marked on the plot. Across the A and B rings we measure $\lambda_{crossing} > 0.53$ $\mu m$ which reduces to about 0.52 $\mu m$ across the C ring to reach a minimum value of about 0.5 $\mu m$ in the CD. The assumption of pure water ice is not valid, especially across C and CD, where darkening agent dominates.

Water ice abundance is derived from the 1.25, 1.5, 2.0 $\mu m$ band depth radial profiles shown in Fig. 18. The band depths are correlated among themselves, showing the same radial variations despite their different intensities, with the 2.0 $\mu m$ more intense than the 1.5 $\mu m$ and the 1.25 $\mu m$ bands. The band depths reach maximum values in the A (from a radius of about 130,000 km to the Encke gap) and B rings (from 104,000 to 116,000 km); a smooth decrease is measured in the inner part of the A ring moving from 130,000 km towards the inner edge of the A ring. This region corresponds to the ballistic transport zone (Cuzzi and Estrada , 1998).

As mentioned Nicholson et al. (2008), the transition zone between the Cassini division and the A ring has peculiar properties: comparing the profiles shown in Fig. 17-18, we observe that both the $S_{0.35-0.55}$ and the BD(1.25, 1.5, 2.0 $\mu m$) profiles decrease with a similar trend moving inwards. This is a consequence of a progressive increase of contaminants bound in water ice grains at the molecular scale. But the effect of this mixing is not seen in the $S_{0.55-0.95}$ profile, which remains constant, because these wavelengths are more sensitive to detect the variations of fractional abundance of the contaminants, nor in the I/F(0.55, 1.82 $\mu m$) profiles, which are more correlated to the optical depth.

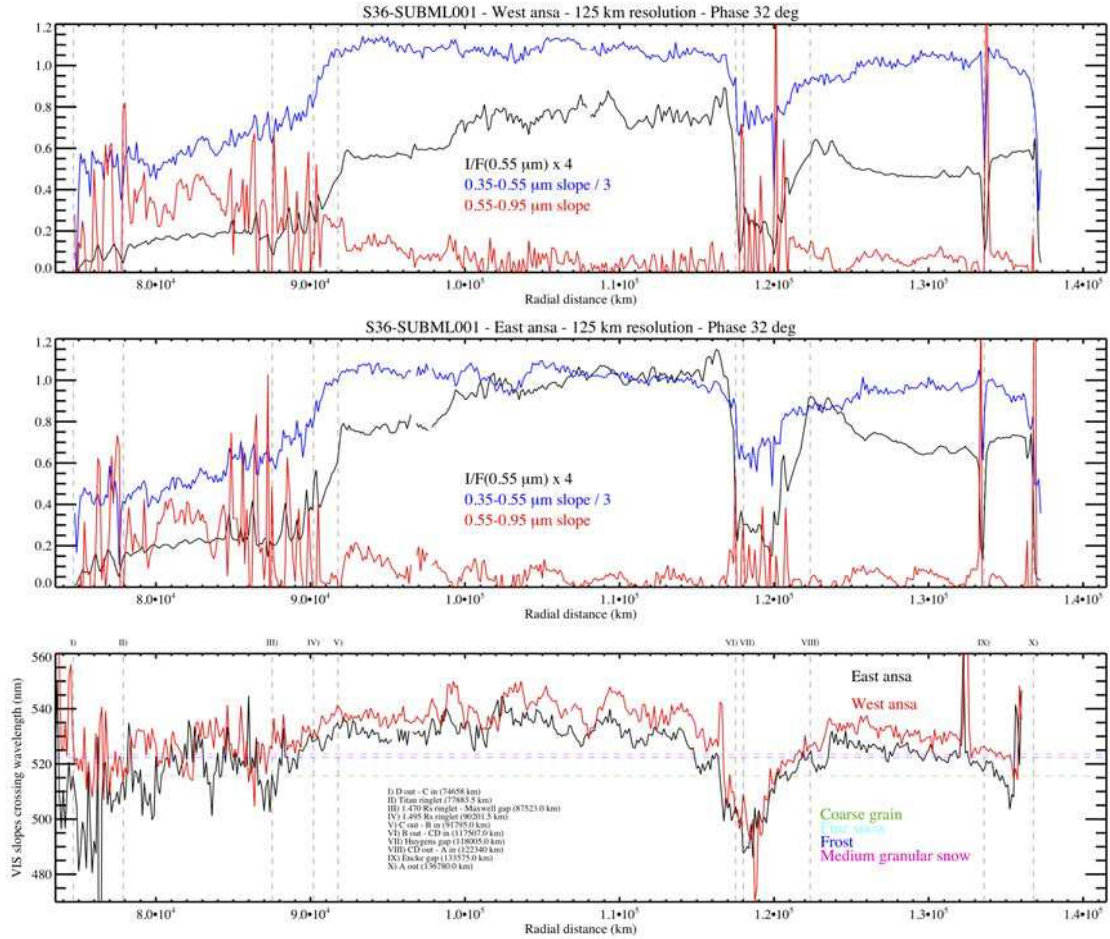

Figure 17: Rings visible slopes radial profiles (0.35 − 0.55 $\mu m$, 0.55 − 0.95 $\mu m$ slopes) and $I/F$(0.55 $\mu m$) for the west (top panel) and east ansae (central panel). The visible slopes crossing wavelength is an indicator of the water ice regolith particle sizes (bottom panel).

Moving across the B ring, three different distinct regions are observed: 1) external B ring (from 104,000 to 117,000 km), where band depths are almost flat with a local minimum at 109,000 km; 2) central B ring (from 98,500 to 104,000 km), where at least 5 ringlets with higher band depth are observed. These features can be explained by the presence of more pure, resurfaced water ice caused by collisional processes; 3) internal B ring (92,000 to 98,500 km), where flat to moderately decreasing band depth profiles are seen moving towards inward and without any evidence of local structures.

Across the C ring the two water ice band depths reach their minimum values; we observe a similarity between the outer part of the C ring (outside Maxwell gap) and the Cassini division.

As discussed in Cuzzi et al. (2009), the regolith grain size is mapped through the $I/F(3.6 \mu m)/I/F(1.822 \mu m)$ ratio (bottom panel in Fig. 18). VIMS is sensitive to the light reflected/scattered by the outer layer of the ring particles allowing us to measure the typical dimensions of the surface regolith grains. Both lab data and models indicate that the reflectance at 3.6 $\mu m$ depends on the water ice grain diameter (higher reflectances are measured for fine grains, as shown in Fig. 5). The 3.6 $\mu m$ reflectance is normalized to 1.822 $\mu m$ in order to remove the illumination and solar phase effects, which are variable across the rings. The ratio indicates regolith particle diameters of about 100 $\mu m$ across the outer part of the A and B rings; about 50 $\mu m$ in the CD and in the B-C rings boundary region and 30-50 $\mu m$ in the C ring.

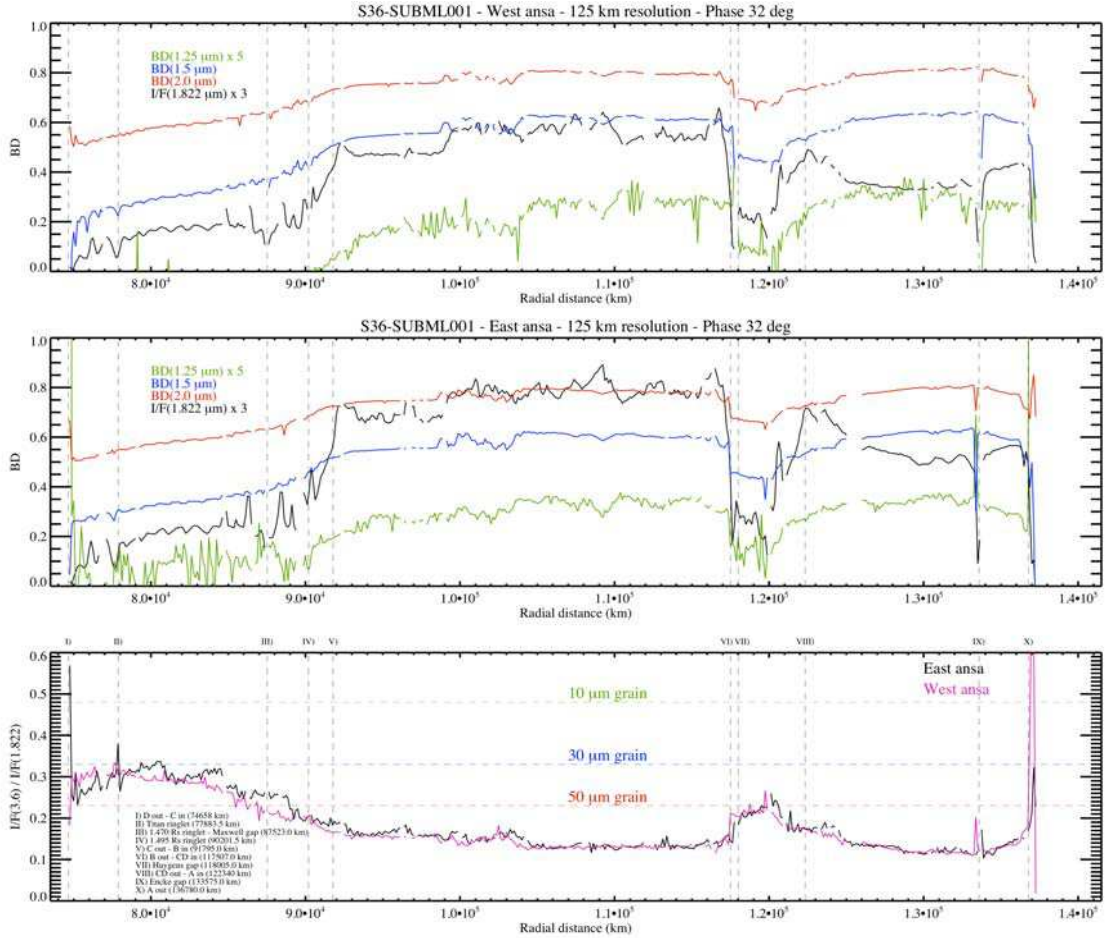

Figure 18: Rings water ice band depths radial profiles (1.25, 1.5, 2.0 $\mu m$) and I/F(1.822 $\mu m$) for the west (top panel) and east ansae (central panel). The ratio $I/F(3.6\mu m)/I/F(1.822\mu m)$ is an indicator of the water ice regolith particle sizes (bottom panel).

## 4. VIS spectral slopes

The classification of the icy satellites surface properties by means of the visible spectral slopes is discussed in two previous papers (Filacchione et al., 2007a, 2010). Here for completeness we use the same method to investigate a much larger dataset, which includes regular, minor satellites and rings observations in order to compare these very different objects orbiting in the Saturnian system. We include also a discussion of the dependency of the slopes on visual $I/F(0.55\ \mu m)$ and illumination conditions (solar phase).

As discussed in section 2.4, moderate negative (or blue) to neutral $S_{0.35-0.55}$ slope and high albedos are associated with fresh water icy surfaces. Moderate to positive (or red) slope is the result of contamination induced by the presence of very different materials, including complex organics produced by irradiation of simple hydrocarbons (Johnson et al. (1983), Moore et al. (1983)), amorphous silicates (Poulet et al., 2003), carbonaceous particles (Cuzzi and Estrada , 1998), nanophase iron or hematite (Clark et al., 2008), tholins intimately mixed in water ice grains (Ciarniello et al., 2011) or combinations of these materials.

Starting from this evidence, the variability of the spectral slopes across this variegated population of objects can be analyzed by using the $S_{0.35-0.55}$ vs. $S_{0.55-0.95}$ vs. $I/F(0.55\ \mu m)$ scatterplot shown in Fig. 19 and the "reduced" $S_{0.35-0.55}$ vs. $S_{0.55-0.95}$ scatterplot displayed in Fig.20. These three variables allow us to trace the effects of the intramolecular mixing of the chromophores in the ices ($S_{0.35-0.55}$), the intimate and areal mixing ($S_{0.55-0.95}$) and the dependence of albedo on phase ($I/F(0.55\ \mu m)$). Adopting such a representation, icy satellites show two distinctive behaviors that separate high albedo from low albedo objects. High albedo satellites (Mimas, Enceladus, Tethys, Dione, Rhea) have a large variability along the $I/F(0.55\ \mu m)$ axis caused by the different illumination conditions: as shown in the top panel of Fig. 19, low phase observations are grouped at higher $I/F(0.55\ \mu m)$ values while high phase points have lower reflectances. However, both $S_{0.35-0.55}$ and $S_{0.35-0.95}$ slopes have a maximum reddening at about 100° (see, for example the Rhea points, indicated as green crosses, for which we report the phase scale). Low albedo objects, like Iapetus' leading hemisphere and Hyperion, show a different trend (bottom panel in Fig. 19) having a lower excursion of albedo with the phase and an almost linear decrease of the $S_{0.35-0.55}$ and $S_{0.35-0.95}$ slopes between 40°-140° phase with no intermediate phase angle maximum (see, for example Iapetus' leading points, indicated as orange crosses, for which we report the phase scale; points similar to Enceladus correspond to trailing hemisphere observations).

The variability induced by solar phase on the two slopes of the principal satellites and Phoebe is shown in Fig. 21 - 22. The two slopes change with illumination conditions, in particular at very low and high phases where the opposition surge and forward scattering effects, respectively, become predominant in the color response of the icy surfaces (Ciarniello et al., 2011). Moreover, the mixing of the chromophores has an influence on these trends as well: while high-albedo objects can be better modeled with an intramolecular mixing of chromophores in the ices (see discussion in section 7), spectra of low-albedo satellites are compatible with areal and intimate mixtures (Clark et al., 2012). In the first case the maximum reddening is observed at phase about 100° while in the second case the peak occurs at about 50°.

We restrict now our discussion to the different colors of the icy satellites and rings without considering the photometric behavior which causes variations in albedo. Since the two spectral slopes are evaluated after the reflectance spectra were normalized at 0.55 $\mu m$ (see Eq. 2 - 3), the distribution of the slopes becomes sensitive only to the color variations. At least 4 different branches are distinguishable in the $S_{0.35-0.55}$ - $S_{0.55-0.95}$ slopes space shown in Fig. 20:

- An upper diagonal branch corresponding to the most organic-rich objects of the system, i.e., Hyperion (lime dots), the leading hemisphere of Iapetus (orange dots) and Phoebe (red dots). Hyperion has the highest $S_{0.55-0.95}$ slope, running between 0.5 to 1.0 $\mu m^{-1}$ while $0.7 \leq S_{0.35-0.55} \leq 1.6\ \mu m^{-1}$. Iapetus (leading hemisphere) observations have $0.1 \leq S_{0.35-0.55} \leq 0.9\ \mu m^{-1}$ and $0 \leq S_{0.55-0.95} \leq 0.5\ \mu m^{-1}$. On both objects we observe a linear decrease of the two slopes at increasing solar phase. Points with the lower values of the slopes correspond to high phase observations where the effects of the forward scattering are predominant (see next Fig. 21 - 22). Finally Phoebe points, taken at phase of about 85°, are scattered around $S_{0.35-0.55} = 0.3$ and $S_{0.55-0.95} = -0.12\ \mu m^{-1}$; the similarity between the Phoebe and Enceladus slope distributions is caused by the normalization at 0.55 $\mu m$ applied to the spectra, which decouples albedo from color.

- An intermediate diagonal branch formed by the trailing hemisphere observations of Iapetus (orange dots), Janus (blue diamonds) and the C ring (purple crosses). Iapetus trailing side points are clustered near $S_{0.35-0.55} = 0.85\ \mu m^{-1}$ and $0 \leq S_{0.55-0.95} \leq 0.12\ \mu m^{-1}$. Janus has a large dispersion, with the majority of the points displaced along a straight line starting from $(S_{0.35-0.55}, S_{0.55-0.95}) = (0.5, 0.5)\ \mu m^{-1}$ to $(1.9, 0.35)\ \mu m^{-1}$. The high-phase observations (#23 − 38) taken at $\phi > 100°$ have the smaller slopes of this dataset. Finally a very sparse distribution, centered approximately at $(S_{0.35-0.55}, S_{0.55-0.95}) = (2.0, 0.4)\ \mu m^{-1}$ corresponds to the C ring spectra taken from the S36-SUBML001 mosaic.

- A bottom diagonal branch includes the iciest objects of the population (except A and B rings):

    - Mimas (black dots in Fig. 20) slopes are similar to those measured on Enceladus, Tethys, and partially Dione. The $S_{0.35-0.55}$ ranges between 0.35 to 1.4 $\mu m^{-1}$ while $-0.25 \leq S_{0.55-0.95} \leq 0.1\ \mu m^{-1}$. With respect to our previous analysis (Filacchione et al., 2010), the more recently acquired data have formed a new horizontal branch at $S_{0.55-0.95} \approx 0.05\ \mu m^{-1}$; these points correspond to observations #199 − 208 in Fig. 7 taken at high solar phase ($\phi = 136° − 140°$). The reduction of the $S_{0.55-0.95}$ slopes at high solar phase angles seems to be a common spectral behavior across the inner regular satellites and is probably caused by forward scattering effects on regolith. The variability of the spectral slopes induced by solar phase is shown in Fig. 21 - 22;

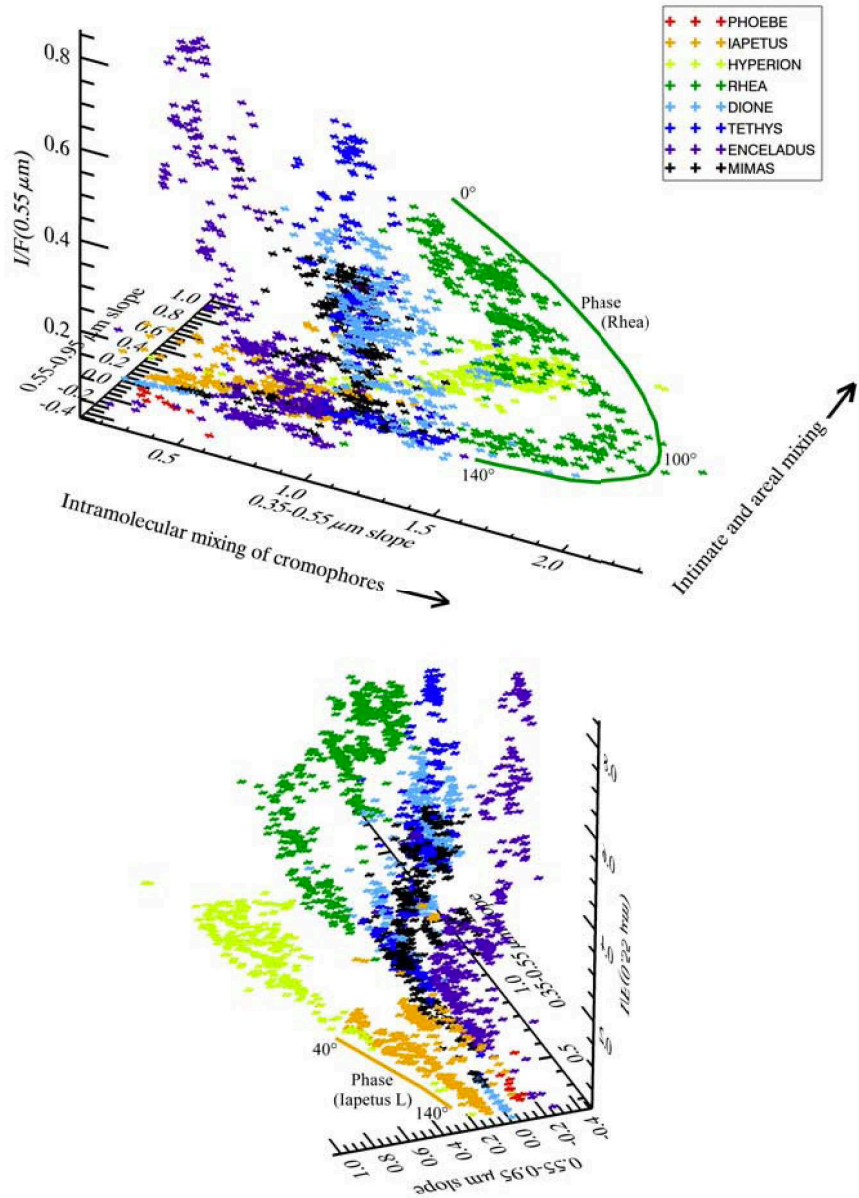

Figure 19: Icy satellites 3D scatterplot: visible spectral slopes $S_{0.35-0.55} - S_{0.55-0.95}$ vs. I/F(0.55). Top panel renders an optimized view of high albedo satellites while bottom panel allows to visualize low albedo objects. Phase scale is indicated for Rhea observations (top panel) and Iapetus-leading (bottom panel).

- Enceladus (purple dots) slopes have the lowest values among the icy satellites, $0.13 \leq S_{0.35-0.55} \leq$ $\mu m^{-1}$ and $-0.25 \leq S_{0.55-0.95} \leq 0.1\ \mu m^{-1}$ resulting in a compact cluster placed at the bottom-left corner of the scatterplot. As mentioned previously for Mimas, a decrease of the two spectral slopes for solar phases $> 110°$ is also seen for Enceladus;

- Tethys (blue dots) and Dione (cyan dots) have similar distributions centered at $(S_{0.35-0.55}, S_{0.55-0.95})$ equal to $(1, 0)\ \mu m^{-1}$, $(1.1, -0.15)\ \mu m^{-1}$, respectively. As shown in Fig. 21, the $S_{0.35-0.55}$ slope of the two satellites is constant across a wide solar phase range $0 \leq \phi \leq 120°$, while they fall at higher phases. The $S_{0.55-0.95}$ slope trends (Fig. 22) increase linearly with phase between $0 \leq \phi \leq 50°$ to become flatter at higher phases ($50 \leq \phi \leq 140°$);

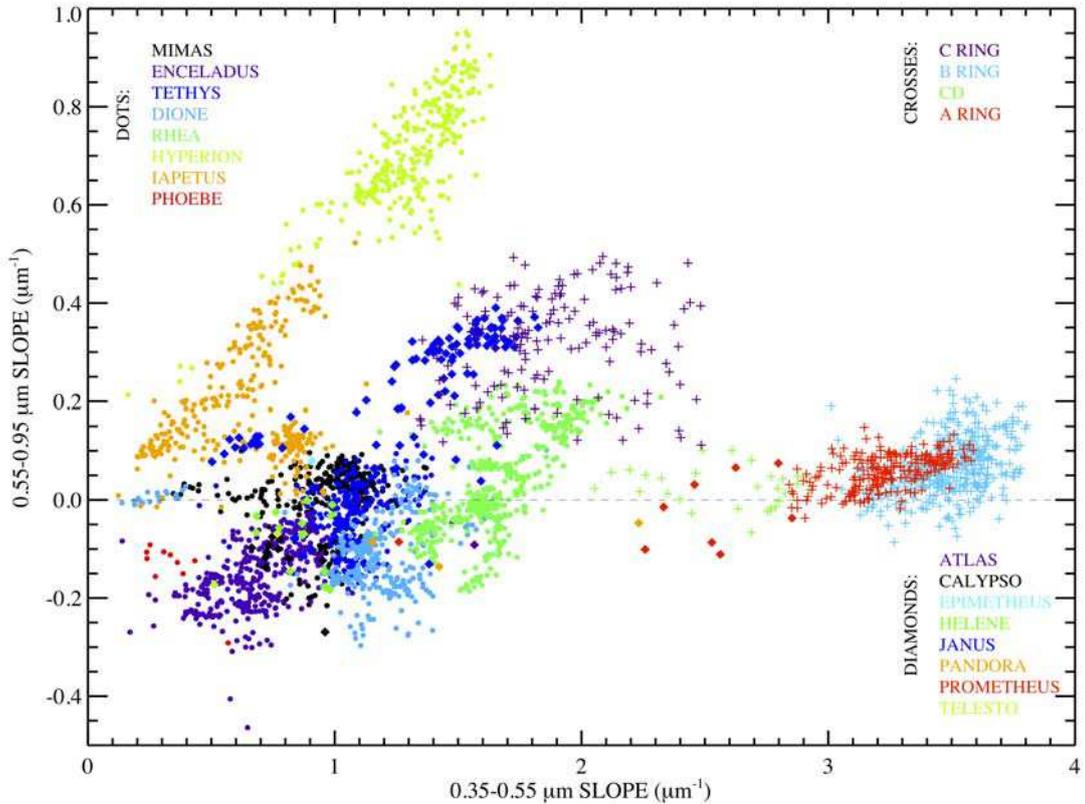

Figure 20: Icy satellites and rings scatterplot of the visible spectral slopes: $S_{0.35-0.55}$ $Vs.$ $S_{0.55-0.95}$.

- Rhea (green dots) has a wide distribution with slopes $S_{0.35-0.55} = (1.2 - 2.2)\ \mu m^{-1}$, $S_{0.55-0.95} = (-0.2 - 0.25)\ \mu m^{-1}$. The $S_{0.35-0.55}$ increases linearly from about 1.5 $\mu m^{-1}$ at phase $\phi = 0°$ to about 2.0 $\mu m^{-1}$ at $\phi = 110°$. Observations taken at higher phases decrease up to 1.2 $\mu m^{-1}$ (Fig. 21). The $S_{0.55-0.95}$ shows a linear increase with the phase in the $0° \leq \phi \leq 150°$ range (Fig. 22);

- Telesto (lime diamonds), Helene (green diamonds), Pandora (orange diamonds), Calypso (black diamonds) and Atlas (purple diamonds) are dispersed across the cluster but the limited number of available observations does not allow us to make an accurate analysis. Atlas is very similar to Rhea, Telesto and Calypso are close to Enceladus while Helene is compatible with Mimas. We find two Pandora observations similar to Dione, but one has a higher slope and is shifted toward the next branch.

• A high $S_{0.35-0.55}$ slope branch contains Prometheus (red diamonds), the Cassini division (green crosses), the A ring (red crosses) and the B ring (cyan crosses). All these objects have null to moderately positive $S_{0.55-0.95}$ slope while maintaining the highest $S_{0.35-0.55}$ slopes in the population. Among the minor satellites Prometheus has the highest reddening, up to $S_{0.35-0.55} = 2.8\ \mu m^{-1}$. It is interesting to observe that the two shepherd satellites, Pandora and Prometheus, have a different behavior at visible wavelengths; in fact, Prometheus has much more intense (thus red) $S_{0.35-0.55}$ slope while Pandora is more neutral. We have to note, however, that this conclusion is affected by the limited number of Pandora observations presently available and needs to be confirmed by future observations. Certainly, Prometheus has spectral properties that fall exactly between the distribution of the icy satellites and the more dense A and B rings, making it, ideally, the spectral bridge between the main rings and the icy satellites. The main ring particles, in fact, are among the reddest objects of the saturnian population: Cassini division points are spread on a wide range $2.0 \leq S_{0.35-0.55} \leq 3.2$ $\mu m^{-1}$ while A and B ring are clustered in the $2.8 \leq S_{0.35-0.55} \leq 3.8$ $\mu m^{-1}$ range, with the B ring having higher slopes with respect to the A ring. These values correspond to A-B rings average properties at 30° solar phase; when phase increases to 135°, we have measured much higher slopes, up to $S_{0.35-0.55} = 4.0$ $\mu m^{-1}$ and $S_{0.55-0.95} = 0.4$ $\mu m^{-1}$ (Filacchione et al., 2011).

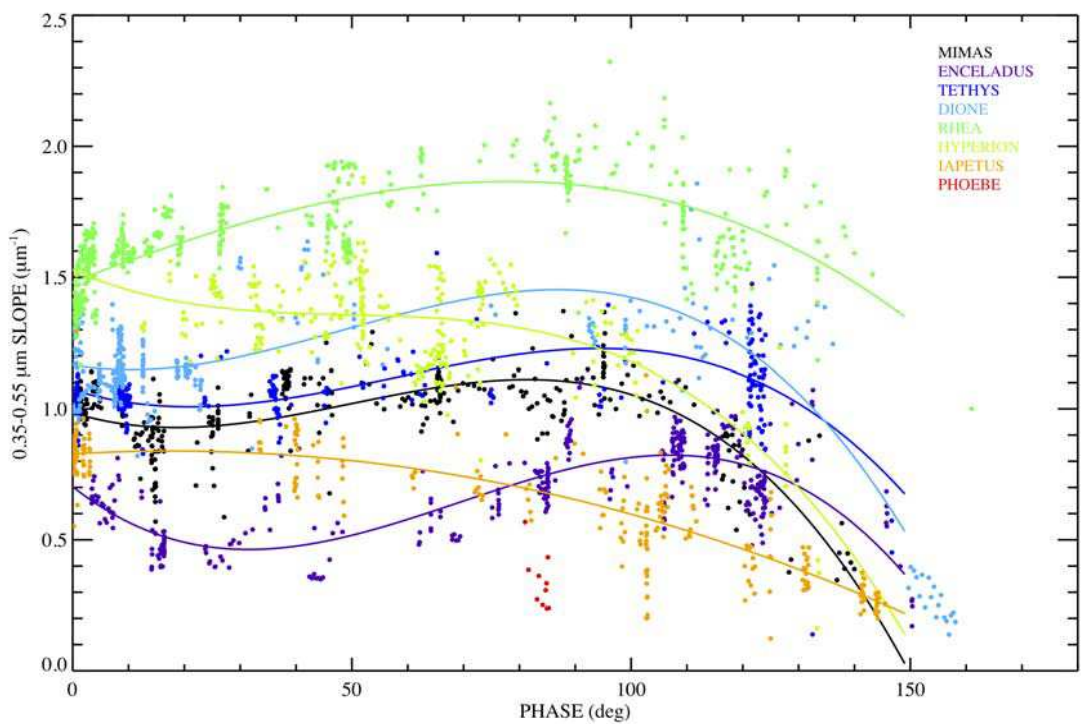

Figure 21: Variability of visible $S_{0.35-0.55}$ spectral slope induced by solar phase for the principal satellites and Phoebe. Solid lines indicate the best fit curve (third degree polynomial) for each satellite.

The trends observed in Fig. 20 have different dispersions depending on the satellite's coverage and illumination conditions. The variability induced by satellite coverage was discussed before in section 3. Concerning the variations of the slopes induced by solar phase we observe that the icy satellites have different responses; some of them, i.e. Enceladus, Tethys, Rhea and Mimas, as shown in Fig. 21, have a $S_{0.35-0.55}$ slope trend which increases with phase from about 20° up to about 100° followed by a decrease at higher phases. At low phases (from 20° to 0°) the slope increases and the spectrum becomes more red; this effect is particularly evident on Enceladus while not characterizing Rhea's points. A different behavior is seen for the more contaminated and red objects, e.g., Hyperion and Iapetus, whose $S_{0.35-0.55}$ slope drops almost linearly with phase.

The effect of phase on $S_{0.55-0.95}$ slope is completely different (Fig. 22): for Enceladus, Tethys, Dione, Mimas and Rhea we observe a trend linearly increasing with phase from 0° to 100 − 120° while Hyperion and Iapetus distributions are peaked at about 50 − 70°.

## 5. IR Water ice band depths

The variability of the 1.25, 1.5, 2.0 $\mu m$ water ice band depths is influenced by surface composition (amount of water ice and contaminants), surface regolith properties (grain size) and illumination conditions (solar phase).

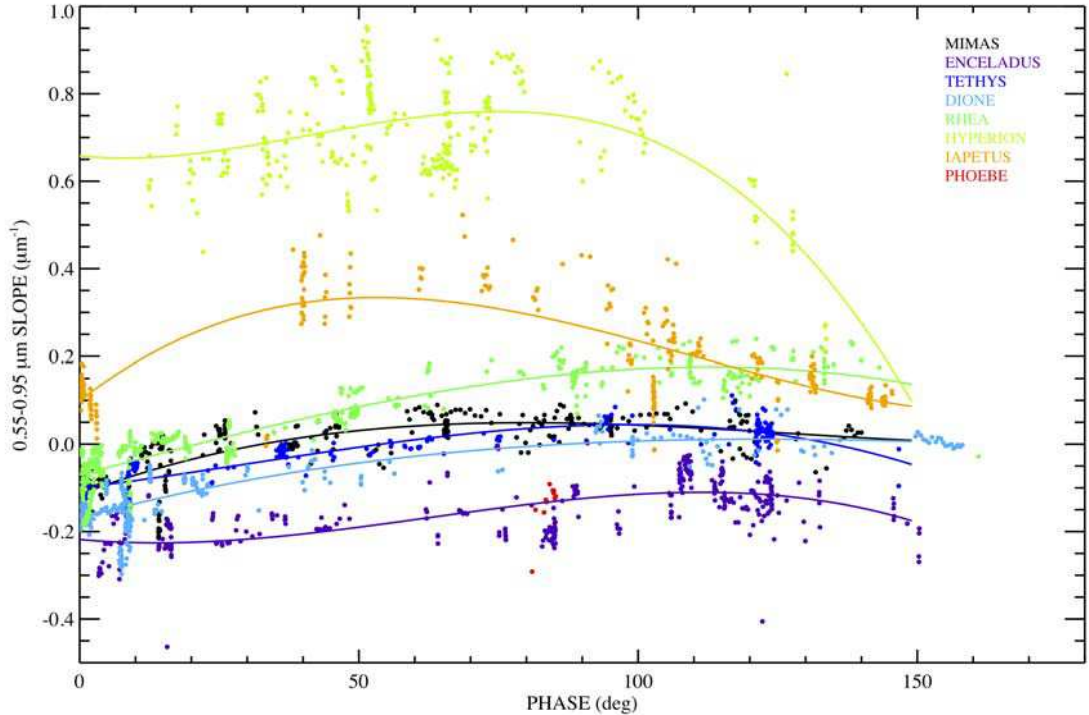

Figure 22: Variability of visible $S_{0.55-0.95}$ spectral slope induced by solar phase for the principal satellites and Phoebe. Solid lines indicate the best fit curve (third degree polynomial) for each satellite.

In Fig. 23 are shown the distribution of the 1.5 vs. 2.0 $\mu m$ band depths for the principal satellites and Phoebe (indicated with dots), minor satellites (diamonds) and rings (crosses). In order to improve visualization, minor satellites and rings points are shifted by +0.1 and +0.2 along the 2.0 $\mu m$ band depth axis, respectively. As we have discussed in previous works (Filacchione et al. (2007a), Cuzzi et al. (2009), Filacchione et al. (2010)), the principal icy satellites have water ice band depths that increase from dark and red objects (Phoebe, Iapetus leading hemisphere, Hyperion) to blue and fresh icy surfaces (Enceladus, Tethys).

Among the icy satellites, Iapetus shows the largest variability; while the majority of the observations correspond to the leading hemisphere, where the water ice band depths are minimal ($BD(1.5\,\mu m) = 0.1$, $BD(2.0\,\mu m) = 0.2$), a few Iapetus points (indicated in red), taken on the trailing hemisphere, have very high band depth values, up to $BD(1.5\,\mu m) = 0.55$, $BD(2.0\,\mu m) = 0.7$ making them very similar to the Enceladus and Tethys groups, which define the high-end extreme of the distribution. Also Phoebe has minimal water ice band depths, making it similar to the Iapetus' trailing hemisphere. These two objects define the lower-end extreme of the distribution.

The Dione points (in green) are spread across a wide range having $BD(1.5\,\mu m) = 0.2-0.4$, $BD(2.0\,\mu m) = 0.4-0.6$; two different clusters are observed in the distribution: the one at $BD(1.5\,\mu m) < 0.3$ corresponds to the trailing hemisphere (observations #37 − 67 and #235 − 259 in Fig. 10), while points having $BD(1.5\,\mu m) > 0.3$ are taken on the leading hemisphere. This evident dichotomy in surface water ice abundance among the two hemispheres points out to exogenic contamination process that impacts preferentially on the trailing side as discussed by Clark et al. (2008) and Stephan et al. (2010)).

Located close to the Dione and Iapetus average groups we find the compact Hyperion cluster (orange points) centered at about $BD(1.5\,\mu m) = 0.3$, $BD(2.0\,\mu m) = 0.55$. A small subgroup of points, corresponding to observations #78 − 87 in Fig. 12 taken at high phase angle ($\approx 130°$), have fainter $BD(2.0\,\mu m)$.

Many water ice-rich satellites, Rhea, Mimas, Tethys and Enceladus, are grouped in the upper branch having $BD(1.5\,\mu m) = 0.4 - 0.6$, $BD(2.0\,\mu m) = 0.55 - 0.7$. Moreover, the general branch slope becomes more flat in correspondence with Enceladus and Tethys points (indicated in blue and cyan, respectively).

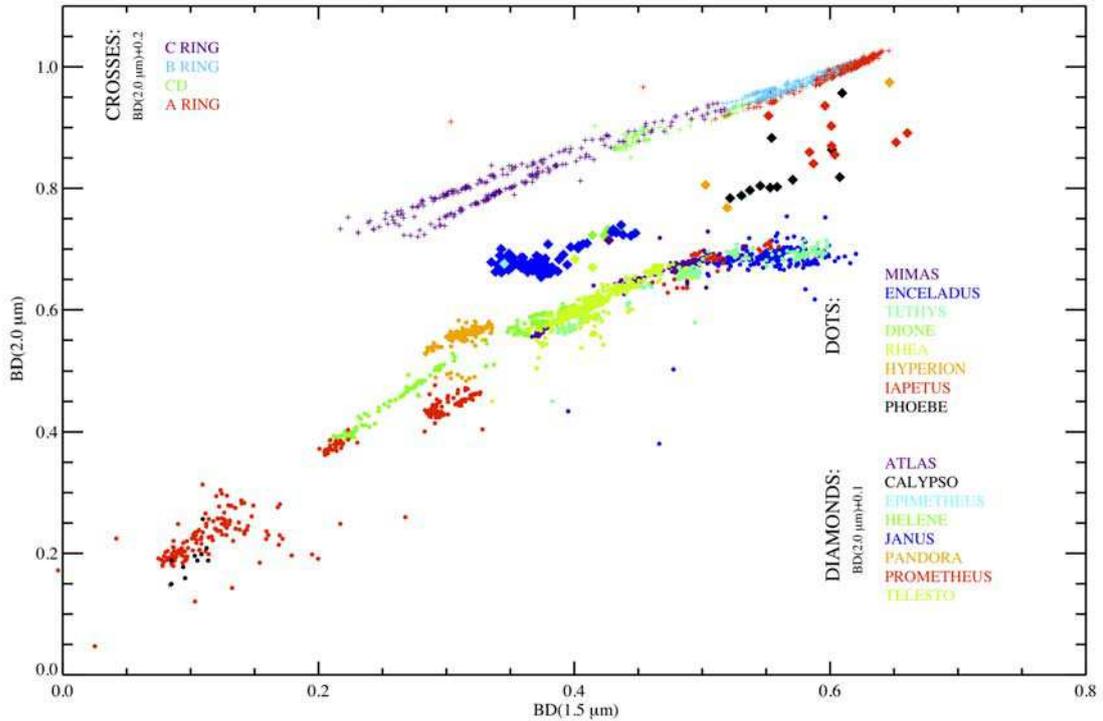

Figure 23: Scatterplot of water ice 1.5 vs. 2.0 $\mu m$ band depths for the principal satellites and Phoebe (dots), minor satellites (diamonds) and rings (crosses). Minor satellites and rings points are plotted with an offset of +0.1 and +0.2, respectively, along the BD(2.0 $\mu m$) axis to improve visualization. Rings points are divided in two branches, more separated on C ring and converging on A ring, with west ansa on the top and east ansa on the bottom.

The distribution of the minor satellites band depths, corresponding to diamonds in Fig. 23 are similar to the principal satellites and for this reason are shown with a +0.1 $BD(2.0\ \mu m)$ offset to avoid overlapping; Atlas, Janus, Epimetheus, Telesto, Helene are grouped together around $BD(1.5\ \mu m) = 0.4$, $BD(2.0\ \mu m) = 0.6$ and appear very similar to Rhea and to the leading hemisphere of Dione. Instead Calypso (black diamonds) and Prometheus (red diamonds) are scattered at much higher values, making them the most water ice-rich satellites in the Saturnian system. The three Pandora's points are intermediate and scattered, with only one observation pointing to high values. On some Prometheus observations we measure a $BD(2.0\ \mu m) = 0.85$. These results point out important consequences about the composition of minor satellites:

- the two F ring shepherd moons Prometheus and Pandora appear similar (in the limits of the three scattered Pandora points) and very rich in water ice particles torn from the nearby F ring;

- the two co-orbital satellites Janus and Epimetheus have similar water ice abundance (but Epimetheus statistics are poor because we have only one available observation, see Table 1);

- Calypso and Telesto, the two lagrangian moons of Tethys, appear very different from one another, with Calypso much more water ice rich than Telesto having a $\Delta BD(1.5\ \mu m) = +0.14$, $\Delta BD(2.0\ \mu m) = +0.05$;

- Helene, one of the lagrangian moons of Dione, is much more water ice-rich than Dione itself.

Finally, the rings points, indicated as crosses in Fig. 23 with a +0.2 $BD(2.0\ \mu m)$ offset, have an almost linear distribution increasing from the C ring and the Cassini division up to the A and B rings. Each point corresponds to the average band depths in a 125 km wide radial region between 73,500 to 141,375 km. The different solar phase angle and saturnshine contamination across the two ansae are responsible for the small divergence seen mainly across the C and B rings distributions, with the west ansa points having band depths slightly greater than the east ansa. The minimum band depths are measured across the C ring (dark blue crosses) where $BD(1.5\ \mu m) = 0.25 - 0.55$, $BD(2.0\ \mu m) = 0.55 - 0.75$. The Cassini division points (green crosses) are clustered in the $BD(1.5\ \mu m) = 0.45 - 0.55$, $BD(2.0\ \mu m) = 0.67 - 0.75$ ranges. Finally the A and B rings (red and light blue crosses respectively) define the upper edge of the distribution, with maxima placed approximately at $BD(1.5\ \mu m) = 0.64$, $BD(2.0\ \mu m) = 0.82$ and minima at $BD(1.5\ \mu m) = 0.52$, $BD(2.0\ \mu m) = 0.75$.

The same analysis for the 1.25 vs. 1.5 $\mu m$ band depths is shown in Fig. 24. Despite the fact that the 1.25 $\mu m$ band is much less intense, in particular for the more contaminated objects, and despite some calibration artifacts affecting this range, we retrieve a distribution that has many similarities with the 1.5 vs 2.0 $\mu m$. Using these band depth distributions and the resulting average values, which are tabulated in Table 6, we can retrieve the typical regolith grain sizes on the assumption of a pure water ice composition. This estimate is made using the band depth values inferred from laboratory measurements shown in Fig. 6.

As Clark et al. (2012) have demonstrated with laboratory measurements, the presence of nanophase particles of hematite mixed in water ice changes the 1.5-2.0 $\mu m$ band depth, center and shape (asymmetry), causing greater uncertainty in the retrieval of the grain size only through the comparison of the two band depths.

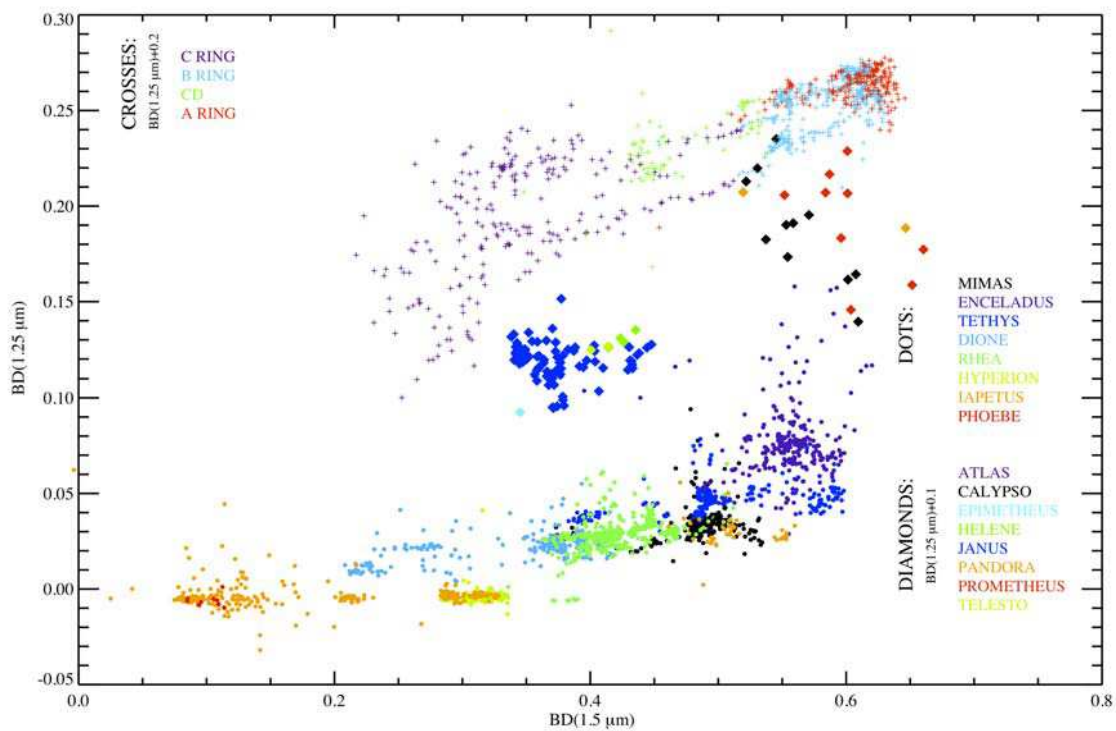

Figure 24: Scatterplot of water ice 1.5 vs. 1.25 $\mu m$ band depths for the principal satellites and Phoebe (dots), minor satellites (diamonds) and rings (crosses). Minor satellites and rings points are plotted with an offset of +0.1 and +0.2, respectively, along the 2.0 $\mu m$ axis to improve visualization. Rings points are divided in two branches with west ansa on the top and east ansa on the bottom.

Finally we discuss the effect of solar phase on the measurement of the water ice 2.0 $\mu m$ band depth as shown in Fig. 25 (top panel): from the theory of phase coefficients described in Gradie and Veverka (1986) is known that the band depth decreases at low and at high phases, while it is almost constant across a wide range of phases, typically between 20-30° to 100-120° (depending on material and grain size). In this range the single scattering regime dominates throughout the entire band. At phases < 15° the band depth decreases as shown in the bottom panel of Fig. 25 because the multiple scattering becomes predominant on the continuum, where $w$ is higher but it is less effective at the band center where $w$ is low and single scattering continue to prevail (Hapke (1993), sect. 11.E.2). In the figure are shown the 2.0 $\mu m$ band depth trends for the inner regular satellites. Since we are considering here different

| | 1.25 μm | | 1.5 μm | | 2.0 μm | | Average grain diameter | |
|---|---|---|---|---|---|---|---|---|
| Target | avg BD | stdev | avg BD | stdev | avg BD | stdev | 1.25-1.5 μm BD | 1.5-2.0 μm BD |
| C Ring (E) | n.a. | n.a. | 0.338 | 0.072 | 0.612 | 0.054 | 8 μm | 1.5 cm |
| C Ring (W) | 0.017 | 0.016 | 0.365 | 0.063 | 0.607 | 0.053 | 8 μm | 1.5 cm |
| B Ring (E) | 0.045 | 0.013 | 0.590 | 0.032 | 0.781 | 0.021 | 60 μm | 80 μm-0.5 cm |
| B Ring (W) | 0.061 | 0.008 | 0.585 | 0.028 | 0.771 | 0.020 | 60 μm | 80 μm-0.5 cm |
| CD (E) | 0.025 | 0.019 | 0.473 | 0.040 | 0.699 | 0.024 | 20 μm | 80 μm-0.5 cm |
| CD (W) | 0.036 | 0.014 | 0.467 | 0.042 | 0.683 | 0.0250 | 20 μm | 80 μm-0.5 cm |
| A Ring (E) | 0.057 | 0.010 | 0.611 | 0.031 | 0.794 | 0.023 | 60 μm | 80 μm-0.5 cm |
| A Ring (W) | 0.065 | 0.009 | 0.596 | 0.040 | 0.778 | 0.026 | 60 μm | 80 μm-0.5 cm |
| Atlas | n.a. | n.a. | 0.427 | n.a. | 0.614 | n.a. | n.a. | 1 cm |
| Prometheus | 0.092 | 0.028 | 0.604 | 0.033 | 0.783 | 0.031 | 60 μm | 80 μm-0.5 cm |
| Pandora | 0.098 | 0.013 | 0.556 | 0.079 | 0.749 | 0.109 | 40 μm | 80 μm-0.5 cm |
| Janus | 0.019 | 0.009 | 0.371 | 0.028 | 0.582 | 0.019 | 8 μm | 2 cm |
| Epimetheus | n.a. | n.a. | 0.345 | n.a. | 0.576 | n.a. | n.a. | 2 cm |
| Mimas | 0.035 | 0.054 | 0.470 | 0.044 | 0.655 | 0.039 | 20 μm | 0.8 cm |
| Enceladus | 0.075 | 0.019 | 0.552 | 0.037 | 0.681 | 0.029 | 40 μm | 80 μm-0.5 cm |
| Tethys | 0.049 | 0.056 | 0.498 | 0.056 | 0.655 | 0.042 | 30 μm | 0.8 cm |
| Calypso | 0.088 | 0.028 | 0.563 | 0.031 | 0.728 | 0.052 | 40 μm | 80 μm-0.5 cm |
| Telesto | 0.026 | 0.001 | 0.407 | 0.010 | 0.577 | 0.009 | 10 μm | 1 cm |
| Dione | 0.021 | 0.007 | 0.342 | 0.063 | 0.541 | 0.071 | 8 μm | 2 cm |
| Helene | 0.030 | 0.004 | 0.425 | 0.009 | 0.627 | 0.006 | 10 μm | 1 cm |
| Rhea | 0.028 | 0.008 | 0.422 | 0.027 | 0.615 | 0.031 | 10 μm | 1 cm |
| Hyperion | n.a. | n.a. | 0.314 | 0.011 | 0.553 | 0.018 | n.a. | 2 cm |
| Iapetus L | n.a. | n.a. | 0.111 | 0.031 | 0.221 | 0.035 | n.a. | < 1 μm |
| Iapetus T | 0.030 | 0.008 | 0.514 | 0.028 | 0.679 | 0.022 | 20 μm | 80 μm-0.5 cm |
| Phoebe | n.a. | n.a. | 0.100 | 0.011 | 0.188 | 0.029 | n.a. | < 1 μm |

Table 6: Statistical results for water ice band depths at 1.25, 1.5, 2.0 μm measured on rings, principal and minor satellites of Saturn. The average grain diameters are retrieved comparing the 1.25-1.5 μm BD and the 1.5-2.0 μm BD with values shown in Fig. 5

observations taken at every sub-spacecraft point longitudes available, the curves can have variable length in phase as well as different intensities: the former effect depends on the number of consecutive observations available while the second depends on the hemispherical variability of each satellite. Despite these limitations, the general trend is clearly evident for these satellites; a particularly well-shaped and continuos curve to illustrate this effect corresponds to Dione's points running from phase 9°, BD(2.0 μm)=0.48 to phase 0.2°, BD(2.0 μm)=0.423. At high phases (> 120°), forward/multiple scattering prevails again on the continuum and as a consequence a significative decrease in band depth is observed.

The theoretical trend matches with the 2.0 μm band depth variation as measured by VIMS on different icy satellites, i.e. Mimas, Enceladus, Tethys, Rhea and Hyperion. The distribution of Iapetus and Dione band depths is more scattered since these two satellites have a much larger spatial variability.

## 6. VIS spectral slopes vs. IR Water ice band depths

In this section we direct our analysis to the correlations between VIS and IR spectral indicators, in particular $S_{0.35-0.55}$ slope and 2.0 μm band depth for icy satellites and the main rings. Since these two quantities are sensitive to contaminants and the presence of water ice, respectively, their correlations allow us to derive general properties among the objects considered in this study.

Referring to the scatterplot shown in Fig. 26 the icy satellites and rings are clustered in four principal clusters departing from $S_{0.35-0.55} \approx 1.5$ $\mu m^{-1}$ and $BD(2.0\ \mu m) \approx 0.55$:

- High albedo icy satellites: these are grouped in the top-left diagonal branch with Enceladus, Tethys, Iapetus trailing hemisphere, Pandora and Calypso as extremes, with a minimum $BD(2.0\ \mu m) \approx 0.57$ and a minimum $S_{0.35-0.55} \approx 0.2$ $\mu m^{-1}$. These objects are the most water ice-rich, having large band depth and low reddening;

- Low albedo icy satellites: these are spread through the bottom-left diagonal branch which includes Phoebe, Iapetus leading hemiphere and Dione trailing hemisphere. These surfaces are characterized by the lower amount of water ice of the population having a minimum $BD(2.0\ \mu m) \approx 0.15$ and a minimum $S_{0.35-0.55} \approx 0.3$ $\mu m^{-1}$;

- A-B rings and Prometheus: these are clustered in the top-right diagonal branch where we observe the maximum $BD(2.0\ \mu m) \approx 0.85$ and maximum reddening $S_{0.35-0.55} \approx 3.8$ $\mu m^{-1}$.

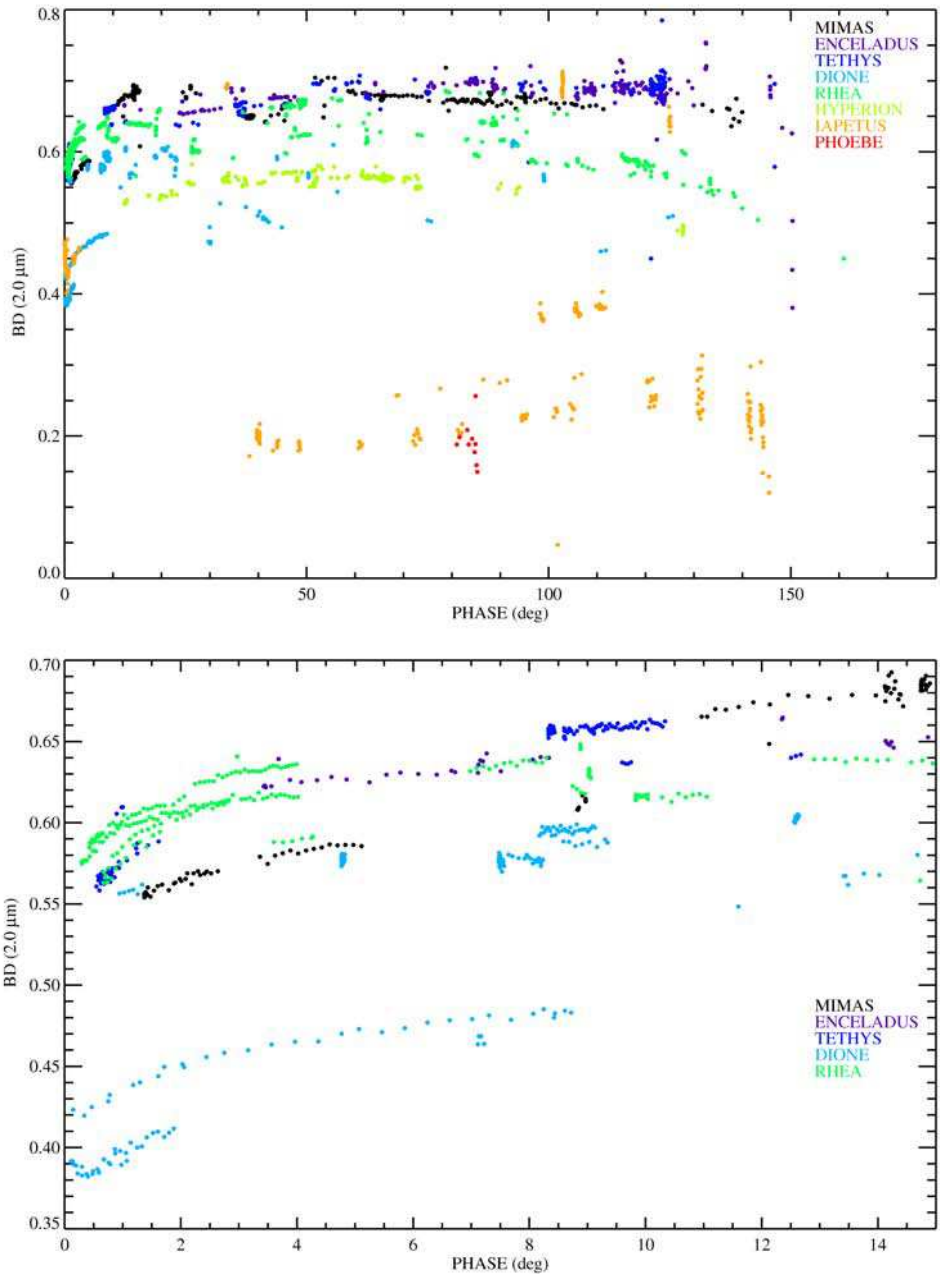

Figure 25: Top panel: variability of water ice 2.0 $\mu m$ band depth with phase for principal satellites and Phoebe (complete dataset). Bottom panel: the decrease of the 2.0 $\mu m$ band depth with solar phase is evident in this zoom spanning across the $0-15°$ phase range for the inner regular satellites.

- Intermediate albedo satellites and rings: these are distributed around the $S_{0.35-0.55} \approx 1.5\,\mu m^{-1}$ and $BD(2.0\,\mu m) \approx 0.55$ values. This group includes Rhea, Dione leading hemisphere, Hyperion, Janus, C ring and CD.

## 7. Comparative Spectral Modeling using Hapke's theory

Hapke (1993) theory is used to perform spectral modeling of icy satellites data presented in the previous sections.

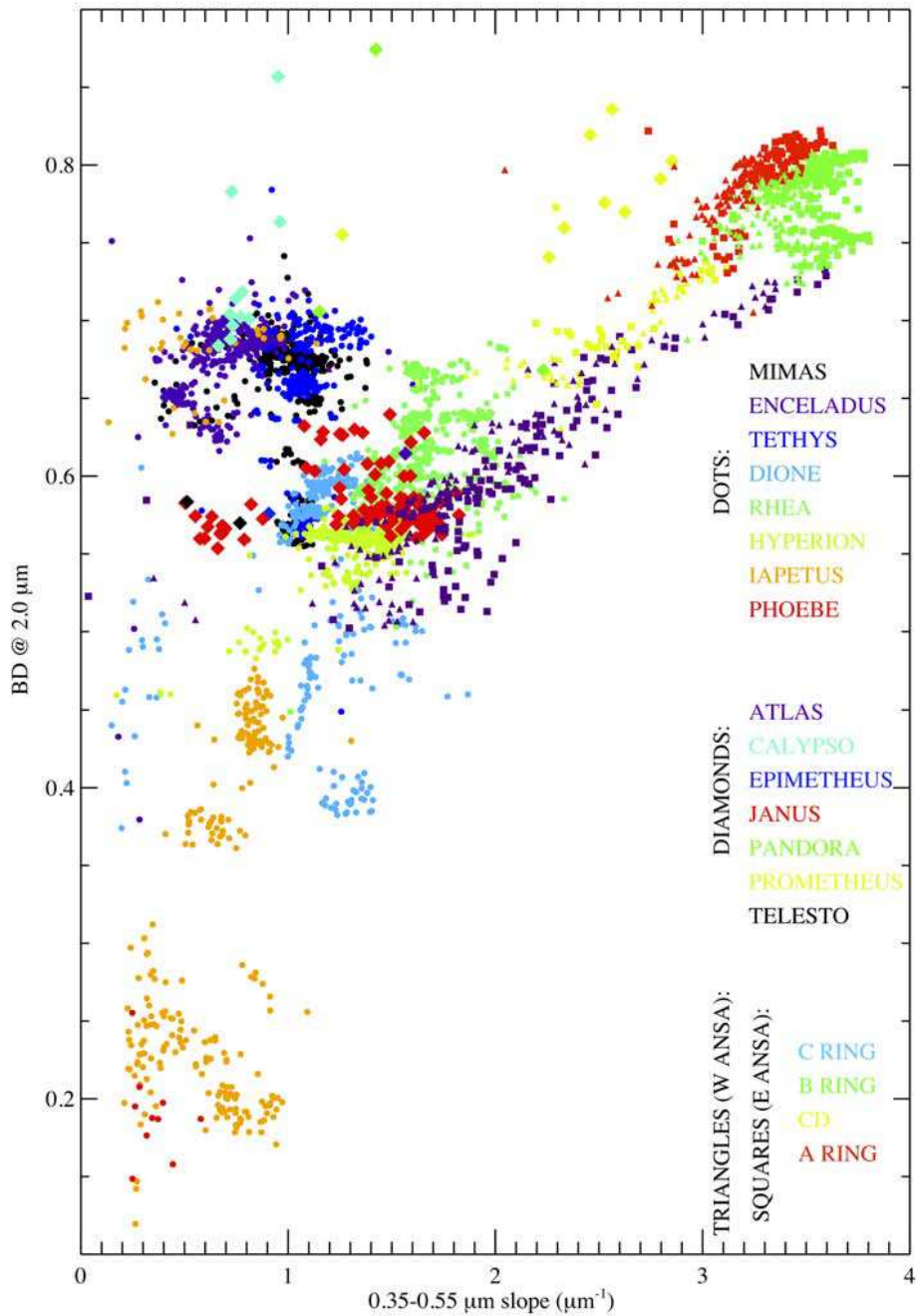

Figure 26: Scatterplot of the visible slope $S_{0.35-0.55}$ $vs.BD(2.0\ \mu m)$ for icy satellites and rings.

In a previous paper, Ciarniello et al. (2011) have described the method used to reproduce the spectral properties of Rhea observed by VIMS in the VIS-IR range. The same method is applied here to Mimas, Enceladus, Tethys and Dione spectra. The results we discuss come from a detailed analysis directed toward identifying compositional similarities and differences among the major satellites of Saturn. Further details about this method are given in Ciarniello (2012).

Satellites surfaces are modeled by means of a mixture of crystalline water ice and one organic contaminant. In the models we have developed tholin (Khare et al., 1993), Triton tholin (McDonald et al. (1994), optical constants from D. Cruikshank, personal communication), Titan tholin (McDonald et al. (1994), Khare et al. (1984), optical constants from D. Cruikshank, personal communication) and hydrogenated amorphous carbon, $ACH_2$ (Zubko et al., 1996) are used as contaminants mixed in water ice.

Optical constants for crystalline water ice are those derived by Warren et al. (1984) (0.35-1.25 $\mu m$, at T=266.15 K), Mastrapa et al. (2008) (1.25-2.5 $\mu m$, at T=120 K), Mastrapa et al. (2009) (2.5-3.20 $\mu m$, at T=120 K) and Clark et al. (2012) (3.20-5.12 $\mu m$, at T=120 K).

Spectral fits for the different satellites are performed applying the same paradigm previously adopted by Ciarniello (2012) for Rhea: 1) VIMS spectra are normalized at 1 $\mu m$ in order to minimize the contribution of the single-particle phase function (assumed isotropic) and to eliminate the shadowing effect caused by surface roughness; 2) Spectra at various phase angles have been best-fitted with the parameters specified in the next paragraphs, e.g., grain sizes and distributions, mixing type and material endmembers.

*Mimas modeling.* The spectrum we analyzed (red line in Fig. 27, top left panel) corresponds to the leading hemisphere of the satellite and was acquired at phase g = 87.5°. The best fit spectrum (black line) has been modeled with an intimate mixture of two type of grains: Ice (I) in a percentage of pi = 98% and Carbon(C) for the remaining pc = 2%. I grains are an intraparticle mixture of water ice (p = 99.9%) and Triton tholin (pt = 0.1%), while C particles are made of amorphous carbon. The grain diameter of both I and C particles in this model is 58 $\mu m$.

*Enceladus modeling.* The best match to the Enceladus spectrum (Fig. 27, top right panel) is obtained at phase g = 107° adopting an intraparticle mixture of water ice (99.992%) and Triton tholin (0.008%) with particle diameter of 63 $\mu m$.

*Tethys modeling.* In (Fig. 27, bottom left panel) are shown the VIMS spectrum corresponding to the leading hemisphere of the satellite at phase g = 84.9° and the best-fit solution, which was obtained with an intimate mixture of two types of grains: I(pi = 99.2%) and C(pc = 0.8%). I grains are an intraparticle mixture of water ice (p = 99.9%) and Triton tholin (pt = 0.1%), while C particles are made of amorphous carbon. The diameter of both I and C grains is 69 $\mu m$.

*Dione modeling.* The spectrum we analyzed (Fig. 27, bottom right panel) corresponds to the leading hemisphere of the satellite and it was acquired at phase g = 84.9°. The best-fit solution consists of an intimate mixture of two types of grains: I(pi = 89%) and C(pc =11%). I grains are an intraparticle mixture of water ice (p = 99.7%) and Triton tholin (pt = 0.3%), while C particles are made of amorphous carbon. The diameter of both I and C grains is 59 $\mu m$.

In general the peak at 0.5 $\mu m$ is well reproduced by the model, especially on Enceladus, but it does not exclude the possibility of Rayleigh scattering by nanophase absorbers. These particles are not explicitly modeled in this work but are largely used in (Clark et al., 2012) to explain the reddening observed across many icy surfaces of the saturnian system. Moreover, we have not applied the diffraction effects in the Hapke model that, according to (Clark et al., 2012), allow a better fit in the 2.4-2.6 $\mu m$ region and to reproduce the asymmetry seen across the 1.5-2.0 $\mu m$ bands.

For the moons considered in this work (Mimas to Rhea), the paradigm represented by an intraparticle mixture of water ice and Triton tholin seems to be a reasonable one to explain the downturn observed by VIMS in the UV range. Some discrepancies between observed data and the models remain, in particular in the 2.4-2.6 and 3.1 $\mu m$ ranges where the simulated spectra are more sensitive to grain size effects and uncertainties in the optical constants. More contaminated objects require the addition of amorphous carbon particles to match the albedo levels. Table 7 shows that the UV reddening, caused in this model by the presence of tholin, is correlated with the orbital distance of the satellites with respect to Enceladus, where the amount of tholin required to fit the VIMS spectra has a minimum. A reasonable explanation for this distribution is that Enceladus is the source that feeds the E ring with almost pure ice particles by cryovolcanic activity (Porco et al., 2006). These particles hit the surface of the satellites that orbit in the E ring (radial distance from Saturn from 180,000 to 480,000 km), with a flux that is maximum at a position corresponding to the orbit of Enceladus (Verbiscer et al., 2007). This flux of particles provides almost entirely uncontaminated ice to the surfaces of the moons. However, as shown in Schenk et al. (2011), other spectral properties of moons orbiting within the E ring can be explained by interaction with energetic electrons that are driven onto the leading side of the satellites by Saturn's magnetic field, providing energy for chemical processes and tholin production (Khare et al. (1984), Khare et al. (1993)). From this point of view, space weathering by energetic particles is more efficient on satellites farther from Saturn, where the flux of icy particles from the E ring is lower and the surfaces are not refreshed by pure ice, and where the electron flux is higher. This may explain why the inner three satellites (Mimas, Enceladus and Tethys) form a distinct clump in Fig. 26, but leaves open the question why the ring connects more nearly to the other clump of icy moons than this one.

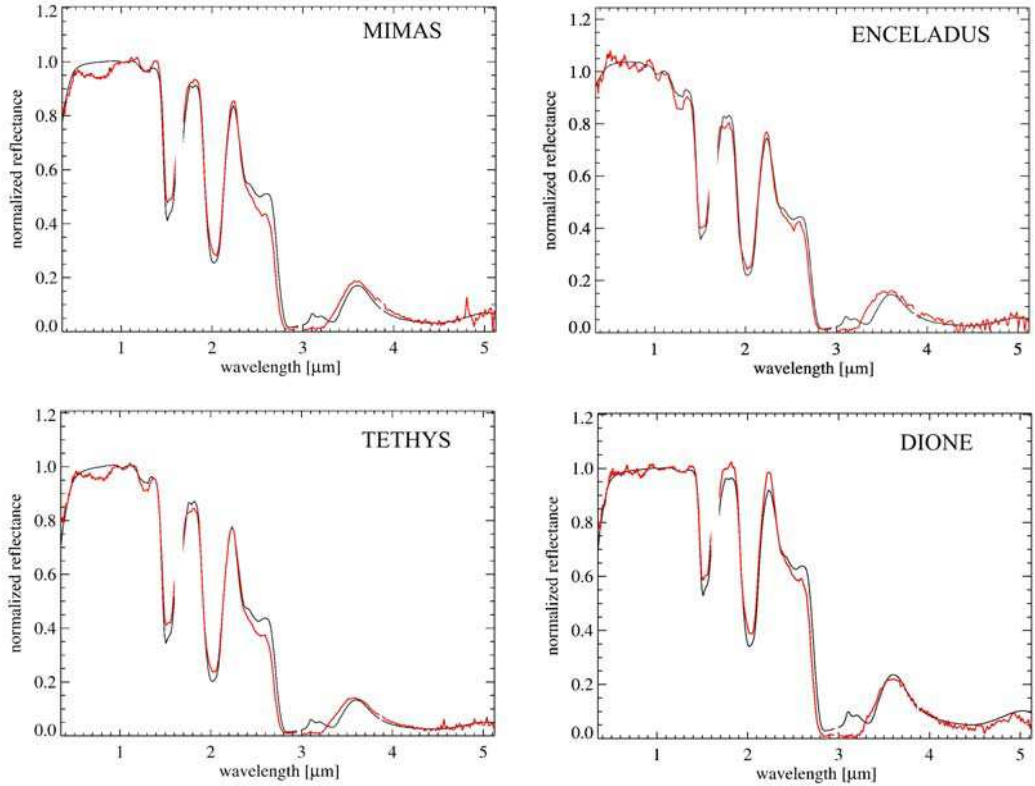

Figure 27: Spectral fits to Mimas (top left), Enceladus (top right), Tethys (bottom left) and Dione (bottom right). VIMS average spectra are the red curves, best fits are the black curves.

| Satellite | Orbital radius ($R_S$) | Grain size ($\mu m$) | Water ice | Tholin | $ACH_2$ |
|---|---|---|---|---|---|
| Mimas | 3.08 | 58 | 97.9% | 0.1% | 2.0% |
| Enceladus | 3.95 | 63 | 99.992% | 0.008% | – |
| Tethys | 4.88 | 69 | 99.1% | 0.1% | 0.8% |
| Dione | 6.26 | 43 | 88.7% | 0.3% | 11.0% |
| Rhea | 8.74 | 38 | 99.6% | 0.4% | – |

Table 7: Icy satellites spectral fit parameters summary. Rhea's parameters from Ciarniello et al. (2011).

# 8. Conclusions

In conclusion, the principal results obtained in this work are summarized here:

1. A statistically significant dataset consisting of 2,298 observations of Saturn's satellites and one ring mosaic are systematically processed in order to derive common spectrophotometric indicators. From these data we have retrieved the average spectral properties of the entire system, spanning from the C ring to Phoebe (Fig. 1, 2, 3).
2. VIMS observations are processed using updated calibration pipeline as discussed in sct. 2.2. Geometric information is retrieved for each hyperspectral cube in order to correlate spectrophotometric parameters with observation conditions (Fig. 7 to 14).
3. Spectra of the icy satellites and rings at the hemispherical scale are dominated by water ice, showing evident 1.5 and 2.05 $\mu m$ absorption features. Overtones at 1.05 and 1.25 $\mu m$ are seen on fresher surfaces (A-B rings, Enceladus, Calypso). The 4.26 $\mu m$ $CO_2$ ice absorption is evident mainly on the three outermost satellites, Hyperion, Iapetus and Phoebe. The absence of any other absorption features makes it difficult to decipher the nature of contaminant/darkening agents.
4. Among the satellites, the surfaces of Enceladus and Calypso appear the most blue and water ice-rich.
5. Iapetus leading hemisphere and Phoebe's surface are redder and metal/organic-rich.
6. The rings are redder than the icy satellites in the VIS range but show more intense 1.5-2.0 $\mu m$ band depths.
7. Janus and Epimetheus have average surface water ice abundances similar to particles in the C ring (Fig. 23) but with much less reddening contaminant (Fig. 20).
8. Prometheus and Pandora, the two F ring shepherd satellites, have very obvious differences in surface composition; Prometheus is rich in water ice but at the same time very red at VIS wavelengths. These properties make it very similar to A-B ring particles, while Pandora appears bluer.
9. The surfaces of all the satellites from Mimas to Rhea are differentially contaminated by E ring particles released by the Enceladus plumes. The dichotomy observed between leading and trailing hemispheres is mainly caused by layers of exogenic material and by interactions with magnetospheric particles.
10. The two lagrangian moons of Tethys show spectral diversity: Calypso is much more water ice-rich and bluer with respect to Telesto. This could be caused by a different initial composition of the two moons or by a preferential deposition of E ring particles occurring on Calypso.
11. Helene, one of Dione's lagrangian moons, appears bluer and hence richer in fresh water ice than Dione.
12. The outer satellites, moving from Hyperion, to Iapetus' leading hemisphere and Phoebe show a linear decrease of water ice and $S_{0.35-055}$ slope (Fig. 26).
13. The correlations among spectral slopes, band depths, visual albedo and phase permit us to cluster the saturnian population in different spectral classes which are detected not only among the principal satellites and rings but among co-orbital minor moons as well (Fig. 19-20 and 26). These bodies are effectively the "connection" elements, both in term of composition and evolution, between the principal satellites and main rings.
14. VIMS data demonstrate that both spectral slopes (Fig. 21-22) and band depth (Fig. 25) are phase-dependent. A significant reduction in band depth at low phases is caused by multiple scattering, which dominates single scattering on the continuum wings.
15. The method used to build the VIS-IR spectrogram (Fig. 16) from VIMS rings mosaics (Fig. 15) is given. This tool allows us to measure the average spectral properties vs. radial distance (Fig. 17 and 18) for the rings. Our analysis points out that the A and B ring spectra have the reddest VIS slopes and deepest water ice band depths, while the C ring and Cassini division reddening is lower, appearing compatible to the values for Janus and Rhea.
16. The comparison of rings 1.25-1.5 $\mu m$ band depth averages with pure water ice bands points to ring particle regolith diameters of 8, 60, 20, 60 $\mu m$ for the C, B, CD, A rings, respectively (Table 6). Fairly larger values are given by the ratio $I/F(3.6\mu m)/I/F(1.822\mu m)$, which indicates regolith particle sizes of about 100 $\mu m$ across the outer part of the A and B rings, about 50 $\mu m$ in the CD and in the B-C rings boundary region and 30-50 $\mu m$ in the C ring.
17. Modeling results using Hapke theory indicate that an intraparticle mixture of water ice and Triton tholin (up to 0.4% for Rhea) and amorphous carbon (up to 11% for Dione) gives a good fit for the inner satellites spectra (from Mimas to Rhea) with typical regolith grain sizes running between 38 to 69 $\mu m$ (Table 7).

In closing, the method we have applied in this and in the previous two papers (Filacchione et al. (2007a), Filacchione et al. (2010)) about Saturn's icy satellites, is mainly focused on retrieving the VIS-IR spectral properties of the very different icy objects populating Saturn's system. The application of spectral modeling codes, e.g., Hapke's model, gives very good fits to observed spectra with different combinations of water ice and non-icy materials (tholins, carbon). When these results are matched with spectrophotometric indicators, e.g., spectral slopes and absorption band parameters, they allow a comparative study of the colors and compositions of icy objects across the system, and enable us to correlate them with the geological history of each satellite. This is an essential step also towards a better evaluation of the size of the particles forming the regoliths of the ring particles.

In the future we will continue this work by deriving the compositional trends measured across the system, from rings to satellites. These results will help us to verify the different transport scenarios of non-icy materials in the system and consequently to understand better the origins and formation processes of the entire system. Finally, we plan to include in this work more recent and future VIMS rings and icy satellites observations in order to enlarge the dataset, to improve the statistical analysis, and to search for possible temporal-seasonal variations.

## Acknowledgments

This paper is dedicated to the memory of our colleague and friend Angioletta Coradini recently passed away who has inspired us to complete this work. The authors thank VIMS technical team at Lunar and Planetary Lab (University of Arizona, Tucson, AZ), the Cassini-Huygens Project, including engineering and operational teams, at JPL (Pasadena, CA) for the hard work and dedication demonstrated during the entire mission. We gratefully thank the Cassini Icy Satellites and Rings Working groups for observations planning and sequencing. Without all them this study would be impossible.

This research was completed thanks to the financial support of the Italian Space Agency (ASI) and NASA through the Cassini project.

## References

Brown, R.H., Baines, K.H., Bellucci, G., Bibring, J.P., Buratti, B.J., Capaccioni, F., Cerroni, P., Clark, R.N., Coradini, A., Cruikshank, D.P., Drossart, P., Formisano, V., Jaumann, R., Langevin, Y., Matson, D.L., McCord, T.B., Mennella, V., Nelson, R.M., Nicholson, P.D., Sicardy, B., Sotin, C., Amici, S., Chamberlain, M.A., Filacchione, G., Hansen, G.B., Hibbitts, C.A., Showalter, M., 2003. Observations with the visual and infrared mapping spectrometer (VIMS) during Cassinis flyby of Jupiter. *Icarus*, 164, 461-470.

Brown, R.H., Baines, K.H., Bellucci, G., Bibring, J.-P., Buratti, B.J., Capaccioni, F., Cerroni, P., Clark, R.N., Coradini, A., Cruikshank, D.P., Drossart, P., Formisano, V., Jaumann, R., Langevin, Y., Matson, D.L., McCord, T.B., Mennella, V., Miller, E., Nelson, R.M., Nicholson, P.D., Sicardy, B., Sotin, C., 2004. The Cassini Visual and Infrared Mapping Spectrometer (VIMS) investigation. *Space Sci. Rev.* 115 (1-4), 111-168.

Brown, R. H., Clark, R. N., Buratti, B. J., Cruikshank, D. P., Barnes, J. W., Mastrapa, R. M. E., Bauer, J. Newman, S., Momary, T., Baines, K. H., Bellucci, G., Capaccioni, F., Cerroni, P., Combes, M., Coradini, A., Drossart, P., Formisano, V., Jaumann, R., Langevin, Y., Matson, D. L., McCord, T. B., Nelson, R. M., Nicholson, P. D., Sicardy, B., Sotin, C., 2006. Composition and Physical Properties of Enceladus' Surface, *Science*, 311, 1425-1428, doi: 10.1126/science.1121031.

Buratti, B.J., Mosher, J.A., Johnson, T.V., 1990. Albedo and color maps of the saturnian satellites. *Icarus*, 87, 339357.

Buratti, B. J., Soderlund, K., Bauer, J., Mosher, J. A., Hicks, M. D., Simonelli, D. P., Jaumann, R., Clark, R. N., Brown, R. H., Cruikshank, D. P., Momary, T., 2008. Infrared (0.83-5.1 $\mu$ m) photometry of Phoebe from the Cassini Visual Infrared Mapping Spectrometer, *Icarus*, 193, 309-322, doi: 10.1016/j.icarus.2007.09.014.

Buratti, B. J., Bauer, J. M., Hicks, M. D., Mosher, J. A., Filacchione, G., Momary, T., Baines, K. H., Brown, R. H., Clark, R. N., Nicholson, P. D., 2010. Cassini spectra and photometry 0.25-5.1 $\mu$m of the small inner satellites of Saturn. *Icarus*, 206, 524-536, doi: 10.1016/j.icarus.2009.08.015.

Buratti, B. J., Brown, R. H., Clark, R. N., Mosher, J. A., Cruikshank, D. P., Filacchione, G., Baines, K. H., Nicholson, P. D., 2010. Peeling the Onion: The Upper Surface of Mimas from Cassini VIMS. American Astronomical Society, DPS meeting #42, #1.08; Bulletin of the American Astronomical Society, Vol. 42, 942.

Buratti,B. J., Brown, R. H. , Clark, R. N., Mosher, J. A., Cruikshank, D. P., Filacchione, G., Baines, K. H., Nicholson, P. D., Sotin, C, 2011. Mimas between 0.35-5.1 microns from Cassini VIMS Observations. Lunar and Planetary Science Conference, Houston, abstract n. 1634.

Carbary, J. F., Mitchell, D. G., Krupp, N., Krimigis, S. M., 2009. L shell distribution of energetic electrons at Saturn. *Journal of Geophysical Research*, 114, Issue A9, CiteID A09210, doi: 10.1029/2009JA014341.

Charnoz, S., Brahic, A., Thomas, P. C., Porco, C. C., 2007. The Equatorial Ridges of Pan and Atlas: Terminal Accretionary Ornaments? *Science*, 318, 1622-1624, DOI: 10.1126/science.1148631

Ciarniello, M., Capaccioni, F., Filacchione, G., Clark, R. N., Cruikshank, D. P., Cerroni, P., Coradini, A., Brown, R. H., Buratti, B. J., Tosi, F., Stephan, K. Hapke modeling of Rhea surface properties through Cassini-VIMS spectra. *Icarus*, 214, 541-555, DOI:10.1016/j.icarus.2011.05.010

Ciarniello, M., 2012. Theoretical models of the spectrophotometric properties of atmophereless bodies surfaces in the Solar System. PhD thesis in Astronomy, Universitá degli studi di Roma la Sapienza.

Clark, R. N., Lucey, P. G., 1984. Spectral properties of ice-particulate mixtures and implications for remote sensing. I - Intimate mixtures *Journal of Geophysical Research*, 89, 6341-6348, doi: 10.1029/JB089iB07p06341.

Clark, R.N., 1999. Spectroscopy of rocks and minerals, and principles of spectroscopy. In: Rencz, A.N. (Ed.), *Remote Sensing for the Earth Sciences, Manual of Remote Sensing*, vol. 3. John Wiley and Sons, New York, pp. 358.

Clark, R. N., Brown, R. H., Jaumann, R., Cruikshank, D. P., Nelson, R. M., Buratti, B. J., McCord, T. B., Lunine, J., Baines, K. H., Bellucci, G., Bibring, J.-P., Capaccioni, F., Cerroni, P., Coradini, A., Formisano, V., Langevin, Y., Matson, D. L., Mennella, V., Nicholson, P. D., Sicardy, B., Sotin, C., Hoefen, T. M., Curchin, J. M., Hansen, G., Hibbitts, K., Matz, K.-D., 2005. Compositional maps of Saturn's moon Phoebe from imaging spectroscopy, *Nature*, 435, 66-69, doi: 10.1038/nature03558.

Clark, R. N., Curchin, J. M., Jaumann, R., Cruikshank, D. P., Brown, R. H., Hoefen, T. M., Stephan, K., Moore, J. M., Buratti, B. J., Baines, K. H., Nicholson, P. D., Nelson, R. M., 2008. Compositional mapping of Saturn's satellite Dione with Cassini VIMS and implications of dark material in the Saturn system, *Icarus*, 193, 372-386, doi: 10.1016/j.icarus.2007.08.035.

Clark, R. N., Cruikshank, D. P., Jaumann, R., Brown, R. H., Curchin, J. M., Hoefen, T. D., Stephan, K., Buratti, B. J., Filacchione, G., Baines, K. H., Nicholson, P. D. The Composition of Iapetus: Mapping Results from Cassini VIMS, 2012. *Icarus*, 218, 831-860, doi:10.1016/j.icarus.2012.01.008

Collins, S. A., Diner, J., Garneau, G. W., Lane, A. L., Miner, E. D., Synnott, S. P., Terrile, R. J., Holberg, J. B., Smith, B. A., Tyler, G. L., 1984. Atlas of Saturn's rings, in Planetary rings, University of Arizona Press, 737-743.

Coradini, A., Tosi, F., Gavrishin, A. I., Capaccioni, F., Cerroni, P., Filacchione, G., Adriani, A., Brown, R. H., Bellucci, G., Formisano, V., D'Aversa, E., Lunine, J. I., Baines, K. H., Bibring, J.-P., Buratti, B. J., Clark, R. N., Cruikshank, D. P., Combes, M., Drossart, P., Jaumann, R., Langevin, Y., Matson, D. L., McCord, T. B., Mennella, V., Nelson, R. M., Nicholson, P. D., Sicardy, B., Sotin, C., Hedman, M. M., Hansen, G. B., Hibbitts, C. A., Showalter, M., Griffith, C., Strazzulla, G., 2008. Identification of spectral units on Phoebe, *Icarus*, 193, 233-251, doi: 10.1016/j.icarus.2007.07.023.

Cruikshank, D. P., Dalton, J. B., Ore, C. M. D., Bauer, J., Stephan, K., Filacchione, G., Hendrix, A. R., Hansen, C. J., Coradini, A., Cerroni, P., Tosi, F., Capaccioni, F., Jaumann, R., Buratti, B. J., Clark, R. N., Brown, R. H., Nelson, R. M., McCord, T. B., Baines, K. H., Nicholson, P. D., Sotin, C., Meyer, A. W., Bellucci, G., Combes, M., Bibring, J.-P., Langevin, Y., Sicardy, B., Matson, D. L., Formisano, V., Drossart, P., Mennella, V., 2007. Surface composition of Hyperion, *Nature*, 448, 54-56, doi: 10.1038/nature05948.

Cruikshank, D. P., Wegryn, E., Dalle Ore, C. M., Brown, R. H., Bibring, J.-P., Buratti, B. J., Clark, R. N., McCord, T. B., Nicholson, P. D., Pendleton, Y. J., Owen, T. C., Filacchione, G., Coradini, A., Cerroni, P., Capaccioni, F., Jaumann, R., Nelson, R. M., Baines, K. H., Sotin, C., Bellucci, G.,Combes, M., Langevin, Y., Sicardy, B., Matson, D. L., Formisano, V., Drossart, P., Mennella, V., 2008. Hydrocarbons on Saturn's satellites Iapetus and Phoebe, *Icarus*, 193, 334-343, doi: 10.1016/j.icarus.2007.04.036.

Cruikshank, D. P., Meyer, A. W., Brown, R. H., Clark, R. N., Jaumann, R., Stephan, K., Hibbitts, C. A., Sandford, S. A., Mastrapa, R. M. E., Filacchione, G., Ore, C. M. D., Nicholson, P. D., Buratti, B. J., McCord, T. B., Nelson, R. M., Dalton, J. B., Baines, K. H., Matson, D. L., 2010. Carbon dioxide on the satellites of Saturn: Results from the Cassini VIMS investigation and revisions to the VIMS wavelength scale, *Icarus*, 206, 561-572, doi: 10.1016/j.icarus.2009.07.012.

Cuzzi, J.N., Estrada, P.R., 1998. Compositional evolution of Saturns rings due to meteoroid bombardment. *Icarus*, 132, 1-35.

Cuzzi, J.N., Clark, R.N., Filacchione, G., French, R., Johnson, R., Marouf, E., Spilker, L., 2009. Ring particle composition and size distribution, in *Saturn from Cassini-Huygens*, edited by M.K. Dougherty, L.W. Esposito, and S.M. Krimigis, Springer, Berlin, 459-509, DOI:10.1007/978-1-4020-9217-6_15.

Cuzzi, J. N., Burns, J. A., Charnoz, S., Clark, R. N., Colwell, J. E., Dones, L., Esposito, L. W., Filacchione, G., French, R. G., Hedman, M. M., Kempf, S., Marouf, E. A., Murray, C. D., Nicholson, P. D., Porco, C. C., Schmidt, J., Showalter, M. R., Spilker, L. J., Spitale, J. N., Srama, R., Sremcevic, M., Tiscareno, M. S., Weiss, J., 2010. An Evolving View of Saturn's Dynamic Rings, *Science*, 327, Issue 5972, 1470-1475, doi: 10.1126/science.1179118.

DAversa, E., Bellucci, G., Nicholson, P. D., Hedman, M. M., Brown, R. H., Showalter, M. R., Altieri, F., Carrozzo, F. G., Filacchione, G., Tosi, F., 2010. The spectrum of a Saturn ring spoke from Cassini/VIMS. *Geophysical Research Letters*, 37, L01203, doi:10.1029/2009GL041427.

Filacchione, G., 2006. Calibrazioni a terra e prestazioni in volo di spettrometri ad immagine nel visibile e nel vicino infrarosso per lesplorazione planetaria (On-ground calibrations and in flight performances of VIS-NIR imaging spectrometers for planetary exploration). PhD dissertation, Universitá degli studi di Napoli Federico II. Available at ftp:\\ftp.iasf-roma.inaf.it\gianrico\phd\Filacchione_PHD_2006.pdf (in Italian).

Filacchione, G., Coradini, A., Capaccioni, F., Cerroni, P., Bellucci, G., Brown, R. H., Baines, K. H., Bibring, J.-P., Buratti, B. J., Clark, R. N., Combes, M., Cruikshank, D. P., Drossart, P., Formisano, V., Jaumann, R., Langevin, Y., Matson, D. L., McCord, T. B., Mennella, V., Nelson, R. M., Nicholson, P. D., Sicardy, B., Sotin, C., Adriani, A., Moriconi, M., D'Aversa, E., Tosi, F., Colosimo, F., 2006. VIS-NIR Spectral Properties of Saturn's Minor Icy Moons. *37th Annual Lunar and Planetary Science Conference*, March 13-17, 2006, League City, Texas, abstract no.1271.

Filacchione, G., Capaccioni, F., McCord, T. B. Coradini, A., Cerroni, P., Bellucci, G., Tosi, F., D'Aversa, E., Formisano, V., Brown, R. H., Baines, K. H., Bibring, J. P., Buratti, B. J., Clark, R. N., Combes, M., Cruikshank, D. P., Drossart, P., Jaumann, R., Langevin, Y., Matson, D. L., Mennella, V., Nelson, R. M., Nicholson, P. D., Sicardy, B., Sotin, C., Hansen, G., Hibbitts, K., Showalter, M., Newman, S., 2007. Saturn's icy satellites investigated by Cassini-VIMS. I. Full-disk properties: 350 5100 nm reflectance spectra and phase curves, *Icarus*, 186, 259-290, doi: 10.1016/j.icarus.2006.08.001.

Filacchione, G., Capaccioni, F., Coradini, A., Cerroni, P., Tosi, F., Bellucci, G., Brown, R. H., Baines, K. H., Buratti, B. J., Clark, R. N., Nicholson, P. D., Nelson, R. M., Cuzzi, J. N., McCord, T. B., Hedman, M. H., Showalter, M. R., 2007. Cassini-VIMS Observations of Saturn's Main Rings. *38th Lunar and Planetary Science Conference*, LPI Contribution No. 1338, p.1513.

Filacchione, G., Capaccioni, F., Clark, R. N., Cuzzi, J. N., Cruikshank, D. P., Coradini, A., Cerroni, P., Nicholson, P. D., McCord, T. B., Brown, R. H., Buratti, B. J., Tosi, F., Nelson, R. M., Jaumann, R., Stephan, K., 2010. Saturn's icy satellites investigated by Cassini-VIMS. II. Results at the end of nominal mission, *Icarus*, 206, 507-523, doi: 10.1016/j.icarus.2009.11.006

Filacchione, G., Capaccioni, F., Ciarniello, M., Nicholson, P. D., Hedmann, M. M., Clark, R. N., Brown, R. H., Cerroni, P., 2011. VIS-IR spectrograms of Saturn's rings retrieved from Cassini-VIMS radial mosaics, *EPSC-DPS Joint Meeting 2011*, EPSC Abstracts, Vol. 6, EPSC-DPS2011-293.

Giese, B., Wagner, R., Neukum, G., Helfenstein, P., Thomas, P. C., 2007. Tethys: Lithospheric thickness and heat flux from flexurally supported topography at Ithaca Chasma. *Geophysical Research Letters*, 34, Issue 21, CiteID L21203, doi: 10.1029/2007GL031467.

Gradie, J., Veverka, J., 1986. The wavelength dependence of phase coefficients. *Icarus*, 66, 455-467, doi: 10.1016/0019-1035(86)90085-0.

Hapke, B., 1993. Theory of Reflectance and Emittance Spectroscopy. Cambridge Univ. Press, Cambridge, UK.

Hedman, M. M., Nicholson, P. D., Cuzzi, J. N., Clark, R. N., Filacchione, G., Capaccioni, F., Ciarniello, M. Correlations between Spectra and Structure in Saturn's Main Rings, submitted to *Icarus*.

Hillier, J.K., Green, S.F., McBride, N., Schwanethal, J.P., Postberg, F., Srama, R., Kempf, S., Moragas-Klostermeyer, G., McDonnell, J.A.M., Grün, E., 2007. The composition of Saturns E ring. *Monthly Notices of the Royal Astronomical Society*, 377 (4), 15881596. doi:10.1111/j.1365-2966.2007.11710.x.

Jaumann, R., Stephan, K., Hansen, G. B., Clark, R. N., Buratti, B. J., Brown, R. H., Baines, K. H., Newman, S. F., Bellucci, G., Filacchione, G., Coradini, A., Cruikshank, D. P., Griffith, C. A., Hibbitts, C. A., McCord, T. B., Nelson, R. M., Nicholson, P. D., Sotin, C., Wagner, R., 2008. Distribution of icy particles across Enceladus' surface as derived from Cassini-VIMS measurements. *Icarus*, 193, 407-419, doi: 10.1016/j.icarus.2007.09.013.

Jaumann, R., Clark, R. N., Nimmo, F., Hendrix, A. R., Buratti, B. J., Denk, T., Moore, J. M., Schenk, P. M., Ostro, S. J., Srama, R., 2009. Icy Satellites: Geological Evolution and Surface Processes, in *Saturn from Cassini-Huygens*, edited by M.K. Dougherty, L.W. Esposito, and S.M. Krimigis, Springer, Berlin, 459-509. DOI: DOI 10.1007/978-1-4020-9217-6_20

Johnson, R. E., Lanzerotti, L. J., Brown, W. L., Augustyniak, W. M., Mussil, C., 1983. Charged particle erosion of frozen volatiles in ice grains and comets. *A&A*, 123, 343-346,

Khare, B. N., Sagan, C., Arakawa, E. T., Suits, F., Callcott, T. A., Williams, M. W., 1984. Optical constants of organic tholins produced in a simulated Titanian atmosphere: From soft x-ray to microwave frequencies. *Icarus*, 60, 127-137.

Khare, B. N., Thompson, W. R., Cheng, L., Chyba, C., Sagan, C., Arakawa, E. T., Meisse, C., Tuminello, P. S., 1993. Production and Optical Constants of Ice Tholin from Charged Particle Irradiation of (1:6) $C_2H_6=H_2O$ at 77 K. *Icarus*, 103, 290-300.

Krupp, N. et al., 2009. Energetic particles in Saturns magnetosphere during the Cassini nominal mission (July 2004July 2008). *Planet. Space Sci.* 57, 1754-1768. doi:10.1016/j.pss.2009.06.010.

Mastrapa, R., Bernstein, M., Sandford, S., Roush, T., Cruikshank, D., Dalle Ore, C., 2008. Optical constants of amorphous and crystalline H2O-ice in the near infrared from 1.1 to 2.6 $\mu m$. *Icarus*, 197, 307-320, doi: 10.1016/j.icarus.2008.04.008.

Mastrapa, R. M., Sandford, S. A., Roush, T L., Cruikshank, D. P., Dalle Ore, C M., 2009. Optical constants of amorphous and crystalline h2o-ice: 2.5 - 22 $\mu m$ (4000 - 455 $cm^{-1}$) optical constants of h2o-ice. *Astrophysical Journal*, 701, 1347-1356, doi: 10.1088/0004-637X/701/2/1347.

Matson, D. L., Spilker, L. J., Lebreton, J-P., 2002. The Cassini/Huygens Mission to the Saturnian System. *Space Science Reviews*, 104, Issue 1, 1-58.

Matson, D. L., Castillo-Rogez, J. C., Schubert, G. Sotin, C. McKinnon, W. B., 2009. The Thermal Evolution and Internal Structure of Saturn's Mid-Sized Icy Satellites, in *Saturn from Cassini-Huygens*, edited by M.K. Dougherty, L.W. Esposito, and S.M. Krimigis, Springer, Berlin, 577-612. ISBN 978-1-4020-9216-9.

McCord, T. B., Coradini, A, Hibbitts, C.A., Capaccioni, F., Hansen, G. B., Filacchione, G., Clark, R. N., Cerroni, P., Brown, R. H., Baines, K. H., Bellucci, G., Bibring, J.-P., Buratti, B. J., Bussoletti, E., Combes, M., Cruikshank, D. P., Drossart, P., Formisano, V., Jaumann, R., Langevin, Y., Matson, D.L., Nelson, R.M., Nicholson, P.D., Sicardy, B., Sotin, C., 2004. CassiniVIMS observations of the Galilean satellites including the VIMS calibration procedure. *Icarus* 172, 104-126, doi: 10.1016/j.icarus.2004.07.001.

McDonald, G. D., Thompson, W. R., Heinrich, M., Khare, B. N., Sagan, C., 1994. Chemical Investigation of Titan and Triton Tholins. *Icarus*, 108, 137-145.

Moore, M. H., Donn, B., Khanna, R., A'Hearn, M. F., 1983. Studies of proton-irradiated cometary-type ice mixtures. *Icarus*, 54, 388-405, doi: 10.1016/0019-1035(83)90236-1.

Moore, J. M., Horner, V. M., Greeley, R., 1985. The geomorphology of Rhea: implications for geologic history and surface processes. *J. Geophys. Res.*, Vol. 90, Suppl., p. C785 - C795.

Nicholson, P. D., Hedman, M. M., Clark, R. N., Showalter, M. R., Cruikshank, D. P., Cuzzi, J. N., Filacchione, G., Capaccioni, F., Cerroni, P., Hansen, G. B., Sicardy, B., Drossart, P., Brown, R. H., Buratti, B. J., Baines, K. H., Coradini, A., 2008. A close look at Saturn's rings with Cassini VIMS, *Icarus*, 193, 182-212, doi: 10.1016/j.icarus.2007.08.036.

Pinilla-Alonso,N., Roush, T. L., Marzo, G., Cruikshank, D. P., Dalle Ore, C. M., 2011. Iapetus surface variability revealed from statistical clustering of a VIMS mosaic: the distribution of $CO_2$. *Icarus*, accepted.

Pitman, K. M., Buratti, B. J., Mosher, J. A., 2010. Disk-integrated bolometric Bond albedos and rotational light curves of saturnian satellites from Cassini Visual and Infrared Mapping Spectrometer. *Icarus*, Volume 206, Issue 2, p. 537-560, doi: 10.1016/j.icarus.2009.12.001.

Porco, C.C., Baker, E., Barbara, J., Beurle, K., Brahic, A., Burns, J.A., Charnoz, S., Cooper, N., Dawson, D.D., Del Genio, A.D., Denk, T., Dones, L., Dyudina, U., Evans, M.W., Giese, B., Grazier, K., Helfenstein, P., Ingersoll, A.P., Jacobson, R.A., Johnson, T.V., McEwen, A., Murray, C.D., Neukum, G., Owen, W.M., Perry, J., Roatsch, T., Spitale, J., Squyres, S., Thomas, P.C., Tiscareno, M., Turtle, E., Vasavada, A.R., Veverka, J., Wagner, R., and West, R., 2005. Cassini imaging science: initial results on Phoebe and Iapetus, *Science*, 307, 12371242, doi:10.1126/science.1107981.

Porco, C. C., Helfenstein, P., Thomas, P. C., Ingersoll, A. P., Wisdom, J., West, R., Neukum, G., Denk, T., Wagner, R., Roatsch, T., Kieffer, S., Turtle, E., McEwen, A., Johnson, T. V., Rathbun, J., Veverka, J., Wilson, D., Perry, J., Spitale, J., Brahic, A., Burns, J. A., Del Genio, A. D., Dones, L., Murray, C. D., Squyres, S., 2006. Cassini Observes the Active South Pole of Enceladus, *Science*, 311, Issue 5766, 1393-1401, doi: 10.1126/science.1123013.

Porco, C. C., Thomas, P. C., Weiss, J. W., Richardson, D. C., 2007. Saturns Small Inner Satellites: Clues to Their Origins. *Science*, 318, 1602-1607, DOI: 10.1126/science.1143977

Poulet, F., Cruikshank, D. P., Cuzzi, J. N., Roush, T. L., French, R. G., 2003. Compositions of Saturn's rings A, B, and C from high resolution near-infrared spectroscopic observations. *A&A*, 412, 305-316, doi: 10.1051/0004-6361:20031123.

Schenk, P., Hamilton, D. P., Johnson, R. E., McKinnon, W. B., Paranicas, C., Schmidt, J., Showalter, M. R., 2011. Plasma, plumes and rings: Saturn system dynamics as recorded in global color patterns on its midsize icy satellites. *Icarus*, 211, 740-757, doi:10.1016/j.icarus.2010.08.016.

Showalter, M.R., Cuzzi, J.N., Larson, S.M., 1991. Structure and particle properties of Saturns E ring. *Icarus* 94, 451473.

Spahn, F., and 15 colleagues, 2006. Cassini dust measurements at Enceladus and implications for the origin of the E ring. *Science*, 311 (5766): 14161418. AAAS. doi:10.1126/science.1121375.

Spencer, J. R., Barr, A. C., Esposito, L. W., Helfenstein, P., Ingersoll, A. P., Jaumann, R., McKay, C. P., Nimmo, F., Hunter Waite, J., 2009. Enceladus: an active cryovolcanic satellite, in *Saturn from Cassini-Huygens*, edited by M.K. Dougherty, L.W. Esposito, and S.M. Krimigis, Springer, Berlin, 683-724. ISBN 978-1-4020-9216-9.

Spencer, J. R., Howett, C. J., Schenk, P., Hurford, T. A., Segura, M. E., Pearl, J. C., 2010. An Unexpected Regional Thermal Anomaly on Mimas. American Astronomical Society, DPS meeting #42, #1.07; Bulletin of the American Astronomical Society, Vol. 42, 942.

Stephan, K., Jaumann, R., Wagner, R., Clark, R. N., Cruikshank, D. P., Hibbitts, C. A. Roatsch, T., Hoffmann, H., Brown, R. H., Filacchione, G., Buratti, B. J., Hansen, G. B., McCord, T. B., Nicholson, P. D., Baines, K. H., 2010. Dione' s spectral and geological properties, *Icarus*, 206, 631-652, doi = 10.1016/j.icarus.2009.07.036.

Stephan, K., Jaumann, R., Wagner, R., Clark, R. N., Cruikshank, D. P., Giese, B., Hibbitts, C. A., Roatsch, T., Matz, K.-D., Brown, R. H., Filacchione, G., Capaccioni, F., Scholten, F., Buratti, B. J., Hansen, G. B., Nicholson, P. D., Baines, K. H., Nelson, R. M., Matson, D. L., 2012. The Saturnian satellite Rhea as seen by Cassini VIMS. *Planetary and Space Science*, 61, 142-160, DOI: 10.1016/j.pss.2011.07.019.

Thekekara, M.P., 1973. Solar energy outside the Earths atmosphere. Sol. Energy 14, 109127.

Tosi, F., Turrini, D., Coradini, A., Filacchione, G., 2010. Probing the origin of the dark material on Iapetus. *Monthly Notices of the Royal Astronomical Society*, 403, 1113-1130, DOI: 10.1111/j.1365-2966.2010.16044.x

Verbiscer, A.J., Veverka, J., 1992. Mimas: Photometric roughness and albedo map. *Icarus*, 99, 6369.

Verbiscer, A. J., French, R. Showalter, M., Helfenstein, P., 2007. Enceladus: Cosmic Graffiti Artist Caught in the Act. *Science*, 315, 815, DOI: 10.1126/science.1134681.

Verbiscer, A.J., Skrutskie, M.F., Hamilton, D.P., 2009. Saturns largest ring. *Nature*, 461, 10981100. DOI: 10.1038/nature08515.

Warren, S. G., 1984. Optical constants of ice from the ultraviolet to the microwave. *Appl. Opt.* 23, 1206-1225.

Wisdom, J., Peale, S. J., Mignard, F., 1984. The chaotic rotation of Hyperion. *Icarus*, 58, 137-152.

Zubko, V. G., Mennella, V., Colangeli, L., Bussoletti, E., 1996. Optical constants of cosmic carbon analogue grains I. Simulation of clustering by a modified continuous distribution of ellipsoids. *Mon. Not. R. Astron. Soc.* 282, 1321-1329.